\documentclass[%
 nofootinbib,
 amsmath,amssymb,
 aps, physrev,
 twocolumn,
prd,
superscriptaddress
]{revtex4}

\usepackage{graphicx}
\usepackage{dcolumn}
\usepackage{bm}
\usepackage{hyperref}
\usepackage{color}    

\usepackage{enumitem}

\newcommand{\hinvmpc}{\,h^{-1}{\rm Mpc}}
\newcommand{\hmpcinv}{\,h\,{\rm Mpc^{-1}}}

\newcommand{\asiaa}{Academia Sinica Institute of Astronomy and Astrophysics (ASIAA), No.~1, Sec.~4, Roosevelt Rd., Taipei 106319, Taiwan}
\newcommand{\ipmu}{Kavli IPMU (WPI), UTIAS, The University of Tokyo, Kashiwa, Chiba 277-8583, Japan}

\begin{document}

\title{
Cosmological inference with halo clustering reconstructed \\ from the redshift-space galaxy distribution
}

\thanks{\textcopyright{} 2026 American Physical Society.
This is the accepted manuscript of the following article:
Ryuichiro Hada and Teppei Okumura,
``Cosmological inference with halo clustering reconstructed from the redshift-space galaxy distribution,''
Phys. Rev. D~\textbf{114}, 103516 (2026).
The final published version is available at
\href{https://doi.org/10.1103/4fr5-9l8j}{10.1103/4fr5-9l8j}.
This manuscript version is posted with permission for non-commercial scholarly use.}

\author{Ryuichiro Hada}
\email{rhada@asiaa.sinica.edu.tw}
\affiliation{\asiaa}

\author{Teppei Okumura}
\email{tokumura@asiaa.sinica.edu.tw}
\affiliation{\asiaa}
\affiliation{\ipmu}

\date{\today}

\begin{abstract}
Accurate modeling of small-scale redshift-space clustering is crucial for full-shape redshift-space distortion (RSD) analyses, where satellite galaxies contribute to one-halo terms and Finger-of-God distortions. We investigate halo reconstruction based on the cylinder grouping method of Okumura et al., \href{https://doi.org/10.1093/mnras/stx850}{Mon.~Not.~R.~Astron.~Soc.~\textbf{469}, 459 (2017)}, which selects an effective halo center tracer from the observed galaxy distribution, and how it impacts cosmological parameter inference.
Using luminous red galaxy mock catalogs from the AbacusSummit simulations at $z=1.1$, we perform effective field theory (EFT)-based full-shape modeling of the power spectrum of the reconstructed-halo sample.
We show that the dominant reconstruction-induced systematics can be described and incorporated within the standard EFT framework.
In particular, a simple multipole-dependent rescaling inferred directly from the data on large scales captures the dominant effect, while residual small-scale changes are absorbed by the standard counterterm and stochastic sector, without introducing additional reconstruction-specific parameters.
The reconstructed-halo sample yields unbiased constraints on cosmological parameters, including the growth rate $f\sigma_{s8}$ and Alcock-Paczynski parameters. Compared to the galaxy sample, it enables both improved robustness and increased statistical precision: at the fiducial limit of $k_{\max}=0.2\hmpcinv$, the uncertainty on $f\sigma_{s8}$ is reduced by more than $20\%$, while the high-$k_{\max}$ diagnostics show comparatively stable goodness of fit and nuisance parameters.
\end{abstract}

\maketitle


\section{Introduction} \label{sec:intro}

Redshift-space distortions (RSD) \citep{Kaiser:1987,Hamilton:1992,Cole:1994,Hamilton:1998} in galaxy clustering provide a powerful route to measuring the growth of cosmic structure, and hence to testing gravity and dark energy (e.g., \cite{Guzzo:2008,Okumura:2016,Alam:2017}) with spectroscopic surveys. 
Full-shape analyses of anisotropic galaxy clustering have become a standard route to cosmological inference from redshift-space two-point statistics. Early work already exploited the full anisotropic correlation function at the linear-theory level \citep{Okumura:2008}. More recent studies have extended these analyses using perturbative modeling of the galaxy power spectrum and emulator-based halo-model approaches \citep{dAmico:2020,Ivanov:2020,Philcox:2020,Kobayashi:2022} (see also \citep{DESI:2024VII} for the latest full-shape RSD results and a comprehensive overview).
In practice, however, extracting this information increasingly relies on modeling modes deep into the quasi-nonlinear regime, where the mapping from real to redshift space is affected by nonlinear evolution, nonlinear bias, and, most importantly for galaxy samples, the virial motions of satellite galaxies.
Satellite contamination boosts the large-scale clustering amplitude (because satellites preferentially reside in massive, highly biased halos) while simultaneously suppressing the small-scale signal through Finger-of-God (FoG) smearing \cite{Jackson:1972}.
These effects complicate the connection between observed galaxy clustering and theoretical predictions, and can limit the robustness of parameter inference when pushing to high $k_{\rm max}$ (e.g., \cite{Okumura:2015}).

A widely used strategy to extend RSD analyses into the quasi-nonlinear regime is the effective field theory (EFT) of large-scale structure and related perturbative frameworks (e.g., \cite{Carrasco:2012,Porto:2014,Vlah:2015}).
In these approaches, the impact of unresolved small-scale physics is parameterized by higher-derivative counterterms and stochastic contributions, allowing the template to remain flexible while preserving a controlled large-scale expansion.
In practice, however, this flexibility comes with an important trade-off: as smaller scales are included, the inference can become increasingly sensitive to the nuisance sector, and degeneracies among stochastic and counterterm parameters can degrade the constraining power on cosmological parameters.
Moreover, such degeneracies in a high-dimensional nuisance sector can lead to non-trivial \emph{projection effects} and increased prior/parameterization sensitivity in marginalized constraints, even when the maximum-likelihood region remains close to the true model (e.g., \cite{Handley:2019,Lemos:2021,Gomez:2022}).
This motivates complementary approaches that reduce the amplitude of problematic small-scale contributions at the level of the tracer field --- in particular, constructing tracers that more closely resemble halo centers, for which one-halo and FoG effects are substantially suppressed.

In simulations, host halos can be identified directly, but in observations the halo-center field must be inferred from the galaxy distribution itself.
This has motivated a variety of approaches aimed at mitigating satellite-driven nonlinearities and virial motions already at the level of the observed tracer field (e.g., \cite{Okumura:2015}).
Early work explored redshift-space friends-of-friends (FoF) style algorithms and related ``FoG compression'' schemes that identify high-density structures and replace member galaxies with group representatives, thereby reducing virial motions and the associated small-scale suppression (e.g., \cite{Reid:2009}).
A complementary line of work instead suppresses the contribution of the most nonlinear regions by applying density-dependent weighting or transformations of the field (e.g., ``clipping''), with the goal of extending the range of scales that can be modeled perturbatively (e.g., \cite{Simpson:2013}).
While these methods can improve the robustness of clustering analyses, their performance depends on sample properties (number density, satellite fraction, typical host-halo mass) and on the details of the adopted procedure.
In particular, redshift-space grouping and FoG-compression schemes can introduce additional line-of-sight (LOS) anisotropies through their inherently anisotropic geometry.

Instead of the FoF method, \citet{Okumura:2017} proposed a similar but simpler cylinder-grouping (CG) method, a practical halo-reconstruction scheme tailored to suppress satellite contamination in redshift space.
The algorithm ranks galaxies by a halo-mass proxy and iteratively identifies central--satellite systems by removing galaxies within a cylinder aligned with the LOS, chosen to encompass FoG displacements.
A key advantage of the CG approach is that it relies only on the observed galaxy distribution and a simple geometric criterion, making it particularly attractive for high-number-density samples.
At the same time, the LOS-aligned cylinder geometry introduces characteristic reconstruction-induced anisotropies: on small scales through an exclusion-like removal of close pairs, and on large scales through an apparent anisotropic modulation that can mimic RSD.
\citet{Okumura:2017} provided a simple semi-analytic description of these systematics,
modeling the small-scale effect with an exclusion window and the large-scale effect as a multipole-dependent rescaling of the linear power spectrum,
and showed that the reconstructed halo power-spectrum multipoles are accurately reproduced over a wide range of scales.

The goal of the present work is to take the next step and assess the practical impact of the CG reconstruction on \emph{cosmological parameter inference}.
This direction is particularly relevant in the context of ongoing and upcoming spectroscopic surveys such as the Dark Energy Spectroscopic Instrument (DESI, \cite{DESI:2016}) and the Prime Focus Spectrograph (PFS, \cite{PFS:2014}).
Specifically, we apply the CG method to galaxy mock catalogs built from cosmological $N$-body simulations and perform RSD fits in an EFT-based framework (using a Lagrangian-PT template).
Our focus is not only on whether the reconstructed clustering can be described at the level of the power-spectrum multipoles, but also whether the reconstruction-induced systematics can be handled consistently within a standard likelihood pipeline.
To this end, we organize the modeling around the two qualitative reconstruction effects highlighted by \citet{Okumura:2017}:
(i) a small-scale exclusion effect associated with cylinder grouping, whose residual impact is treated within the EFT stochastic and counterterm sector, and
(ii) a large-scale ``directional masking'' effect that primarily rescales the clustering amplitude in a multipole-dependent manner, which we correct by applying large-scale ratios measured from the mocks as a multiplicative renormalization of the large-scale deterministic part of the model.
This separation provides a simple and practical way to relate reconstruction systematics to the components of the EFT-based template used in the inference.

This paper is organized as follows.
In Sec.~\ref{sec:reconstruction} we summarize the cylinder-grouping reconstruction and discuss its small- and large-scale systematic effects.
Section~\ref{sec:mocks} describes the simulations and mock catalogs used in our tests.
In Sec.~\ref{sec:inference} we present our power spectrum modeling and fitting methodology.
Section~\ref{sec:results} presents the main results for cosmological constraints and their scale dependence, and we conclude in Sec.~\ref{sec:conclusion} with a discussion of implications and future directions.
In Appendices~\ref{app:other_params}--\ref{app:highk}, we present supplementary posterior constraints, validate the large-scale rescaling, test the sensitivity of the CG reconstruction to imposed Alcock--Paczynski distortions, and provide additional high-$k_{\max}$ diagnostics, respectively.

\section{Halo power spectrum reconstruction} \label{sec:reconstruction}

\subsection{Cylinder-grouping method} \label{subsec:CGmethod}

The Cylinder-grouping (CG) method \citep{Okumura:2017} provides a practical approach for reconstructing a halo-like field directly from the observed galaxy distribution. The primary goal of the method is to mitigate small-scale nonlinear effects in redshift space, in particular the FoG distortions produced by the virial motions of satellite galaxies within dark matter halos. In redshift space, the velocity dispersion of satellite galaxies elongates galaxy systems along the line-of-sight (LOS) direction, producing the characteristic FoG structures that strongly suppress clustering power on small scales. The CG method aims to identify these satellite systems and replace them with a single effective tracer corresponding to the host halo center. By removing most satellite galaxies, the reconstructed sample more closely traces the underlying halo population and therefore exhibits reduced small-scale nonlinearities.

A key feature of the CG method is that the grouping window is anisotropic, taking the form of a cylinder aligned with the LOS. This geometry is motivated by the physical origin of FoG distortions: satellite motions primarily produce displacements along the LOS while leaving transverse separations largely unaffected. By adopting a grouping region elongated along the LOS, the CG algorithm efficiently captures galaxies that belong to the same halo despite their apparent LOS displacements in redshift space. Importantly, the CG procedure relies only on the observed galaxy distribution, without requiring external halo catalogs or detailed dynamical modeling. Following the implementation described in \citet{Okumura:2017}, the CG reconstruction proceeds through the following steps.

\begin{enumerate}[label=(\roman*)]
  \item \emph{Ranking of galaxies.} \label{CG:ranking} 
  Galaxies are first sorted in descending order of an indicator strongly correlated with halo mass, such as galaxy luminosity, stellar mass, or an environment-based quantity such as local galaxy density. In this work, we adopt the number of neighboring galaxies within a radius of $10\hinvmpc$ as the ranking variable, which serves as a practical estimator of halo mass.

  \item \emph{Identification of central and satellite galaxies.} \label{CG:identification} 
  Starting from the highest-ranked galaxy, assumed to be the central galaxy of its host halo, we search for neighboring galaxies within a cylindrical region centered on this galaxy.  The cylinder is aligned with the LOS and characterized by a transverse radius of $R=1.5\hinvmpc$ and a LOS height of $L=20\hinvmpc$ as motivated by \citet{Okumura:2017}. All galaxies located inside this cylinder are tagged as satellite galaxies associated with the central. Both the identified central and its satellites are removed from the ranked list.

  \item \emph{Iterative grouping.} \label{CG:iteration} 
  The process described in step \ref{CG:identification} is repeated iteratively. At each iteration, the highest-ranked galaxy among the remaining objects is selected as a new central candidate, and neighboring galaxies inside the corresponding cylinder are assigned as satellites. The algorithm continues until all galaxies in the sample have been classified either as centrals or satellites.\footnote{Because both the radius used to define the environmental ranking and the grouping cylinder are specified in fiducial comoving coordinates, the resulting CG assignments can depend on the assumed coordinate mapping. We test this sensitivity under imposed Alcock--Paczynski distortions in Appendix~\ref{app:ap_distortions}.}
\end{enumerate}

The CG procedure thus yields a reconstructed halo field, consisting of the galaxies identified as centrals. Because the majority of satellite galaxies are removed, the resulting tracer population approximates the distribution of halo centers and exhibits substantially reduced FoG distortions.
In practice, however, the reconstruction is not perfect. 
The ranking variable provides only an approximate proxy for halo mass, and the grouping relies on a fixed cylindrical geometry that cannot capture the full diversity of halo environments. As a consequence, several types of imperfections can occur, including misidentification of centrals and satellites, fragmentation of groups, and merging of neighboring halos into a single reconstructed object. These effects introduce characteristic signatures in the clustering statistics of the reconstructed sample.
To quantify how these imperfections propagate into two-point statistics, we introduce in the following subsection a decomposition of the tracer field and a set of CG-specific labels that track the contributions of correctly and incorrectly identified galaxies.

\subsection{Tracer decomposition and CG labels}
\label{subsec:decomp}

Following the notation introduced in \citet{Okumura:2017}, we begin by decomposing the original galaxy sample into true centrals and satellite populations,
\begin{equation}
N_g = N_c + N_s,
\label{eq:Ng_decomp}
\end{equation}
where $N_g$, $N_c$, and $N_s$ denote the numbers of galaxies, true centrals, and true satellites, respectively. 
Applying the CG procedure yields reconstructed (post-CG) subsamples. To distinguish the reconstructed subsamples from the original populations, we denote post-CG quantities by a tilde, $\tilde Q$, for any quantity $Q$.
In particular, we define $\tilde Q_c$ and $\tilde Q_s$ as objects \emph{retained} as centrals and \emph{removed} as satellites after CG, respectively.
Because the reconstruction is not perfect, we further introduce two misidentification classes defined after applying CG:
$\tilde Q_{\bar c}$ (true centrals incorrectly removed) and $\tilde Q_{\bar s}$ (true satellites incorrectly retained as centrals).
These definitions lead to simple counting identities for the reconstructed sample. The number of objects retained as centrals and the number of removed objects are, respectively, 
\begin{equation}
\tilde N_c = N_c - \tilde N_{\bar c} + \tilde N_{\bar s},\qquad
\tilde N_s = N_s + \tilde N_{\bar c} - \tilde N_{\bar s}.
\label{eq:counting_identities}
\end{equation}
These relations imply $\tilde N_c + \tilde N_s = N_g$ by construction. 
We define (global) composition ratios relative to the true number of centrals,
$\tilde a_X \equiv \tilde N_X/N_c$, so that Eq.~\eqref{eq:counting_identities} implies
\begin{equation}
\tilde a_c = 1 - \tilde a_{\bar c} + \tilde a_{\bar s}.
\label{eq:ac_relation}
\end{equation}

To connect these labels to clustering statistics, we associate with each subsample $A$ an overdensity field,
$\delta_A(\mathbf{x}) \equiv n_A(\mathbf{x})/\bar n_A - 1$, and define the corresponding two-point correlation function, $\xi_{AB}(\mathbf{r}) \equiv \langle \delta_A(\mathbf{x})\,\delta_B(\mathbf{x}+\mathbf{r})\rangle$, and its Fourier counterpart, the power spectrum, $P_{AB}(\mathbf{k})$. 
Since the cosmological analysis in the following is based on redshift-space power-spectrum multipoles, it is useful to also introduce the multipole moments of the anisotropic correlation function,
\begin{align}
    \xi_{AB,\ell}(r) = \frac{2\ell+1}{2}\int_{-1}^{1} d\mu_r\, \xi_{AB}(r,\mu_r)\,\mathcal{L}_\ell(\mu_r),
\end{align}
and the corresponding power-spectrum multipoles,
\begin{align}
    P_{AB,\ell}(k) = 4\pi (-i)^\ell \int_{0}^{\infty} dr\, r^2\, \xi_{AB,\ell}(r)\, j_\ell(kr),
\end{align}
where $r \equiv |\mathbf{r}|$ and $\mu_r \equiv \hat{\mathbf{r}}\cdot\hat{\mathbf{n}}$ denotes the cosine of the angle between the separation vector and the LOS. Here $\mathcal{L}_\ell$ is the Legendre polynomial of order $\ell$ and $j_\ell$ is the spherical Bessel function. In what follows, we use the full anisotropic correlation function for schematic decompositions, while the quantitative analysis is carried out in terms of the corresponding power-spectrum multipoles.
Using the composition ratios defined above, the reconstructed autocorrelation function can be expanded as
\begin{align}
\tilde a_c^2\,\tilde\xi_{cc}(\mathbf{r})
&=
\xi_{cc}(\mathbf{r})
+ \tilde a_{\bar c}^2\,\tilde\xi_{\bar c\bar c}(\mathbf{r})
+ \tilde a_{\bar s}^2\,\tilde\xi_{\bar s\bar s}(\mathbf{r})
\nonumber \\
&- 2 \tilde a_{\bar c}\,\tilde\xi_{c\bar c}(\mathbf{r})
+ 2 \tilde a_{\bar s}\,\tilde\xi_{c\bar s}(\mathbf{r})
- 2\tilde a_{\bar c} \tilde a_{\bar s}\,\tilde\xi_{\bar c\bar s}(\mathbf{r}).
\label{eq:xi_expand}
\end{align}
The leading contribution in this expression is the correlation of true central galaxies, $\xi_{cc}(\mathbf{r})$, which represents the clustering signal that the reconstruction aims to recover. 
Hereafter, it is referred to as the baseline term. 
The remaining terms encode the effect of reconstruction imperfections arising from central–satellite misidentification.

To isolate the characteristic scale dependence of these corrections, it is useful to split the $c$--$\bar c$ cross-correlation term, $\tilde\xi_{c\bar c}$, into contributions from separations inside and outside the grouping cylinder,
\begin{equation}
\tilde\xi_{c\bar c}(\mathbf{r})=\tilde\xi^{<{\rm cyl}}_{c\bar c}(\mathbf{r})+\tilde\xi^{>{\rm cyl}}_{c\bar c}(\mathbf{r}),
\label{eq:xi_split_cyl}
\end{equation}
which yields a schematic ``roadmap'' for the reconstructed correlation function, 
\begin{align}
\tilde a_c^2\,\tilde\xi_{cc}(\mathbf{r})
&=
\underbrace{\xi_{cc}(\mathbf{r})}_{\rm baseline}
\ \ \underbrace{
\;-\;
2 \tilde a_{\bar c}\,\tilde\xi^{<{\rm cyl}}_{c\bar c}(\mathbf{r})}_{\text{(\ref{subsec:exclusion}) exclusion}}
\nonumber \\
&
\underbrace{
\;-\;
2 \tilde a_{\bar c}\,\tilde\xi^{>{\rm cyl}}_{c\bar c}(\mathbf{r})
\;+\;
2 \tilde a_{\bar s}\,\tilde\xi_{c\bar s}(\mathbf{r})
\;+\;
\tilde\xi_{\rm quad}(\mathbf{r})}_{\text{(\ref{subsec:masking}) masking}}.
\label{eq:roadmap}
\end{align}
Here $\tilde\xi_{\rm quad}(\mathbf{r})$ collects the quadratic mis-identification terms involving
$\tilde\xi_{\bar c\bar c}$, $\tilde\xi_{\bar s\bar s}$, and $\tilde\xi_{\bar c\bar s}$.

Equation~\eqref{eq:roadmap} highlights that CG imperfections affect the clustering signal through two qualitatively different channels, separated by the cylinder scale.
The $\tilde\xi^{<{\rm cyl}}_{c\bar c}$ term corresponds to pairs removed inside the grouping cylinder. This represents an additional short-range suppression induced by the CG algorithm. We refer to this contribution as the ``exclusion'' effect, discussed in detail in Sec.~\ref{subsec:exclusion}.

The remaining contributions, $-2\tilde a_{\bar c}\tilde\xi^{>{\rm cyl}}_{c\bar c}$, $2\tilde a_{\bar s}\tilde\xi_{c\bar s}$, and $\tilde\xi_{\rm quad}$, primarily affect larger separations and can generate a smooth, potentially line-of-sight--anisotropic response, which we refer to as ``masking'' and discuss in Sec.~\ref{subsec:masking}.
It is useful to comment on the terms entering this masking category.
By construction, $\tilde\xi_{c\bar s}$ pairs true centrals with satellites that survive the grouping, so such pairs are expected to lie predominantly outside the cylinder scale relevant for exclusion.\footnote{Strictly, $\tilde\xi_{c\bar s}$ also includes pairs between removed true centrals and retained satellites. Their contribution is proportional to $\tilde a_{\bar c}\tilde a_{\bar s}$ and is therefore expected to be small; the subset arising within a single halo is expected to be further suppressed.}
Moreover, the quadratic mis-identification contributions collected in $\tilde\xi_{\rm quad}$ are expected to be dominated by inter-halo pairs. 
The misidentification factors are expected to be small, $\tilde a_{\bar c},\tilde a_{\bar s}\ll \tilde a_c$ \cite{Okumura:2017}, so multiple mis-identification events within a single halo should be further suppressed.
We therefore treat these terms as contributing primarily to the large-scale response, while any residual short-scale pieces are absorbed by the standard EFT stochastic and counterterm sectors in our later modeling.

In the following subsections, we describe these two effects and how we incorporate them in our modeling.
While \citet{Okumura:2017} identified the physical origin of both systematics and derived a theoretical description relating the reconstructed and true-halo power spectra, here we build on their formulation and provide an implementation tailored to our EFT-based RSD template.
Specifically, we express the reconstructed power spectrum in terms of the true-halo contribution (i.e., $\xi_{cc}$) together with a controlled set of CG-induced corrections, which enables a transparent interpretation of the fitted EFT and cosmological parameters.
We then account for the large-scale anisotropic response using multipole ratios that are directly measurable from the catalogs themselves, avoiding the introduction of additional free functions.

\subsection{Small-scale effect: Exclusion}
\label{subsec:exclusion}

In this subsection, we focus on the ``exclusion'' contribution identified in Eq.~\eqref{eq:roadmap}, namely the term proportional to $\tilde\xi^{<{\rm cyl}}_{c\bar c}$, which captures the small-scale suppression of pairs induced by the CG procedure.
This effect originates from imperfect grouping: when a true central galaxy is misidentified and removed as a satellite, pairs that would otherwise contribute to the baseline central–central clustering, $\xi_{cc}$, are artificially eliminated.
We describe this reconstruction-induced \emph{additional} exclusion (e.g., \cite{Baldauf:2013,vdBosch:2013}) as an effective hard-core window acting on the dominant baseline term $\xi_{cc}$.
The discussion here corresponds to the first line of Eq.~\eqref{eq:roadmap}, where the dominant baseline term $\xi_{cc}$ is modified by an additional close-pair loss induced by the CG procedure.\footnote{
Real halos already exhibit physical halo exclusion at small separations (e.g., \cite{Baldauf:2013}).
The scale $R_{\rm excl}$ introduced here is an \emph{effective} radius meant to idealize the \emph{additional} close-pair loss induced by imperfect CG grouping, rather than the physical exclusion scale of the underlying halo population.
}

To make the underlying mechanism transparent, we first consider a simplified isotropic toy model in real space. 
In this case, the reconstructed correlation function can be modeled using the effective exclusion radius $R_{\rm excl}$ as
\begin{eqnarray}
  \tilde a_c^2\tilde\xi_{cc}(r)
  \simeq
  \begin{cases}
    -1, & r < R_{\rm excl}, \\[4pt]
    \xi_{cc}(r), & r \ge R_{\rm excl}.
  \end{cases}
  \label{eq:xi_excl_piecewise}
\end{eqnarray}
The equation states that pairs within $R_{\rm excl}$ are completely removed, whereas pairs at larger separations follow the baseline true-central correlation.\footnote{We use the symbol ``$\simeq$’’ in Eq.\eqref{eq:xi_excl_piecewise} to emphasize that this is a schematic effective description of the additional close-pair loss, retaining only the baseline and exclusion terms in the first line of Eq.~(\ref{eq:roadmap}). 
The remaining terms in the second line are treated separately in Sec.~\ref{subsec:masking}.}
In this simplified spherical case, the additional pair removal acts as a sharp top-hat mask in configuration space.

Now we generalize this below to the anisotropic cylindrical geometry relevant for CG reconstruction in redshift space. Introducing an anisotropic window function, $\Theta(\mathbf{r})$, we write the reconstructed anisotropic correlation function as 
\begin{eqnarray}
  \tilde a_c^2\tilde\xi_{cc}(\mathbf{r})
  \simeq - \Theta(\mathbf{r}) + \left[1 - \Theta(\mathbf{r})\right]\,\xi_{cc}(\mathbf{r}) .
  \label{eq:xi_excl_window}
\end{eqnarray}
If we ignore RSD and adopt a Heaviside step function $H(x)$ as the window function, $\Theta(\mathbf{r}) \equiv H\!\left(R_{\rm excl}-|\mathbf{r}|\right)$, this coincides with Eq.~\eqref{eq:xi_excl_piecewise}.
The first term of Eq.~(\ref{eq:xi_excl_window}) represents the ``pure'' exclusion contribution (corresponding to $\xi=-1$ inside the exclusion region), while the second term multiplies the baseline correlation by a mask that removes contributions from pairs inside the exclusion region.

Fourier-transforming Eq.~\eqref{eq:xi_excl_window}, we obtain
\begin{eqnarray}
  \tilde a_c^2\tilde P_{cc}(\mathbf{k})
  &\simeq& P_{cc}(\mathbf{k}) + \Delta P_{\rm excl}(\mathbf{k}),
  \label{eq:Pk_excl_master}
\end{eqnarray}
with
\begin{equation}
\Delta P_{\rm excl}(\mathbf{k})\equiv
-\,V_{\rm excl}\,W(\mathbf{k})\;-\;V_{\rm excl}\,[\,W*P_{cc}\,](\mathbf{k}).
\label{eq:DeltaPexcl_def}
\end{equation}
Here $P_{cc}$ and $\tilde P_{cc}$ are the power spectra associated with $\xi_{cc}$ and $\tilde\xi_{cc}$, respectively, $V_{\rm excl} = \int d\mathbf{r}\,\Theta(\mathbf{r})$ is the exclusion volume, $W(\mathbf{k})$ is the normalized Fourier transform of the exclusion window, and ``$*$'' denotes convolution in Fourier space.
Equation~\eqref{eq:DeltaPexcl_def} shows that the exclusion correction, $\Delta P_{\rm excl}$, naturally separates into two pieces: the first term, $-V_{\rm excl}W$, is the ``pure'' exclusion contribution corresponding to $\xi=-1$ inside the exclusion region, while the second term, $-V_{\rm excl}[W*P_{cc}]$, describes how the exclusion window removes the part of the baseline power sourced by pairs within the excluded region.

In the CG reconstruction, the exclusion region is cylindrical, aligned with the LOS direction.
This reflects the fact that FoG distortions are LOS-directed.
Adopting a cylinder of transverse radius $R_\perp$ and LOS length $L_\parallel$, the exclusion volume is
\begin{eqnarray}
  V_{\rm excl} = \pi R_\perp^2 L_\parallel,
\end{eqnarray}
and the normalized Fourier-space window takes the form
\begin{eqnarray}
  W(k,\mu)
  = \frac{2 J_1(k_\perp R_\perp)}{k_\perp R_\perp}\,
    \frac{\sin(k_\parallel L_\parallel/2)}{k_\parallel L_\parallel/2},
  \label{eq:WR_cyl_exact}
\end{eqnarray}
where $\mu \equiv \hat{\mathbf{k}}\cdot\hat{\mathbf{n}}$
denotes the cosine of the angle between the wavevector and the LOS, $k_\parallel = k\mu$ and $k_\perp = k\sqrt{1-\mu^2}$ are the LOS and transverse components of $\mathbf{k}$, and $J_1$ is the Bessel function of the first kind. 

It is useful to briefly comment on the expected scale dependence of the exclusion correction $\Delta P_{\rm excl}$, following the discussion in \citet{Baldauf:2013}.
For the isotropic (spherical) exclusion model with a top-hat window, the normalized Fourier-space window satisfies
$W(k)\to 1$ as $k\to 0$, with the leading departure scaling as $W(k)=1+\mathcal{O}(k^2R_{\rm excl}^2)$.
Moreover, the convolution term $[W*P]$ has the generic interpretation of a local (in $k$-space) smoothing of an underlying power spectrum $P$ over a bandwidth $\Delta k\sim 1/R_{\rm excl}$.
As a result, in the low-$k$ limit the convolution approaches a constant plus corrections of order $k^2R_{\rm excl}^2$ set by moments of the underlying power spectrum.
For the cylindrical CG window considered here, the same qualitative behavior is expected, except that the leading $k^2$ corrections are anisotropic and can be organized into $k_\perp^2$ and $k_\parallel^2$ (equivalently $\mu$-dependent) pieces.
Consequently, on sufficiently large scales the leading impact of $\Delta P_{\rm excl}$ is smooth and has the form of stochastic and higher-derivative contributions, which are expected to be largely absorbed by the standard EFT stochastic and counterterm sectors (see Secs.~\ref{subsec:velocileptors} and \ref{subsec:incorp_recon_sys}).

\subsection{Large-scale effect: Directional masking}
\label{subsec:masking}
We now turn to the remaining terms in Eq.~\eqref{eq:roadmap},
namely $\tilde\xi^{>{\rm cyl}}_{c\bar c}$, $\tilde\xi_{c\bar s}$, and $\tilde\xi_{\rm quad}$,
which together describe the large-scale effect of imperfect CG grouping.
This contribution arises whenever the reconstruction is not perfect: central--satellite misidentification and group merging/fragmentation correlate the reconstructed field with the underlying density in a way that leaves a residual imprint on the two-point statistics even on scales much larger than the exclusion region.
Such a term would be non-zero irrespective of whether the grouping geometry is isotropic or anisotropic (an isotropic, real-space grouping would yield an isotropic response), but the LOS-aligned cylindrical grouping used in CG makes the residual response anisotropic.
Following \citet{Okumura:2017}, we interpret this as an apparent anisotropic clustering induced by misidentification in the CG grouping, and refer to it as \emph{directional masking}.

Physically, directional masking reflects an environment- and direction-dependent selection of tracers caused by imperfect grouping.
Because the misidentification probability depends on the local density, the reconstructed tracer abundance is biased toward specific environments, which on large scales manifests as an overall change in the clustering amplitude (i.e., an effective modification of the large-scale bias).
In redshift space, the LOS-dependent nature of the grouping makes this bias change orientation-dependent, thereby inducing a smooth $\mu$-dependence.
As a result, on sufficiently large scales the effect is well captured as a multipole-dependent rescaling of the clustering amplitude, without introducing new scale-dependent features.

\citet{Okumura:2017} have shown that this apparent large-scale anisotropy can be accurately absorbed into an effective rescaling of the linear power spectrum.
Specifically, on sufficiently large scales, the reconstructed power spectrum can be corrected by
\begin{eqnarray}
\Delta P(\mathbf{k}) = \alpha(\mu)\, P_{\rm lin}(k),
\label{eq:directional_masking_scaling}
\end{eqnarray}
where $\alpha(\mu)$ encapsulates the anisotropic large-scale response associated with the imperfect CG reconstruction.
Since this effect reflects a large-scale, orientation-dependent modification of the tracer bias, the angular dependence of $\alpha(\mu)$ is expected to be smooth and dominated by the lowest even multipoles.

In this work, we implement this correction in a data-driven manner.
Rather than introducing additional nuisance parameters for $\alpha(\mu)$, we characterize the large-scale response by the multipole ratios
\begin{equation}
R^{\rm obs}_\ell \equiv
\left.\frac{\tilde P_{cc,\ell}(k)}{P_{gg,\ell}(k)}\right|_{k \to 0}, \label{eq:ratio_R}
\end{equation}
defined from the reconstructed halo and parent-galaxy power-spectrum multipoles.
Both $\tilde P_{cc,\ell}$ and $P_{gg,\ell}$ are directly measurable from the reconstructed and parent galaxy catalogs, respectively;
in this analysis we estimate $R^{\rm obs}_\ell$ from mocks in a controlled setting, but the same procedure can in principle be applied directly to survey data.
Since $R^{\rm obs}_\ell$ is nearly scale-independent at low $k$, we define a constant normalization for each multipole by averaging $R^{\rm obs}_\ell$ over the largest scales used in the analysis, $k < 0.03\,h\,{\rm Mpc}^{-1}$.
We then apply these constants as multiplicative normalizations to the corresponding theoretical model multipoles, effectively absorbing directional masking into a redefinition of the large-scale clustering amplitude without modifying the scale dependence governed by perturbation theory.

\section{Mock Catalog} \label{sec:mocks}

To test the performance of halo power spectrum reconstruction and to assess its impact on redshift-space clustering and cosmological parameter inference, we construct mock galaxy catalogs based on high-resolution cosmological $N$-body simulations. Our mock setup is designed to closely resemble DESI-like galaxy samples while retaining full control over the underlying halo population, enabling a clean comparison between galaxy clustering, true halo clustering, and reconstructed halo clustering within a unified framework.

\subsection{N-body simulations}

We make use of the \textsc{AbacusSummit}\footnote{\url{https://abacussummit.readthedocs.io}} suite of cosmological $N$-body simulations \cite{Maksimova:2021}, which provides accurate nonlinear evolution of dark matter in large periodic volumes. From these simulations, we extract halo catalogs at redshift $z = 1.1$. 
The simulations are run in cubic volumes of $(2\,h^{-1}\mathrm{Gpc})^{3}$ with a fixed fiducial cosmology, $h = 0.6736$, $\omega_b = 0.02237$, $\omega_{\rm cdm} = 0.12$, $A_s = 2.0830 \times 10^{-9}$, and $n_s = 0.9649$,
corresponding to a sound horizon at the baryon drag epoch of $r_d = 99.08\,h^{-1}\mathrm{Mpc}$. 
We analyze an ensemble of 25 independent realizations with different initial conditions, which allows clustering statistics to be measured over a wide range of scales and significantly reduces sample variance when averaged over realizations.

Throughout this work, we analyze individual snapshots in a periodic box and neglect light-cone and redshift-evolution effects, in order to isolate nonlinear dynamics and reconstruction-induced systematics from observational complications.

\subsection{Halo occupation distribution}

Dark matter halos are populated with galaxies using a halo occupation distribution (HOD) model calibrated from hydrodynamical simulations. Specifically, we adopt HOD parameters inferred for DESI-like luminous red galaxy (LRG) samples using the \textsc{IllustrisTNG} simulations \cite{Yuan:2022}. In that work, DESI target selection is applied directly to simulated galaxies, and the resulting galaxy populations are matched to their host halos in the corresponding dark-matter-only simulation, yielding an HOD model that incorporates realistic baryonic effects and satellite populations.

For LRGs, the mean occupation numbers of central and satellite galaxies in a halo with mass $M$ are modeled following the five-parameter baseline HOD introduced by \citet{Zheng:2007},
\begin{eqnarray}
\langle N_{\mathrm{cen}}(M) \rangle 
    &=&\frac{f_{\mathrm{ic}}}{2}\,
\mathrm{erfc}\!\left[
\frac{\log_{10}(M_{\mathrm{cut}}/M)}{\sqrt{2}\,\sigma}
\right], 
    \nonumber \\
\langle N_{\mathrm{sat}}(M) \rangle 
    &=&\left( \frac{M - \kappa M_{\mathrm{cut}}}{M_{1}} \right)^{\alpha}
\langle N_{\mathrm{cen}}(M) \rangle ,
\end{eqnarray}
where
$M_{\mathrm{cut}}$ sets the minimum halo mass for hosting a central galaxy, $M_{1}$ characterizes the typical halo mass hosting one satellite galaxy, $\sigma$ controls the width of the central transition, $\alpha$ is the satellite power-law slope, and $\kappa M_{\mathrm{cut}}$ defines the minimum halo mass for hosting satellites. The incompleteness parameter $f_{\mathrm{ic}}$ modulates the overall number density of the sample.

Following the fiducial values inferred from the \textsc{IllustrisTNG}-based DESI LRG mocks \cite{Yuan:2022}, we adopt
\begin{eqnarray}
&&\log_{10} M_{\mathrm{cut}} = 12.7, \quad
\log_{10} M_{1} = 13.6, \quad
    \nonumber \\
&&\sigma = 0.2, \quad
\alpha = 1.15, \quad
\kappa = 0.08, \quad
f_{\mathrm{ic}} = 0.8 .
\end{eqnarray}
These parameters correspond to a mean halo mass of $\bar{M}_{\mathrm{h}} \simeq 2.4 \times 10^{13}\,h^{-1}M_{\odot}$ and a satellite fraction of approximately $20\%$ in the calibration simulation. 
The actual satellite fraction in the AbacusSummit mocks analyzed here is reported in Sec.~\ref{subsec:catalog_number}.

In this work, we apply this HOD model to the \textsc{AbacusSummit} halo catalogs at $z = 1.1$. Although the HOD parameters are calibrated at $z \simeq 0.8$, this modest redshift mismatch does not affect our primary goal, which is to assess the relative impact of halo reconstruction on clustering statistics and cosmological inference under controlled conditions.

\begin{table}[t]
\centering
\caption{
Summary of the three tracer catalogs at $z = 1.1$ used in this work.
For each sample, $\bar n$ and $\langle N\rangle$ denote the mean number density and the mean number of objects across the 25 realizations, respectively. Each simulation volume is
$(2\,h^{-1}\mathrm{Gpc})^{3}$.
}
\label{tab:tracer_catalogs}
\vspace{15pt}
\begin{tabular}{lcc}
\hline\hline
Tracer
& $10^{4}\bar n$
& $\langle N\rangle/10^{6}$ \\
& $(h\,\mathrm{Mpc}^{-1})^{3}$
& \\
\hline
Galaxies            & 5.11 & 4.09 \\
Host halos          & 4.49 & 3.59 \\
Reconstructed halos & 3.82 & 3.05 \\
\hline\hline
\end{tabular}
\renewcommand{\arraystretch}{1.0}
\setlength{\tabcolsep}{6pt}
\end{table}

\subsection{Tracer catalogs and number densities} \label{subsec:catalog_number}

For each simulation realization, we construct three tracer catalogs that are used throughout this analysis:
(i) a galaxy catalog populated via the HOD model,
(ii) a host-halo catalog containing all dark matter halos that host at least one galaxy, and
(iii) a reconstructed halo catalog obtained by applying the cylinder-grouping (CG) algorithm to the galaxy distribution. 
The galaxy and reconstructed-halo catalogs correspond directly to the total galaxy population $N_g$ and the reconstructed-central population $\widetilde N_c$, respectively. The host-halo catalog instead provides a satellite-free reference for halo-center clustering. It closely traces, but is not strictly identical to, the true-central galaxy population $N_c$, because the selections of host halos and true central galaxies are defined at the halo and galaxy levels, respectively.

For notational continuity with Sec.~\ref{subsec:decomp}, we denote the corresponding number densities by
$n_g$ for galaxies, $n_c$ for host halos, and $\tilde n_c$ for reconstructed halos. These quantities are measured directly from the mock catalogs. Their mean number densities, together with the corresponding mean numbers of objects per simulation box, are summarized in Table~\ref{tab:tracer_catalogs}.
The reduction in number density from $n_{c}$ to $\tilde n_{c}$ reflects the fact that the CG reconstruction operates on the galaxy distribution and does not recover all host halos. 
In particular, close galaxy systems may be grouped into a single reconstructed object, and some true centrals may be missed due to the imperfect identification of central galaxies inherent to the CG algorithm. These three tracer populations are analyzed in parallel in the following sections to isolate the impact of halo reconstruction on redshift-space clustering and cosmological parameter inference.

Because the mock catalogs retain the true central--satellite labels, we directly quantify the performance of the reconstruction. We define the central purity as the fraction of reconstructed objects that are true centrals and the central completeness as the fraction of input true centrals retained by the reconstruction. Across the 25 AbacusSummit realizations, the input satellite fraction is $14.805\%\pm0.021\%$, while the reconstructed sample has a residual satellite fraction of $8.511\%\pm0.087\%$. The corresponding central purity and completeness are $91.489\%\pm0.087\%$ and $80.157\%\pm0.074\%$, respectively, and the reconstruction removes $57.089\%\pm0.414\%$ of the input satellites. At the host level, almost all reconstructed objects represent distinct occupied host halos: the number of distinct represented hosts is $99.291\pm0.005\%$ of the total number of reconstructed objects. Consistently, only $0.555\pm0.004\%$ of the occupied host halos are split into multiple reconstructed objects.
Here, the uncertainties denote the standard deviations among the realizations.

\section{Cosmological inference} \label{sec:inference}

In this section, we describe the theoretical modeling and fitting methodology used to quantify the impact of halo reconstruction on redshift-space clustering.
Our analysis closely follows the framework developed in \citet{Maus:2025}, who performed a systematic comparison of full-modeling approaches using the {\tt velocileptors} perturbation theory code.
In the present work, we adopt a similar setup but extend it to the reconstructed-halo field, with particular emphasis on (i) the treatment of reconstruction-induced effects within the EFT template and (ii) the stability of the parameter inference.

\subsection{Power spectrum measurements, data vector, and covariance}  \label{subsec:power_measure}

The redshift-space power spectrum multipoles used as the data vector are measured using a grid-based fast Fourier transform (FFT) estimator implemented in \texttt{nbodykit}.\footnote{\url{https://nbodykit.readthedocs.io}}
In this analysis, we focus on the monopole and quadrupole moments ($\ell = 0, 2$), which capture the dominant information on the clustering amplitude and redshift-space anisotropy.
For the monopole, we subtract the Poisson shot-noise contribution, $1/\bar{n}$, from the measured power spectrum.
Any residual non-Poissonian contribution is absorbed into the stochastic sector of the EFT model (see Sec.~\ref{subsec:velocileptors}).

To characterize the large-scale anisotropic response induced by the reconstruction, we directly measure the multipole-dependent ratios $R^{\rm obs}_\ell$ introduced in Sec.~\ref{subsec:masking}, which quantify the relative amplitude of reconstructed and original clustering in each multipole. 
These ratios are estimated using the two-point correlation function multipoles with \texttt{pycorr},\footnote{\url{https://py2pcf.readthedocs.io}} which are then Fourier transformed to obtain the corresponding power-spectrum multipoles.
For each realization, we determine a constant $R_\ell^{\mathrm{obs}}$ from the four wavenumber bins at $k<0.03\hmpcinv$, which is used only as the normalization factor for the corresponding theoretical model multipole.
The use of configuration-space estimators mitigates discretization effects that can affect the lowest-$k$ modes. It provides a cleaner estimate of the large-scale deterministic response by avoiding the explicit shot-noise subtraction.

To reduce sample variance and isolate systematic trends, the data vector used for parameter inference is constructed as the mean of the power-spectrum multipoles over 25 independent N-body realizations.
Unless otherwise stated, all results presented in this work are based on these averaged measurements.

The covariance matrix is estimated directly from the scatter among the 25 realizations.
In this work, we adopt a covariance matrix that is block diagonal in $k$: correlations between distinct $k$ bins are neglected, while the full $2\times2$ covariance of $(P_0,P_2)$ is retained within each $k$ bin.
As in \citet{Maus:2025}, we do not rescale the covariance by the number of realizations. This choice avoids unrealistically tight constraints that may be sensitive to residual theoretical modeling uncertainties (see Section~2 of that work for details).

\subsection{Modeling with {\tt velocileptors}} \label{subsec:velocileptors}

We model redshift-space power spectrum multipoles using the {\tt velocileptors}\footnote{\url{https://github.com/sfschen/velocileptors}} code \citep{Chen:2020,Chen:2021}, which is based on Lagrangian perturbation theory (LPT).
In this framework, the redshift-space galaxy (or halo) power spectrum is written as a sum of perturbative contributions describing nonlinear gravitational evolution, biasing, and redshift-space distortions, together with effective field theory (EFT) counterterms and stochastic contributions \citep{Carrasco:2012,Porto:2014,Vlah:2015}.

We denote the Lagrangian bias parameters entering the model template by $b_{n\mathrm{L}}$ and the Eulerian linear bias by $b_{1\mathrm{E}}=1+b_{1\mathrm{L}}$. The redshift-space power spectrum with the EFT parametrization adopted in \citet{Maus:2025} is given by
\begin{eqnarray}
P(k,\mu)&=&
P^{\mathrm{LPT}}(k,\mu)
\nonumber \\
&& \hspace{-4em} + (b_{1\mathrm{E}}+f\mu^2)
\left(
b_{1\mathrm{E}}\,\alpha_{0\mathrm{p}}
+ f\alpha_{2\mathrm{p}}\,\mu^2
+ f\alpha_{4\mathrm{p}}\,\mu^4
\right)
k^2 P_{b_{1\mathrm{L}}^2}(k)
\nonumber \\
&& \hspace{-4em} + \left(
\mathrm{SN}_0
+ \mathrm{SN}_2\,k^2\mu^2
+ \mathrm{SN}_4\,k^4\mu^4 
\right),
\label{eq:velocileptors_master}
\end{eqnarray}
where $P^{\mathrm{LPT}}(k,\mu)$ denotes the one-loop LPT prediction including nonlinear bias and redshift-space distortions, and $f$ is the linear growth rate.
The term $P_{b_{1\mathrm{L}}^2}(k)$ is the tree-level redshift-space basis spectrum associated with the $b_{1\mathrm{L}}^2\,\xi_{\rm lin}$ term in the LPT expansion, evaluated outside the exponential in the IR-resummed expression.
It is therefore closely related to the linear power spectrum, but is not identical to $P_{\rm lin}(k)$, as it retains the isotropic IR-resummation structure of the LPT framework.
The coefficients $\alpha_{n\mathrm{p}}$ parametrize higher-derivative EFT counterterms that describe fractional corrections to this leading-order contribution.
The subscript ``$\mathrm{p}$'' labels the parametrization in terms of fractional corrections adopted in \citet{Maus:2025} and distinguishes it from the original counterterm coefficients $\alpha_n$ used in the default \texttt{velocileptors} parametrization \citep{Chen:2021}. The relation between the two parametrizations is summarized in a notebook provided in the \texttt{velocileptors} repository.\footnote{\url{https://github.com/sfschen/velocileptors/blob/master/notebooks/LPT_RSD_Optimal_Settings.ipynb}}
The stochastic terms $\mathrm{SN}_n$ capture shot noise and other short-scale stochastic effects.
This flexible but physically motivated framework allows us to model nonlinear redshift-space clustering up to moderately high wavenumbers while marginalizing over small-scale uncertainties in a controlled manner.

\subsection{Fitting methodology: ShapeFit}

To quantify the impact of halo reconstruction on the broadband shape and amplitude of the redshift-space power spectrum, we adopt the ShapeFit parameter-compression framework introduced by \citet{Brieden:2021}.
Rather than directly fitting a full cosmological model (“full modeling”), ShapeFit compresses the information contained in the power spectrum multipoles into a small set of physically interpretable parameters that capture late-time geometry, growth, and deviations in the power-spectrum shape relative to a fiducial cosmology.

In the ShapeFit framework, Alcock--Paczynski (AP) distortions \cite{Alcock:1979} are parametrized by the scaling parameters \cite{Ballinger:1996}
\begin{eqnarray}
q_{\parallel} \equiv \frac{H^{\rm ref}(z)}{H(z)}, \qquad
q_{\perp} \equiv \frac{D_A(z)}{D_A^{\rm ref}(z)},
\end{eqnarray}
which rescale the LOS and transverse components of the wavevector, respectively.
These parameters control anisotropic distortions of the power spectrum arising from the mismatch between the true and fiducial distance--redshift relations.

Redshift-space distortions are captured by the compressed amplitude parameter $f\sigma_{s8}$, the product of the linear growth rate $f$ and the normalization of matter fluctuations $\sigma_{s8}$, where
\begin{eqnarray}
\sigma_{s8}\equiv \sigma\!\left(R=8s\hinvmpc, z\right),
\quad
s\equiv \frac{r_d}{r_d^{\mathrm{ref}}}.
\end{eqnarray}
This $f\sigma_{s8}$ parameter primarily controls the relative amplitude of the monopole and quadrupole moments of the power spectrum.
The quantity $\sigma_{s8}$ should be distinguished from the conventional $\sigma_8$, defined at $R=8\hinvmpc$. At $s=1$, as is the case for the fiducial cosmology of the mock catalogs analyzed here, $\sigma_{s8}=\sigma_8$.

To allow for controlled deviations in the broadband shape of the linear power spectrum, ShapeFit introduces additional shape parameters that modify the fiducial linear power spectrum according to
\begin{eqnarray}
P_{\rm lin}(k) = P_{\rm lin}^{\rm ref}(k)\,
\exp\!\left[
\frac{m}{a}\,
\tanh\!\left(
a \ln \frac{k}{k_p}
\right)
+ n \ln \frac{k}{k_p}
\right],
\nonumber \\
\end{eqnarray}
where $P_{\rm lin}^{\rm ref}(k)$ is the linear power spectrum of the fiducial cosmology, and $(a, k_p)$ are fixed constants chosen to mimic the effect of varying early-universe parameters such as $\omega_m$, $\omega_b$, and $n_s$.
The parameter $m$ controls smooth, scale-dependent modifications of the broadband shape, while the parameter $n$ captures an overall tilt-like distortion. 
We adopt the standard ShapeFit choices $a = 0.6$ and $k_p = 0.03\,h\,\mathrm{Mpc}^{-1}$ \citep{Brieden:2021}.
For simplicity, we fix $n=0$ as in \citet{Maus:2025}$,$ since the associated spectral-index freedom (e.g., $n_s$) is not considered here, and retain $m$ as the only broadband-shape degree of freedom.
The modified linear power spectrum $P_{\rm lin}(k)$ is then used as the input to the {\tt velocileptors} framework to compute the full one-loop redshift-space power spectrum prediction for a given set of ShapeFit parameters $(q_{\parallel}, q_{\perp}, m, f\sigma_{s8})$.

\subsection{Incorporating reconstruction systematics}
\label{subsec:incorp_recon_sys}

The reconstruction-induced systematics described in Secs.~\ref{subsec:exclusion} and \ref{subsec:masking} affect distinct components of the theoretical model and are incorporated in complementary ways within the EFT-based modeling framework.

Within the {\tt velocileptors} template introduced in Sec.~\ref{subsec:velocileptors}, it is convenient to separate the model into deterministic and stochastic contributions.
Concretely, we identify the first two lines in Eq.~\eqref{eq:velocileptors_master}, namely the one-loop LPT prediction $P^{\mathrm{LPT}}(k,\mu)$ plus the higher-derivative counterterm corrections proportional to the $\alpha_{n\mathrm{p}}$, as the deterministic component.
The remaining contribution, parametrized by $\mathrm{SN}_n$, defines the stochastic sector. 
The former describes the correlated clustering signal (and its smooth EFT corrections), whereas the latter captures uncorrelated short-scale contributions that are effectively noise-like on large scales.
Accordingly, we write the model multipoles for a tracer $X$ as
\begin{equation}
P_{X,\ell}(k)=P^{\rm det}_{X,\ell}(k)+P^{\rm stoch}_{X,\ell}(k)\, ,
\end{equation}
where $P^{\rm det}_{X,\ell}(k)$ and $P^{\rm stoch}_{X,\ell}(k)$ denote the deterministic and stochastic contributions, respectively.

To incorporate reconstruction systematics consistently within the {\tt velocileptors} framework, it is useful to apply the same deterministic--stochastic decomposition to the exclusion model introduced in Eq.~\eqref{eq:Pk_excl_master}.
In particular, we treat the exclusion correction $\Delta P_{\rm excl}$ as contributing to both the deterministic and stochastic sectors, which we denote by $\Delta P^{\rm det}_{{\rm excl},\ell}$ and $\Delta P^{\rm stoch}_{{\rm excl},\ell}$, respectively. By contrast, directional masking is a large-scale response of the \emph{correlated} clustering signal and should therefore be applied to the deterministic part, which controls the large-scale amplitude, rather than to the stochastic sector.
We thus model the power spectrum multipoles of reconstructed halos as
\begin{eqnarray}
\tilde P_{cc,\ell}(k)&=&R_\ell\Bigl[P^{\rm det}_{cc,\ell}(k)+\Delta P^{\rm det}_{{\rm excl},\ell}(k)\Bigr]
\nonumber \\
&& \qquad +\frac{1}{\tilde a_c^2}\Bigl[P^{\rm stoch}_{cc,\ell}(k)+\Delta P^{\rm stoch}_{{\rm excl},\ell}(k)\Bigr],
\label{eq:recon_master_detstoch}
\end{eqnarray}
where $R_\ell$ denotes the large-scale multipole rescaling induced by directional masking, defined as
$R_\ell \equiv [\tilde P_{cc,\ell}(k)/P_{cc,\ell}(k)]|_{k\to 0}$,
and the prefactor $1/\tilde a_c^2$ accounts for the change in the central abundance in the reconstructed sample (Sec.~\ref{subsec:decomp}).

The factor $R_\ell$ effectively rescales different multipoles by different amounts, thereby altering their relative amplitudes.
Such a reconstruction-induced, multipole-dependent rescaling is not part of the standard perturbation-theory template for the correlated signal, which motivates treating $R_\ell$ as an explicit large-scale response.
In practice, we estimate the catalog-measurable ratio, $R^{\rm obs}_\ell=[\tilde P_{cc,\ell}(k)/P_{gg,\ell}(k)]|_{k\to 0}$ (Eq.~\eqref{eq:ratio_R}), and use it to implement the multiplicative rescaling $R_\ell$ in Eq.~\eqref{eq:recon_master_detstoch}.
On sufficiently large scales, the difference between $R^{\rm obs}_\ell$ and the underlying deterministic response is expected to be smooth and is well approximated by the Kaiser-level ratio between the galaxy and central multipoles in redshift space; it therefore primarily changes the normalization of the inferred linear bias.
We test the use of the galaxy-referenced response $R_\ell^{\mathrm{obs}}$ as a practical proxy for the central-referenced response in Appendix~\ref{app:large_scale_rescaling}, where we compare the two calibrations and assess realization-to-realization fluctuations using all 25 mocks.

Once the large-scale response $R_\ell$ is specified, Eq.~\eqref{eq:recon_master_detstoch} implies that (up to the overall abundance normalization encoded by $\tilde a_c$) the reconstructed multipoles can be described in terms of the baseline true-central power spectrum, $P_{cc,\ell}$.
In addition, the exclusion correction can, in principle, be written explicitly using the window function and its convolution with the baseline power, as described in Sec.~\ref{subsec:exclusion}.
In the likelihood implementation, however, we do not include the full window convolution explicitly.
Instead, we rely on the flexibility of the EFT nuisance sector in Eq.~\eqref{eq:velocileptors_master} to absorb the residual reconstruction-induced effects encoded by $\Delta P_{\rm excl}$.

This expectation is supported by the qualitative behavior of the relevant terms.
First, since $P_{cc,\ell}$ is based on halo centers, it is largely free of satellite-driven FoG stochasticity, and we therefore expect the stochastic contribution $P^{\rm stoch}_{cc,\ell}$ to be significantly reduced relative to the galaxy case.
Second, as discussed in Sec.~\ref{subsec:exclusion}, the window-induced correction $\Delta P_{\rm excl}$ takes the form of smooth stochastic and higher-derivative contributions on sufficiently large scales, while at higher $k$ it acts as a broadband distortion of the multipoles.
Both behaviors are expected to be captured by the stochastic sector and higher-derivative counterterms of the EFT template over the fitting range considered here (e.g., $k \lesssim 0.2\,h\,{\rm Mpc}^{-1}$), enabling robust marginalization over residual reconstruction systematics.

\begin{table*}[!t]
\caption{
Parameters and inference priors used for all three tracer samples.
Here $\mathcal{U}[a,b]$ denotes a uniform prior on $[a,b]$, and $\mathcal{N}(\mu,\sigma)$ denotes a Gaussian prior with mean $\mu$ and standard deviation $\sigma$.
For $\mathrm{SN}_{2\mathrm{p}}$, ``flat'' denotes $p(\mathrm{SN}_{2\mathrm{p}})\propto1$ over the full real line.
The counterterm prior widths are given in units of $h^{-2}\mathrm{Mpc}^{2}$.
}
\label{tab}
\begin{ruledtabular}
\begin{tabular}{llll}
Sector & Parameter(s) & Treatment & Prior or fixed value \\
\colrule
ShapeFit
& $q_\parallel$, $q_\perp$
& MCMC sampled
& $\mathcal{U}[0.8,1.2]$ \\

ShapeFit
& $dm$
& MCMC sampled
& $\mathcal{U}[-3,3]$ \\

ShapeFit
& $df$
& MCMC sampled
& $\mathcal{U}[0,2]$ \\

ShapeFit
& $dn\equiv n-n^{\mathrm{ref}}$
& Fixed
& $0$ \\


Bias
& $b_{1\mathrm{E}}\sigma_8$
& MCMC sampled
& $\mathcal{U}[0,3]$ \\

Bias
& $b_{2\mathrm{L}}\sigma_8^2$, $b_{s\mathrm{L}}\sigma_8^2$
& MCMC sampled
& $\mathcal{N}(0,5)$ \\

Bias
& $b_{3\mathrm{L}}\sigma_8^3$
& Fixed
& $0$ \\

Counterterm
& $\alpha_{0\mathrm{p}}$, $\alpha_{2\mathrm{p}}$
& Analytically marginalized
& $\mathcal{N}(0,12.5)$ \\

Counterterm
& $\alpha_{4\mathrm{p}}$
& Fixed
& $0$ \\

Stochastic
& $\mathrm{SN}_{0\mathrm{p}}$
& Analytically marginalized
& $\mathcal{N}(0,2)$ \\

Stochastic
& $\mathrm{SN}_{2\mathrm{p}}$
& Analytically marginalized
& Flat, $p(\mathrm{SN}_{2\mathrm{p}})\propto1$ \\

Stochastic
& $\mathrm{SN}_{4\mathrm{p}}$
& Fixed
& $0$ \\
\end{tabular}
\end{ruledtabular}
\end{table*}


\begin{figure*}[t]
\centering
\begin{minipage}[t]{0.49\textwidth}
    \centering
    \includegraphics[width=\textwidth]{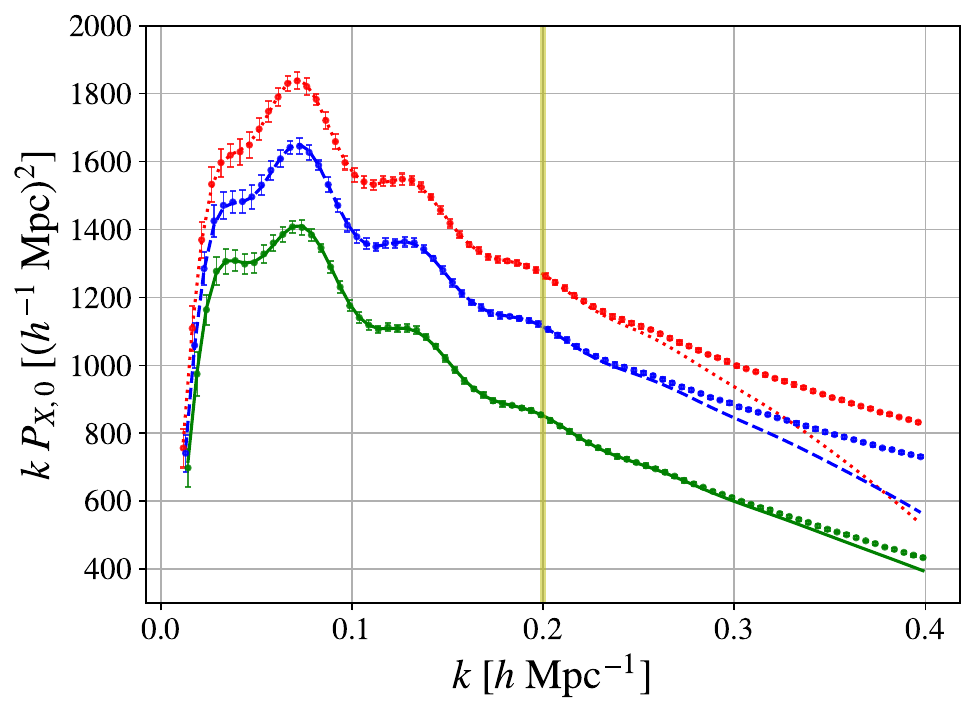} 
\end{minipage}
\hfill
\begin{minipage}[t]{0.49\textwidth}
    \centering
    \includegraphics[width=\textwidth]{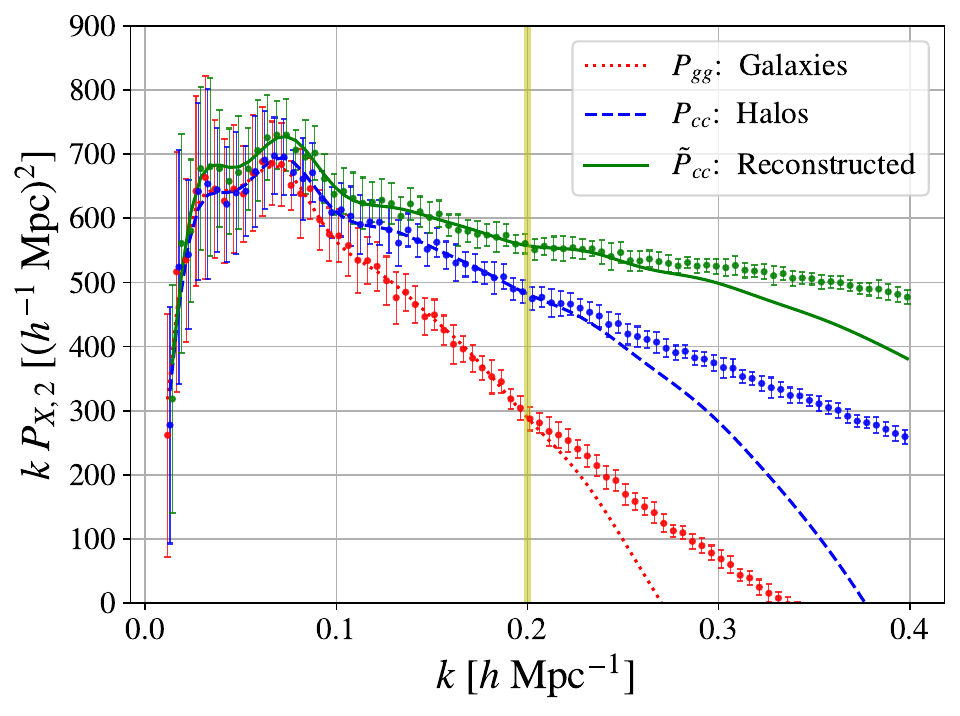}
\end{minipage}
\caption{
Redshift-space power spectrum multipoles at redshift $z=1.1$ for galaxies ($P_{gg,\ell}$; red), host halos ($P_{cc,\ell}$; blue), and reconstructed halos ($\tilde P_{cc,\ell}$; green). The two panels show the monopole (left) and quadrupole (right); circles with error bars denote the mean and standard deviation measured from the 25 mock realizations, while lines denote the corresponding best-fit model predictions for fits with $k_{\max}=0.2\,h\,\mathrm{Mpc}^{-1}$, indicated by the vertical yellow line. For visual clarity, the data points and the corresponding model curves are slightly shifted horizontally for each tracer.}
\label{fig:kP_vs_k}
\end{figure*}

\subsection{Parameter inference details}
Parameter inference is performed using the \texttt{desilike}\footnote{\url{https://desilike.readthedocs.io}} framework.
The parameter treatment and inference priors adopted in this analysis are summarized in Table~\ref{tab}.
To enable a fair comparison among the analyses of galaxies, host halos, and reconstructed halos, we use identical inference priors for all three tracers, including an unbounded flat prior on the normalized stochastic parameter $\mathrm{SN}_{2\mathrm{p}}$.
Given this prior choice and the inclusion of relatively high values of $k_{\max}$ in some fits, we adopt a slightly relaxed Gelman--Rubin convergence criterion of $|\widehat R-1|<0.05$ for the Markov Chain Monte Carlo (MCMC) sampling.

We distinguish the ShapeFit variables sampled in the chains from the quantities derived from them. The sampled variables $dm$ and $df$ are defined as
\begin{eqnarray}
dm &\equiv& m-m^{\mathrm{ref}},
\nonumber\\
df &\equiv&
\frac{f\sqrt{A_{sp}}}
{\left(f\sqrt{A_{sp}}\right)^{\mathrm{ref}}},
\end{eqnarray}
where $A_{sp}\equiv s^{-3}P_{\mathrm{nw}}^{\mathrm{lin}}(k_p/s)$
denotes the amplitude of the no-wiggle linear power spectrum at the ShapeFit pivot scale. Their fiducial values are $dm=0$ and $df=1$. The reported growth parameter is obtained from these sampled quantities in post-processing according to
\begin{eqnarray}
f\sigma_{s8}
&=&
\left(f\sigma_{s8}\right)^{\mathrm{ref}}
df
\exp\left[
\frac{dm}{2a}
\tanh\left(
a\ln\frac{r_d^{\mathrm{ref}}}{R_8}
\right)
\right],
\nonumber \\
\end{eqnarray}
where $a=0.6$ is the ShapeFit transition parameter, and $R_8 \equiv 8\hinvmpc$ \citep{Brieden:2023}. Consequently, no independent prior is assigned to $f\sigma_{s8}$; its prior is induced by the joint priors on $dm$ and $df$.

For the bias sector, the sampled combinations are
$b_{1\mathrm{E}}\sigma_8$,
$b_{2\mathrm{L}}\sigma_8^2$, and
$b_{s\mathrm{L}}\sigma_8^2$, while
$b_{3\mathrm{L}}\sigma_8^3$ is fixed to zero.
Here $\sigma_8\equiv\sigma(R=R_8,z)$ denotes the conventional fluctuation amplitude at the analysis redshift, evaluated at $R=8\hinvmpc$ for the reference template, and should not be confused with $\sigma_{s8}$.
The stochastic coefficients $\mathrm{SN}_n$ appearing in Eq.~(\ref{eq:velocileptors_master}) are related to the dimensionless parameters, $\mathrm{SN}_{n\mathrm{p}}$, used in the inference by
\begin{eqnarray}
\mathrm{SN}_0
&=&
\frac{1}{\bar n}\,
\mathrm{SN}_{0\mathrm{p}},
\nonumber\\
\mathrm{SN}_2
&=&
\frac{f_{\mathrm{sat}}}{\bar n}\,
\sigma_v^2\,
\mathrm{SN}_{2\mathrm{p}},
\nonumber\\
\mathrm{SN}_4
&=&
\frac{f_{\mathrm{sat}}}{\bar n}\,
\sigma_v^4\,
\mathrm{SN}_{4\mathrm{p}},
\end{eqnarray}
Here $\bar n$ denotes the mean number density across the 25 realizations for each tracer, summarized in Table~\ref{tab:tracer_catalogs}.
The quantities $f_{\mathrm{sat}}$ and $\sigma_v$ represent an assumed satellite fraction and a characteristic satellite velocity dispersion expressed in comoving length units, respectively.
We adopt the \texttt{desilike} LRG preset,
$f_{\mathrm{sat}}=0.15$ and $\sigma_v=6.19388\,h^{-1}\mathrm{Mpc}$, as common fixed normalization values for all three samples.
Since $\mathrm{SN}_{4\mathrm{p}}$ is fixed to zero, these two settings enter only the normalization of the free coefficient $\mathrm{SN}_2$. The unbounded flat prior on $\mathrm{SN}_{2\mathrm{p}}$ also implies an unbounded flat prior on $\mathrm{SN}_2$, so the adopted nonzero normalization values do not affect the cosmological constraints.

Following \citet{Maus:2025}, the four linear nuisance parameters
$\alpha_{0\mathrm{p}}$, $\alpha_{2\mathrm{p}}$,
$\mathrm{SN}_{0\mathrm{p}}$, and $\mathrm{SN}_{2\mathrm{p}}$
are analytically marginalized over with the priors listed in
Table~\ref{tab} to improve sampling efficiency.
The remaining higher-order coefficients $\alpha_{4\mathrm{p}}$ and
$\mathrm{SN}_{4\mathrm{p}}$ are fixed to zero.
Theoretical predictions are accelerated using an emulator based on a fourth-order Taylor expansion applied only to the common perturbative basis. The tracer-dependent bias, counterterm, and stochastic contributions, together with the reconstruction-induced multipole rescaling, are evaluated separately when constructing the full tracer power-spectrum prediction. This allows the same perturbative emulator to be used consistently for all three tracers.
Unless otherwise stated, we quote parameter constraints using the median of the marginalized posterior distribution as the central value, with symmetrized uncertainties defined as half the width of the equal-tailed 68\% credible interval.

\section{Results} \label{sec:results}

In this section, we present the main results of our analysis, focusing on how the cylinder-grouping reconstruction modifies redshift-space clustering and how these modifications propagate into cosmological parameter inference. 
First, in Sec.~\ref{sec:VA}, we compare the measured redshift-space power spectrum multipoles of the three tracer populations---galaxies, host halos, and reconstructed halos---and assess the ability of the EFT model to reproduce these measurements.
We then examine the resulting constraints in the compressed parameter space in Sec.~\ref{sec:VB} and quantify the gain in constraining power relative to the galaxy sample in Sec.~\ref{subsec:degeneracy_params}.
Finally, we study the dependence of parameter constraints on the maximum wavenumber, $k_{\max}$, in Sec.~\ref{sec:VD}.

\subsection{Measured and modeled impact of the reconstruction}
\label{sec:VA}

Figure~\ref{fig:kP_vs_k} shows the redshift-space monopole and quadrupole of the galaxy, host-halo, and reconstructed-halo samples at $z=1.1$.
The galaxy sample shows the expected signature of satellite contamination. In particular, the quadrupole shows a clear suppression of power on small scales relative to the host-halo sample, reflecting the FoG damping induced by virial motions of satellite velocities. 
In the monopole, the impact of satellites is more subtle: while the presence of satellites enhances the overall clustering amplitude, at high $k$ the galaxy power is more strongly suppressed than the halo power due to FoG smearing \cite{Okumura:2017}. The observed small-scale behavior therefore reflects a competition between the two effects.

The reconstruction shifts the clustering signal toward that of the host-halo sample over a wide range of scales. 
In particular, the small-scale suppression in the quadrupole is substantially reduced, demonstrating that the CG procedure successfully removes a substantial fraction of the satellite-induced FoG contribution.
The remaining differences between the power spectra of the reconstructed- and host-halo samples likely arise from a combination of small-scale nonlinear redshift-space contributions from halo pairs within the grouping cylinder, which remain in the host-halo sample but are preferentially suppressed by the grouping, and residual reconstruction systematics.

Importantly, these residuals exhibit a smooth, broadband scale dependence, without introducing sharp features, consistent with the discussion in Secs.~\ref{subsec:exclusion} and~\ref{subsec:masking}: the small-scale exclusion effect induces smooth contributions resembling higher-derivative counterterms and stochastic terms, while the large-scale effect was expected to reduce primarily to a smooth amplitude rescaling.
As we see in the following subsections, this behavior lies within the functional flexibility of the EFT template, indicating that the reconstruction-induced systematics to the clustering signal can be modeled consistently within the standard EFT framework. This provides the basis for the parameter-level tests presented below.

\subsection{Constraints in the compressed parameter space}
\label{sec:VB}


\begin{figure}[t]
\centering
    \includegraphics[width=0.48\textwidth]{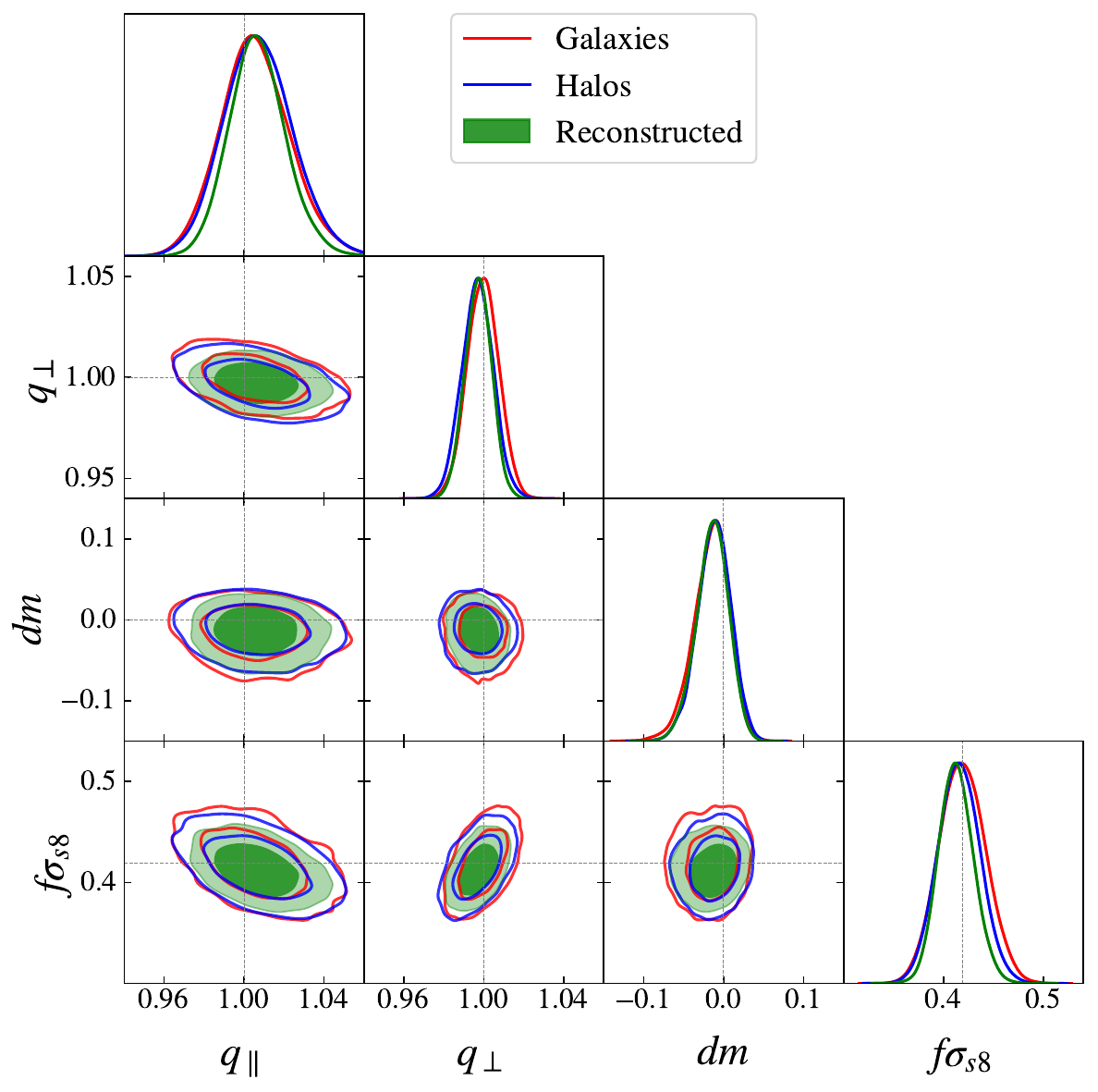}
\caption{
Cosmological constraints in the compressed parameter space, $(q_\parallel, q_\perp, dm, f\sigma_{s8})$, for galaxy (red), host-halo (blue), and reconstructed-halo (green) samples. The contours show the $68\%$ and $95\%$ confidence regions from inner to outer.
The results are shown for a representative choice of $k_{\max}=0.2\,h\,\mathrm{Mpc}^{-1}$.
The dashed lines indicate the fiducial values.
}
\label{fig:posterior_compressed}
\end{figure}

Figure~\ref{fig:posterior_compressed} shows the posterior distributions in the compressed parameter space
($q_\parallel, q_\perp, dm, f\sigma_{s8}$) at $z=1.1$ for a representative choice of $k_{\max}=0.2\,h\,\mathrm{Mpc}^{-1}$.
Here, the parameter $dm$ quantifies deviations in the 
broadband shape of the power spectrum relative to the fiducial cosmology.
All three tracer samples yield constraints that are consistent with the fiducial cosmology, indicating that no significant bias is introduced by the halo reconstruction within our modeling framework.

Quantitatively, the reconstructed-halo sample achieves a constraint on the growth parameter of $f\sigma_{s8} = 0.412 \pm 0.017$, which is tighter than that obtained from galaxies
($0.419 \pm 0.023$) and marginally tighter than that from host halos
($0.416 \pm 0.021$).
A similar trend is observed for the AP parameters; the reconstructed sample yields uncertainties of $\sigma(q_\parallel) = 0.014$ and $\sigma(q_\perp) = 0.007$, corresponding to an improvement of approximately $\sim 10-20\%$ relative to the galaxy sample at the same $k_{\max}$.

For the shape parameter $dm$, the reconstructed halos give $dm = -0.013 \pm 0.020$, which is consistent with both the galaxy ($dm = -0.014 \pm 0.022$) and host-halo ($dm = -0.011 \pm 0.021$) results.
In contrast to $f\sigma_{s8}$ and the AP parameters, the improvement in $dm$ is relatively modest.
Overall, the reconstructed-halo sample provides cosmological constraints that are competitive with those obtained from the true host halos, while offering a clear improvement over the galaxy sample.

We compare the best-fitting model predictions with the measured power spectrum multipoles in Fig.~\ref{fig:kP_vs_k}. The figure shows that the EFT-based template provides an excellent description of the monopole and quadrupole for all three tracers within the fitted range, $k \leq k_{\max}=0.2\,h\,\mathrm{Mpc}^{-1}$.

The cosmological constraints presented here are obtained from a catalog reconstructed using the fiducial coordinate mapping. In Appendix~\ref{app:ap_distortions}, we separately test the sensitivity of the CG assignments to imposed AP distortions and find that the bulk reconstruction diagnostics remain stable.

\subsection{Parameter degeneracies with stochastic parameters} \label{subsec:degeneracy_params}

\begin{figure}[bt]
\centering
    \centering
    \includegraphics[width=0.49\textwidth]{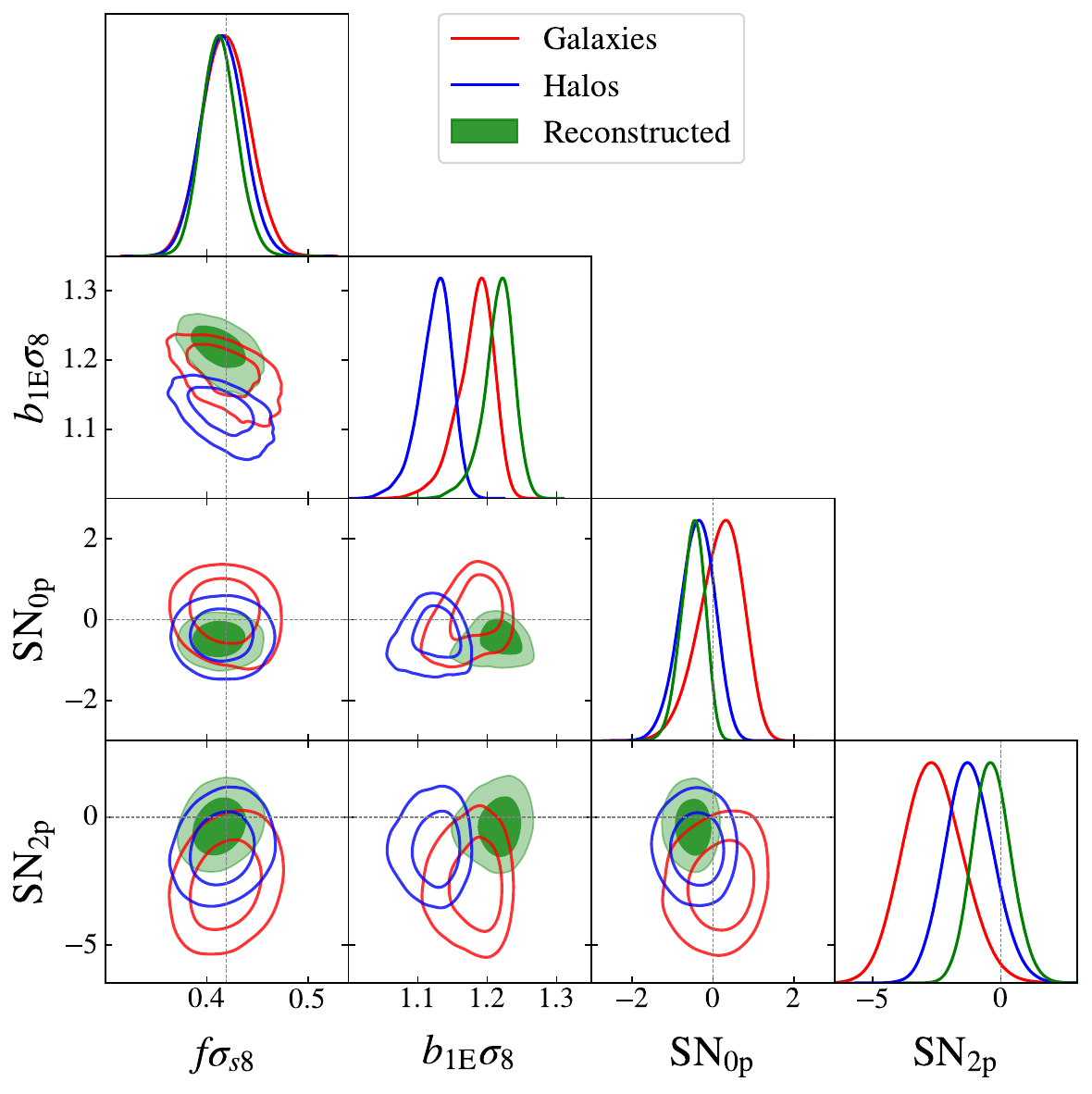}
\caption{
Same as Fig.~\ref{fig:posterior_compressed}, but showing constraints on the growth-rate parameter $f\sigma_{s8}$ and key bias and stochastic parameters ($b_{1\mathrm{E}}\sigma_8,\mathrm{SN}_{0\mathrm{p}},\mathrm{SN}_{2\mathrm{p}}$).
The posterior median of $b_{1\mathrm{E}}\sigma_8$ for the reconstructed halos is closer to that for galaxies than to that for the true halos because of how we correct for the large-scale masking effect in Sec.~\ref{subsec:incorp_recon_sys}; see the main text for the interpretation.
}
\label{fig:fsig_sn}
\end{figure}

To further understand the origin of the improved constraints obtained with the reconstructed-halo sample, we examine the parameter degeneracies involving the stochastic sector of the model. Figure~\ref{fig:fsig_sn} shows the joint posterior distributions of $f\sigma_{s8}$, the linear bias parameter $b_{1\mathrm{E}}\sigma_8$, and the stochastic parameters $\mathrm{SN}_{0\mathrm{p}}$ and $\mathrm{SN}_{2\mathrm{p}}$ at $z=1.1$ for a representative choice of $k_{\max}=0.2\hmpcinv$.
Because the multipole-dependent rescaling adopted in our analysis (see Sec.~\ref{subsec:incorp_recon_sys}) is defined relative to the galaxy power-spectrum multipoles, the posterior median of $b_{1\mathrm{E}}\sigma_8$ for the reconstructed halos ($b_{1\mathrm{E}}\sigma_8=1.218 \pm 0.021$) is closer to that for galaxies ($b_{1\mathrm{E}}\sigma_8=1.186 \pm 0.026$) than to that for the true host halos ($b_{1\mathrm{E}}\sigma_8=1.127 \pm 0.023$).
As shown in Appendix~\ref{app:large_scale_rescaling}, when the rescaling is instead calibrated relative to the host-halo power spectrum, the inferred reconstructed-halo bias shifts to $b_{1\mathrm{E}}\sigma_8=1.138\pm0.022$, consistent with the true host-halo value. This behavior is expected: the inferred bias is closer to the galaxy or host-halo value depending on the reference used to calibrate $R_\ell$.


In linear theory, the growth-rate parameter $f\sigma_8$ is intrinsically degenerate with the linear bias $b_{1\mathrm{E}}\sigma_8$.
This degeneracy is further complicated by FoG effects, which are largely captured by the higher-order stochastic parameter $\mathrm{SN}_{2\mathrm{p}}$.
As a result, the galaxy sample exhibits degeneracies among $f\sigma_{s8}$, $b_{1\mathrm{E}}\sigma_8$, and $\mathrm{SN}_{2\mathrm{p}}$. In contrast, the reconstructed-halo sample shows a substantially reduced degeneracy within the stochastic sector, particularly between $\mathrm{SN}_{0\mathrm{p}}$ and $\mathrm{SN}_{2\mathrm{p}}$.
This reduction leads to weaker correlations between $f\sigma_{s8}$ (and $b_{1\mathrm{E}}\sigma_8$) and $\mathrm{SN}_{2\mathrm{p}}$.
Consequently, $f\sigma_{s8}$ can be constrained more tightly.
This simplification of the stochastic sector also explains why the reconstruction yields more visible gains for the FoG-sensitive parameters (e.g., $f\sigma_{s8}$ and $q_\parallel$) than for $dm$ (see Fig.~\ref{fig:params_vs_kmax} in Appendix~\ref{app:other_params}), whose constraints are only weakly degenerate with $\mathrm{SN}_{2\mathrm{p}}$.

The host-halo sample exhibits an intermediate behavior, with partial mitigation of these degeneracies but a less pronounced improvement compared to the reconstructed-halo sample. 
A key reason for this difference is that the LOS-aligned cylinders used in the reconstruction efficiently suppress galaxy FoG effects, while also reducing the impact of strongly nonlinear clustering on scales below the cylinder size.
As a result, the reconstructed sample can retain small-scale information that is less entangled with stochastic velocity effects than in the galaxy case and less directly affected by sub-cylinder nonlinearity than in the host-halo case, leading to improved constraints on the growth rate.

Taken together, the improvement in the growth-rate constraint is driven by a qualitative change in how the model parameters are correlated.  
By suppressing satellite-induced FoG effects, the reconstruction reduces the coupling between velocity dispersion and stochastic contributions.
At the same time, as discussed in Sec.~\ref{subsec:incorp_recon_sys}, the residual effects induced by the cylinder grouping are expected, at least on large scales, to be sufficiently smooth that they can be absorbed by the standard EFT nuisance sector over the range of scales considered here. As a result, they do not appear to generate severe additional parameter degeneracies beyond those already present in the model. 
The reduced nuisance-parameter degeneracies, combined with the smooth residual reconstruction effects, provide a natural explanation for the tighter constraint on $f\sigma_{s8}$ at the fiducial fitting limit.

It is worth noting that halo reconstruction corresponds to a nonlinear, LOS-anisotropic transformation of the galaxy field \cite{Okumura:2017}.
Such transformations can induce effective modifications to the large-scale redshift-space response (sometimes discussed as a ``velocity bias''; see, e.g., \citet{Seljak:2012}).
In our case, these effects are entangled with the directional masking discussed in Sec.~\ref{subsec:masking}.
While we do not attempt to separate these contributions, our empirical correction based on $R^{\rm obs}_\ell$ appears to capture the combined large-scale anisotropic response over the fitted range.

\begin{figure}[bt]
\centering
    \includegraphics[width=0.49\textwidth]{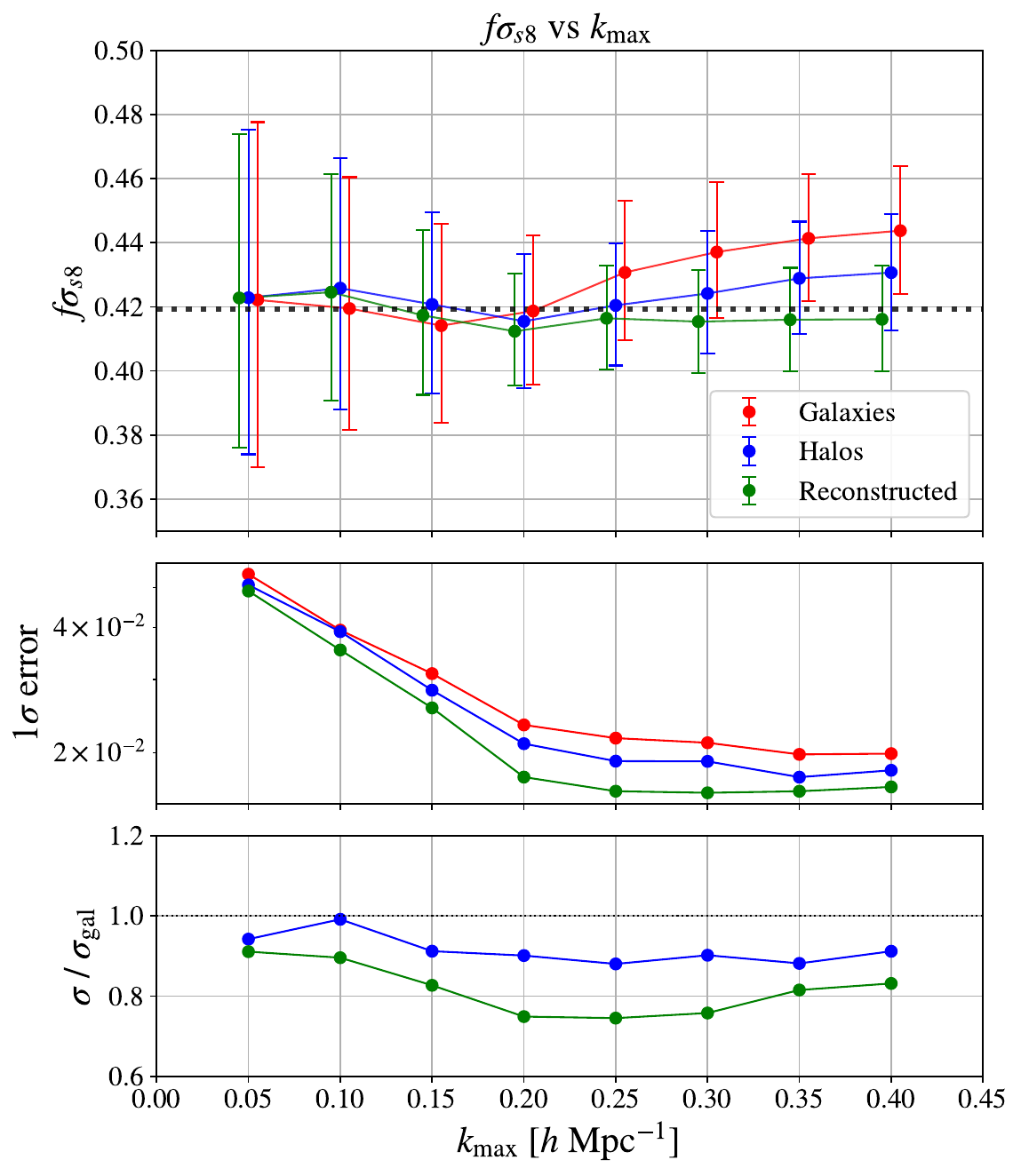} 
\caption{
    Constraints on the growth-rate parameter $f\sigma_{s8}$ as a function of the maximum wavenumber $k_{\max}$ obtained for galaxy (red), host-halo (blue), and reconstructed-halo (green) samples. The top row shows the posterior median together with the $68\%$ uncertainties, with the horizontal dotted line indicating the fiducial value; the middle row shows these uncertainties alone; and the bottom row shows their ratio to the galaxy case, $\sigma/\sigma_{\rm gal}$.
}
\label{fig:fsig_kmax}
\end{figure}

\subsection{Stability of the growth-rate constraint with respect to $k_{\max}$}
\label{sec:VD}

Figure~\ref{fig:fsig_kmax} shows the inferred constraints on the growth-rate parameter $f\sigma_{s8}$ as a function of the maximum wavenumber $k_{\max}$ used in the fit at $z=1.1$.
For the reconstructed-halo and host-halo samples, the posterior medians remain consistent with the fiducial prediction over the full range of scales considered.
In contrast, the galaxy sample exhibits a systematic upward shift in the inferred $f\sigma_{s8}$ when extending the fit beyond $k_{\max}\simeq 0.20\,h\,\mathrm{Mpc}^{-1}$, indicating the onset of scale-dependent modeling systematics.

In addition to improved robustness, the reconstructed-halo sample achieves a more rapid reduction in statistical uncertainties with increasing $k_{\max}$.
At the fiducial limit of $k_{\max}=0.2\hmpcinv$, the reconstructed sample yields more than a $20\%$ reduction in the $1\sigma$ uncertainty on $f\sigma_{s8}$ relative to the galaxy case, whereas the host-halo sample shows a smaller improvement of approximately $10\%$.
At lower $k_{\max}$, where the constraints are dominated by large-scale modes, the differences between tracers are small, indicating that the gain from reconstruction arises primarily from the more reliable use of additional small-scale information.

A similar level of stability is observed for the AP parameters $(q_\parallel,q_\perp)$, as summarized in Fig.~\ref{fig:params_vs_kmax} in Appendix~\ref{app:other_params}. The reconstructed-halo sample consistently yields tighter constraints than the galaxy sample.
The improvement is more pronounced for $q_\parallel$ than for $q_\perp$, reflecting the fact that FoG smearing preferentially degrades LOS information, which the reconstruction mitigates by suppressing satellite-driven velocity dispersion.
By contrast, the shape parameter $dm$ exhibits only weak dependence on the tracer choice (at least up to $k_{\max}\simeq 0.2\,h\,\mathrm{Mpc}^{-1}$). This behavior is consistent with the interpretation that $dm$ is primarily sensitive to the broadband shape of the power spectrum, which is largely unaffected by FoG-like small-scale anisotropies.

To assess whether this apparent stability also corresponds to an adequate description of the power-spectrum multipoles, we present additional diagnostics of the goodness of fit and the behavior of the nuisance parameters in Appendix~\ref{app:highk}. At larger $k_{\max}$, the goodness of fit deteriorates for the galaxy and host-halo samples. The clearest change among the nuisance parameters is a negative shift in $\alpha_{2\mathrm{p}}$ for galaxies, accompanied by a stronger dependence of $\mathrm{SN}_{2\mathrm{p}}$ on the fitting
range. For the galaxy sample, this behavior indicates that the apparent stability of the cosmological parameters can partly reflect the absorption of small-scale modeling residuals by the nuisance sector. By contrast, the reconstructed-halo sample shows comparatively stable goodness of fit and nuisance parameters, suggesting that the stability of its cosmological constraints is not primarily maintained by increasingly large compensating shifts in the nuisance sector. We therefore retain $k_{\max}=0.2\hmpcinv$ as the fiducial modeling limit and regard the fits at larger $k_{\max}$ as diagnostic comparisons among the three tracers.

\section{Conclusion} \label{sec:conclusion}

We have investigated how halo reconstruction based on the cylinder-grouping (CG) method \cite{Okumura:2017} impacts redshift-space clustering analyses and cosmological parameter inference.
Using DESI-like LRG mock catalogs from the \textsc{AbacusSummit} $N$-body simulations at $z=1.1$, we applied the CG procedure to construct a reconstructed-halo tracer from the galaxy distribution, and performed EFT-based full-shape fits to the redshift-space power spectrum multipoles using {\tt velocileptors} to generate the perturbative templates.

Our analysis demonstrates that the main reconstruction-induced systematics can be consistently incorporated within the standard EFT framework without introducing additional reconstruction-specific parameters. This is achieved by exploiting a scale-separated description of the reconstruction effects.
On large scales, the CG procedure induces a smooth, multipole-dependent rescaling (``directional masking''), which we calibrate directly using the observable ratios $R^{\rm obs}_\ell \equiv [\tilde P_{cc,\ell}(k)/P_{gg,\ell}(k)]|_{k\to 0}$ and implement as a multiplicative correction to the deterministic component of the EFT model.
On smaller scales, the exclusion-like suppression of close pairs leads to smooth, broadband modifications that are effectively absorbed by the standard EFT counterterms and stochastic sectors over the range of scales considered.

Within this framework, we find that the reconstructed-halo sample yields unbiased cosmological constraints in the compressed parameter space $(q_\parallel, q_\perp, dm, f\sigma_{s8})$, consistent with the fiducial cosmology. Compared to galaxy-based analyses, the halo reconstruction leads to a clear improvement in both robustness and precision. 
In particular, at the fiducial limit of $k_{\max}=0.2\hmpcinv$, the reconstructed-halo sample yields an uncertainty on $f\sigma_{s8}$ that is more than $\sim 20\%$ smaller than that obtained from the galaxy sample. 
We trace this improvement to the reduced $\mathrm{SN}_{0\mathrm{p}}$--$\mathrm{SN}_{2\mathrm{p}}$ degeneracy and weaker correlations between the stochastic parameters and $f\sigma_{s8}$.
The diagnostics of the goodness of fit and the behavior of the nuisance parameters in Appendix~D further suggest that the stability of the reconstructed-halo constraints at higher $k_{\max}$ is not primarily maintained by compensating shifts in the nuisance sector. 

These results demonstrate that CG-based halo reconstruction provides a practical route to mitigating satellite-driven nonlinearities and FoG effects at the field level, thereby enabling more robust cosmological inference from quasi-nonlinear scales within standard EFT pipelines.
The treatment adopted here also clarifies how reconstruction systematics map onto EFT ingredients and suggests that halo reconstruction and EFT modeling are naturally complementary.

Several extensions are left for future work.
It will be important to 
(i) explore the dependence of the results on the cylinder geometry and its optimization for different tracer populations and redshifts---an avenue that can be further investigated within the same geometric window-function framework used here to characterize exclusion---
(ii) revisit and optimize priors for the stochastic sector in the reconstructed field, for example, by adopting {\it physically motivated} priors as in standard EFT analyses, 
(iii) incorporate a full covariance treatment, including non-Gaussian correlations between distinct $k$ bins and additional multipoles where relevant; if reconstruction suppresses mode coupling and reduces the covariance between distinct $k$ bins, its relative gain over galaxy analyses may be even more pronounced when using the full covariance, and (iv) apply the method to observational data, with realistic survey geometries and selection effects, while assessing whether reconstruction can mitigate practical observational limitations such as fiber collisions.
A fully self-consistent test away from the fiducial cosmology, including both catalog reconstruction and parameter inference, is also left for future work.
Together, these steps will clarify the ultimate potential of halo reconstruction as a component of next-generation RSD analyses.

\begin{acknowledgments}
We thank Kazuyuki Akitsu for helpful discussions and comments.
T.~O. acknowledges support from the Taiwan National Science and Technology Council under Grants No.~NSTC 114-2112-M-001-004- and No.~NSTC 115-2112-M-001-059-MY3, and the Academia Sinica Investigator Project Grant No.~AS-IV-114-M03 for the period of 2025--2029.
The authors acknowledge the access to high-performance computing facilities (Theory cluster and storage) provided by Academia Sinica Institute of Astronomy and Astrophysics (ASIAA).
\end{acknowledgments}

\section*{Data Availability}
There are no publicly available research data or software supporting this manuscript.
Requests for further information or data should be sent to the authors.

\appendix

\section{Supplementary posterior constraints}
\label{app:other_params}

\begin{figure*}
  \centering
  \includegraphics[width=.95\textwidth]{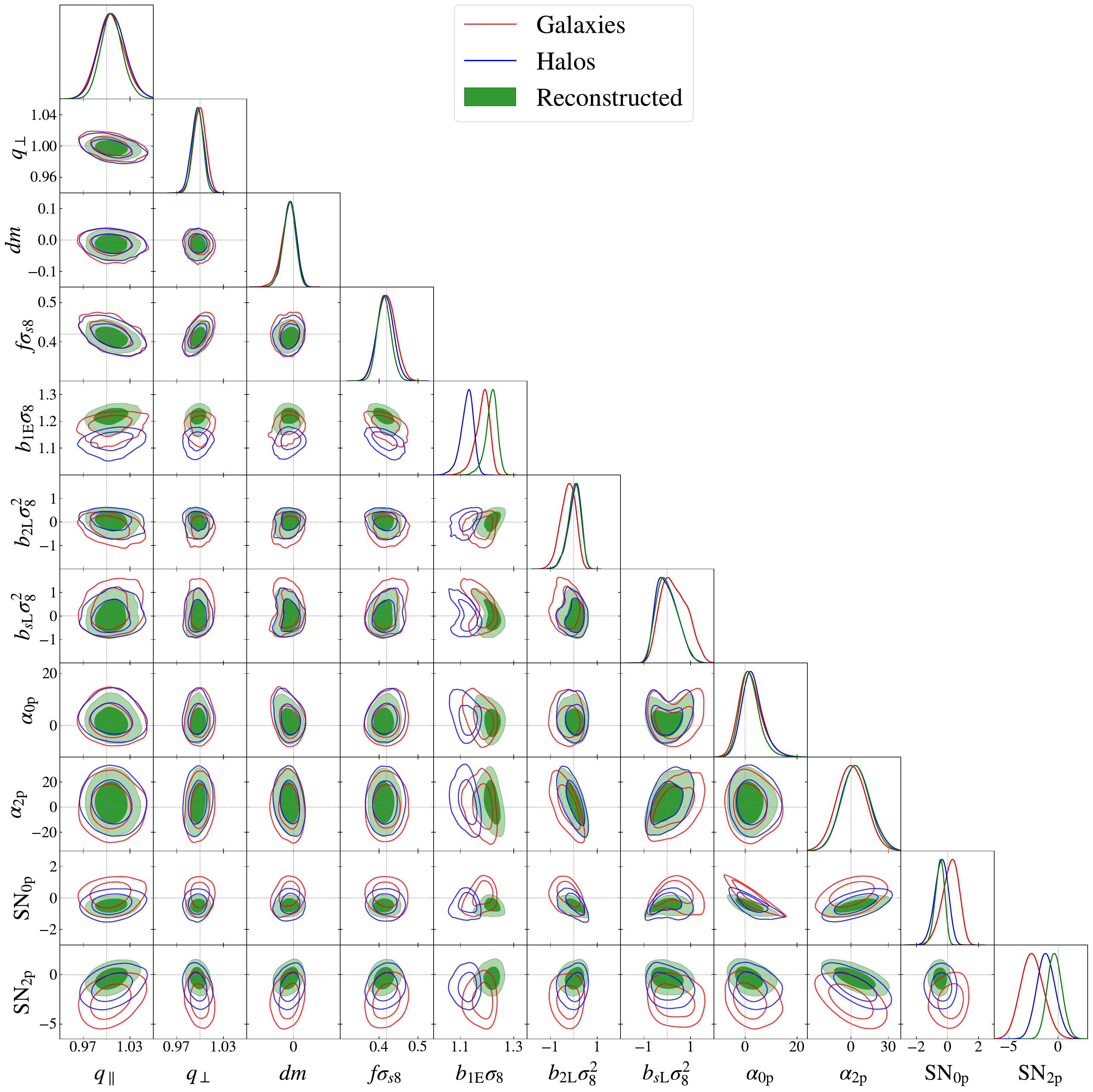}
  \caption{
  Full posterior distributions in the complete parameter space for galaxies
  (red), halos (blue), and reconstructed halos (green).
  Diagonal panels show marginalized 1D posteriors, and off-diagonal panels
  show the corresponding 2D confidence regions.
  }
  \label{fig:posterior_all_params}
\end{figure*}

\begin{figure*}
  \centering

%
%

    \includegraphics[width=0.325\textwidth]{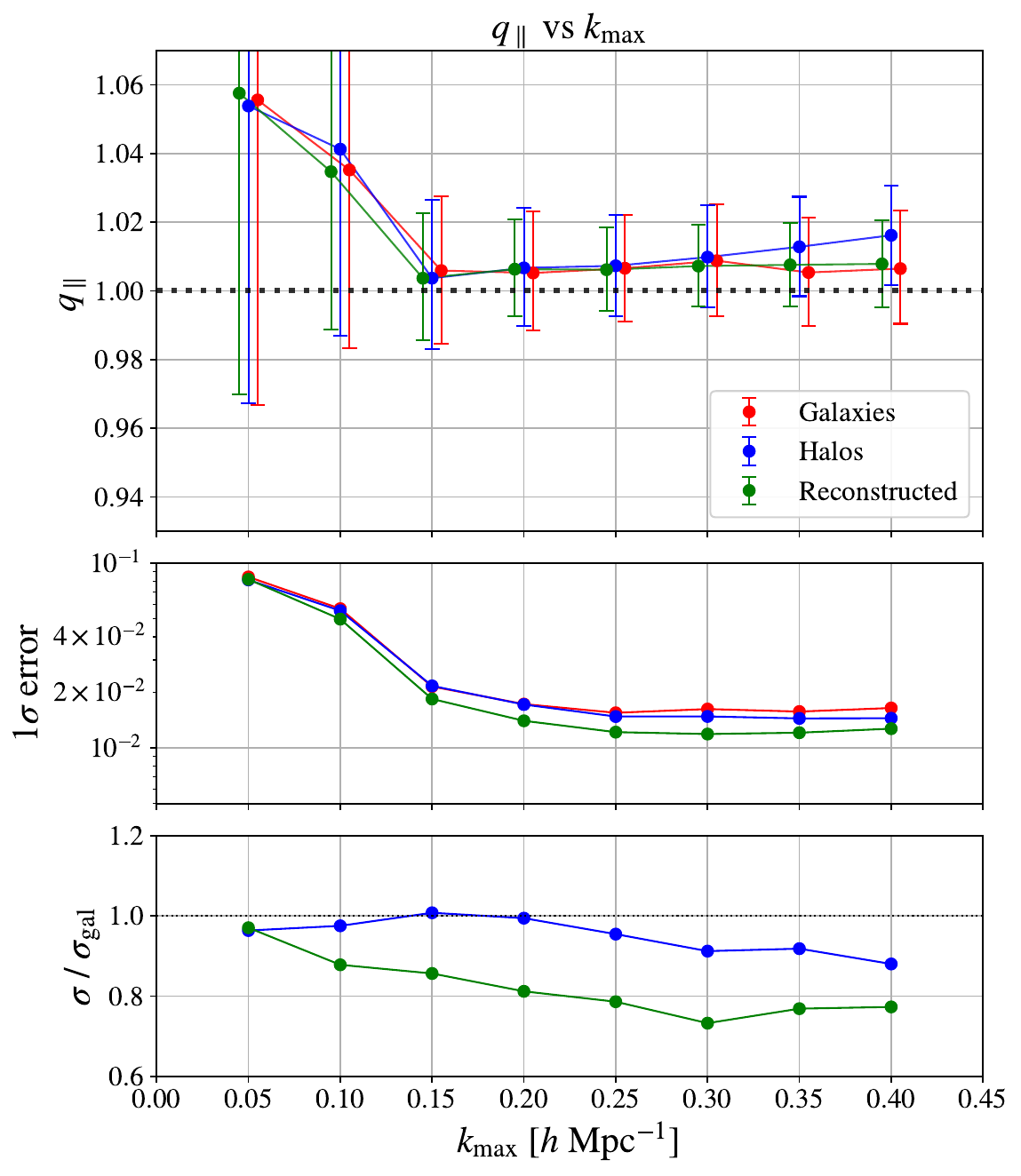}
    \includegraphics[width=0.325\textwidth]{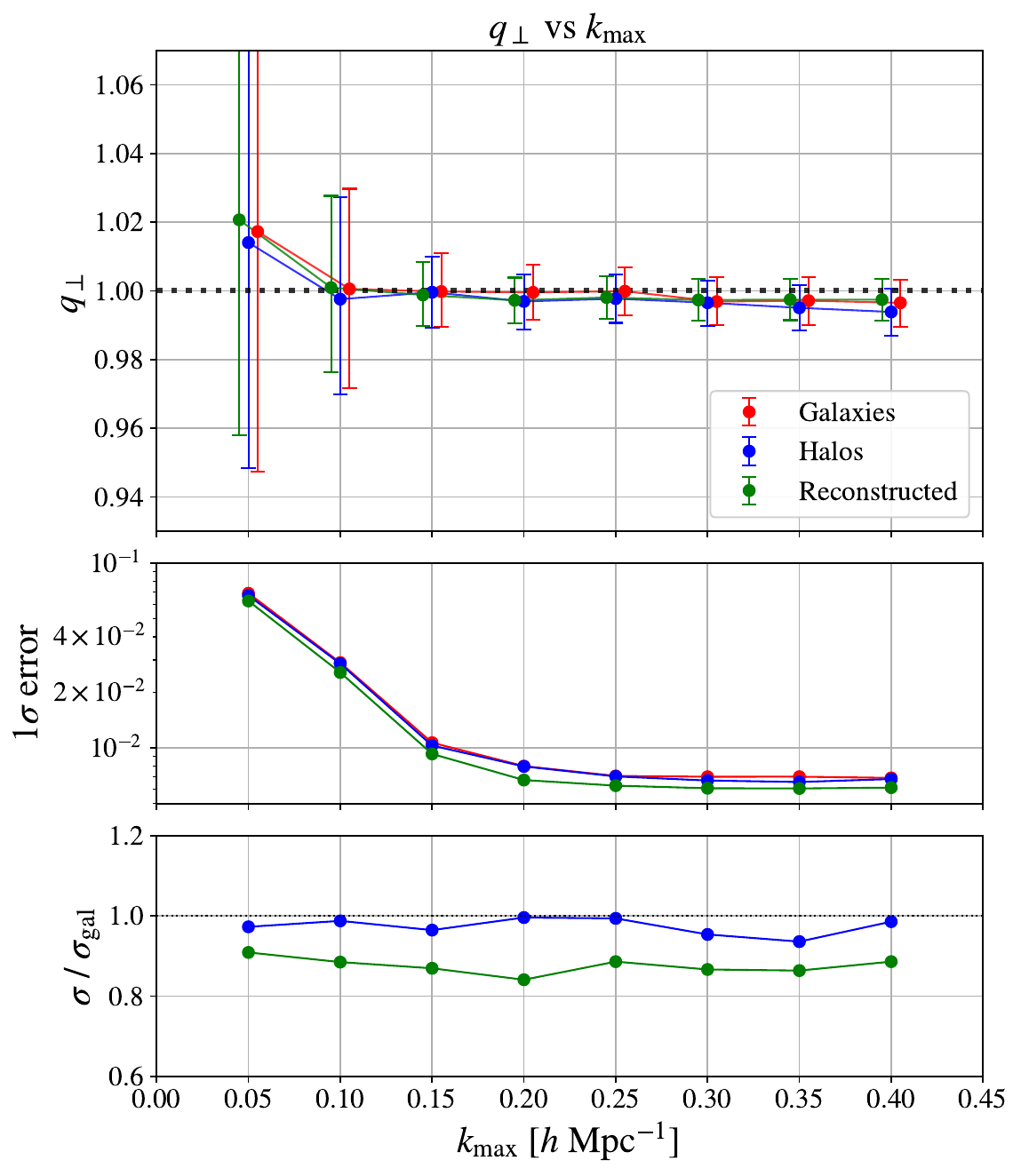}
    \includegraphics[width=0.325\textwidth]{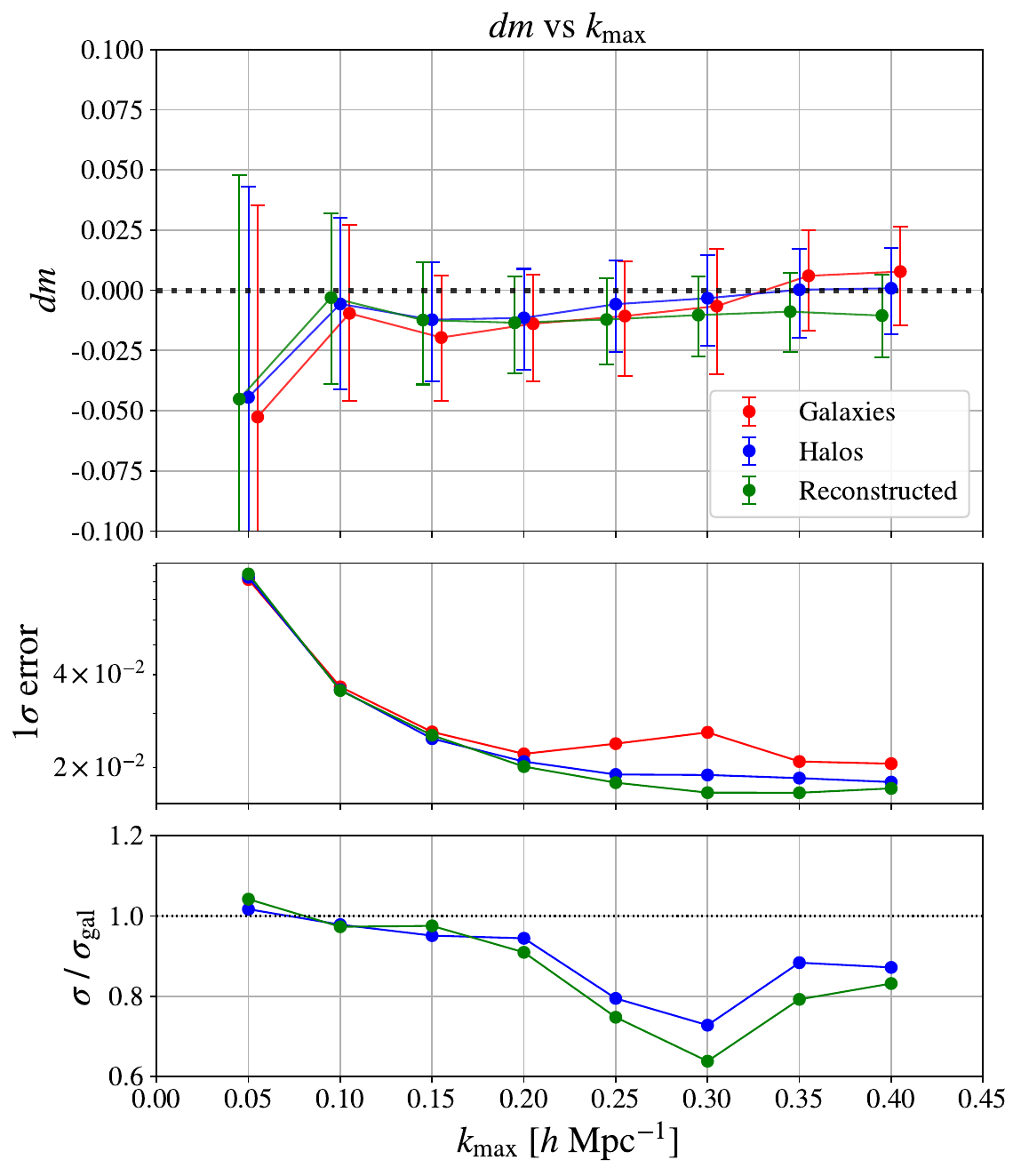}

  \caption{
  Same as Fig.~\ref{fig:fsig_kmax}, but for the AP parameters $q_\parallel$ and $q_\perp$, and the shape parameter $dm$.
  }
  \label{fig:params_vs_kmax}
\end{figure*}

In this appendix we provide supplementary results for parameters that are not highlighted in the main text.
Figure~\ref{fig:posterior_all_params} shows the full posterior distributions (including cross-correlations) in the complete parameter space for the three tracers considered in this work: galaxies (red), halos (blue), and reconstructed halos (green).  The diagonal panels show marginalized 1D posteriors, while the off-diagonal panels show the joint constraints.
This figure is intended as a compact reference for assessing parameter degeneracies beyond the compressed parameter set discussed in the main text.

To further illustrate the robustness of the inferred cosmological parameters with respect to the fitting range, Fig.~\ref{fig:params_vs_kmax} summarizes the $k_{\max}$ dependence of the AP parameters ($q_\parallel$ and $q_\perp$) and the shape parameter ($dm$).
Each panel shows the posterior median as a function of $k_{\max}$ (top), the corresponding $1\sigma$ uncertainty (middle), and the ratio of the uncertainty relative to the galaxy case (bottom). Consistent with the behavior of $f\sigma_{s8}$ discussed in Sec.~\ref{sec:results}, the reconstructed-halo sample yields stable posterior medians for $q_\parallel$, $q_\perp$, and $dm$ as $k_{\max}$ is increased, while achieving reduced uncertainties compared to galaxies.
In particular, at the fiducial limit of $k_{\max}=0.2\hmpcinv$, the $1\sigma$ errors on $q_\parallel$ and $q_\perp$ are reduced by approximately $10$--$20\%$ relative to the galaxy sample, in line with the compressed-parameter constraints shown in Fig.~\ref{fig:posterior_compressed}.

Interestingly, the improvement is slightly more pronounced for the LOS dilation parameter $q_\parallel$ than for $q_\perp$.
This trend is naturally expected if the dominant limiting systematic in the galaxy sample is FoG smearing, which preferentially degrades information along the LOS and therefore propagates more strongly into $q_\parallel$.
By construction, the CG reconstruction suppresses satellite-driven FoG distortions at the field level, and the enhanced gain in $q_\parallel$ reflects the fact that the recovered LOS information is less entangled with non-perturbative velocity effects than in the galaxy case.

Unlike the other parameters, $dm$ shows little sensitivity to the choice of tracer: up to $k_{\max}\simeq 0.2\,h\,\mathrm{Mpc}^{-1}$, the three samples exhibit nearly identical central values and uncertainties.
This behavior is expected because $dm$ is primarily determined by the broad-band shape of the power spectrum and is only very weakly degenerate with the anisotropic stochastic contribution $\mathrm{SN}_{2\mathrm{p}}$, indicating that it is comparatively insensitive to FoG smearing and other small-scale anisotropies.
Consequently, even if satellite-driven distortions are reduced, as is the case for the halo and reconstructed-halo samples, there is limited additional room for improvement in $dm$, explaining why reconstruction yields only marginal gains relative to galaxies.

\section{Validation of the large-scale rescaling}
\label{app:large_scale_rescaling}

\begin{figure}
\centering
\includegraphics[width=.48\textwidth]{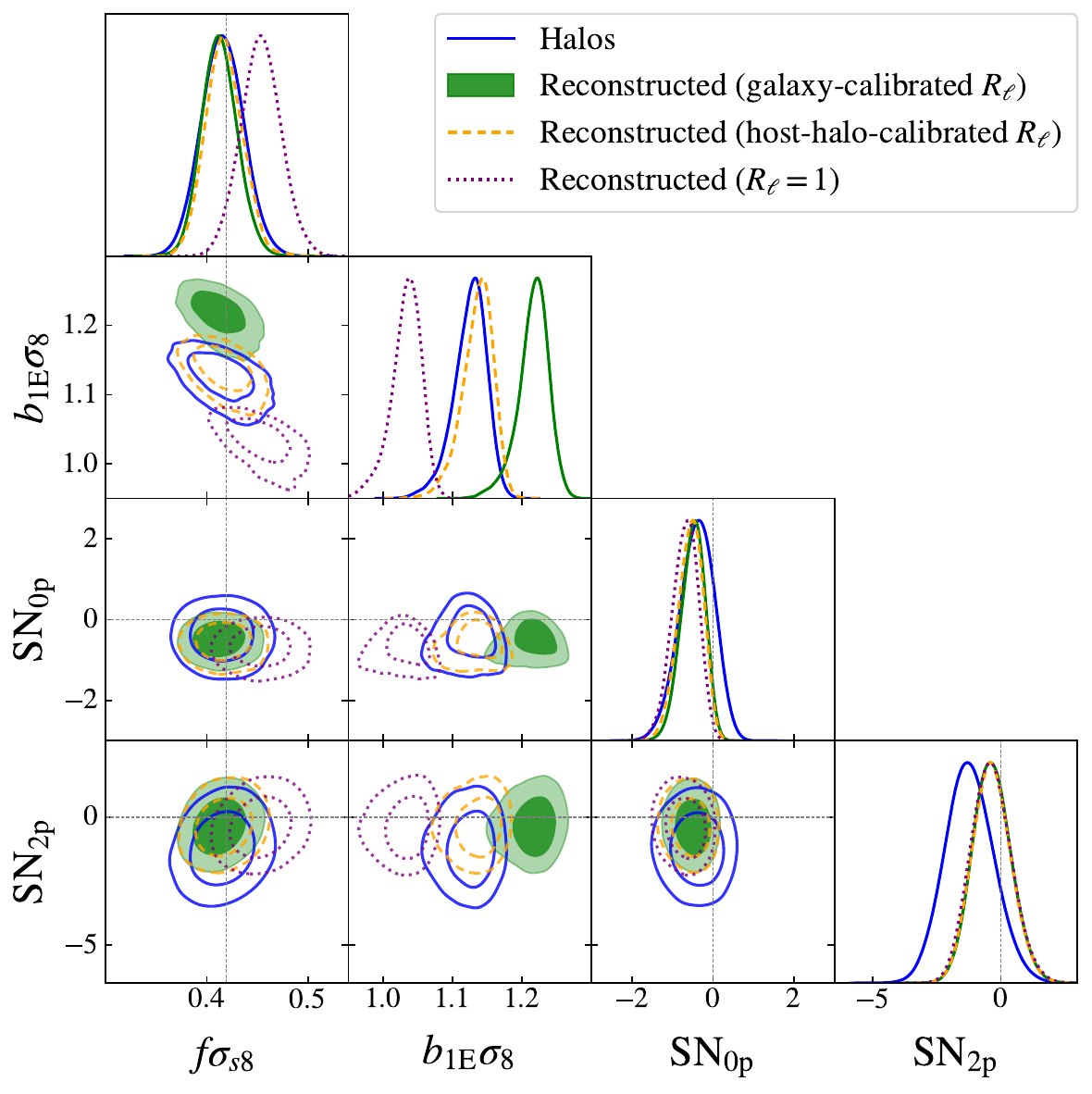}
\caption{
Same as Fig.~\ref{fig:fsig_sn}, but showing the host-halo sample (blue) together with the reconstructed-halo sample analyzed using three treatments of the large-scale rescaling: the fiducial galaxy-calibrated $R_\ell$ (green), the host-halo-calibrated $R_\ell$ (orange dashed), and no rescaling, $R_\ell=1$ (purple dotted).
}
\label{fig:rescaling_posterior}
\end{figure}

\begin{figure*}
\centering

\includegraphics[width=.98\textwidth]{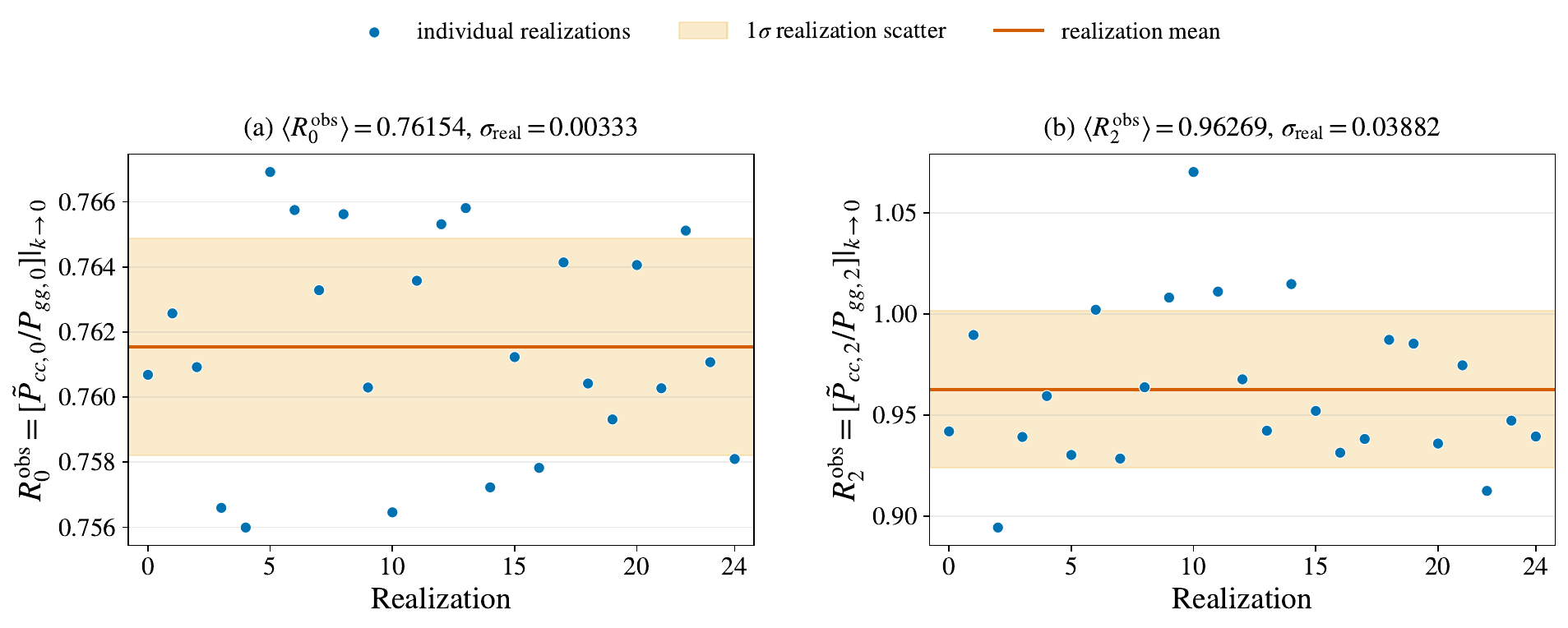}

\vspace{0.5em}

\includegraphics[width=.50\textwidth]{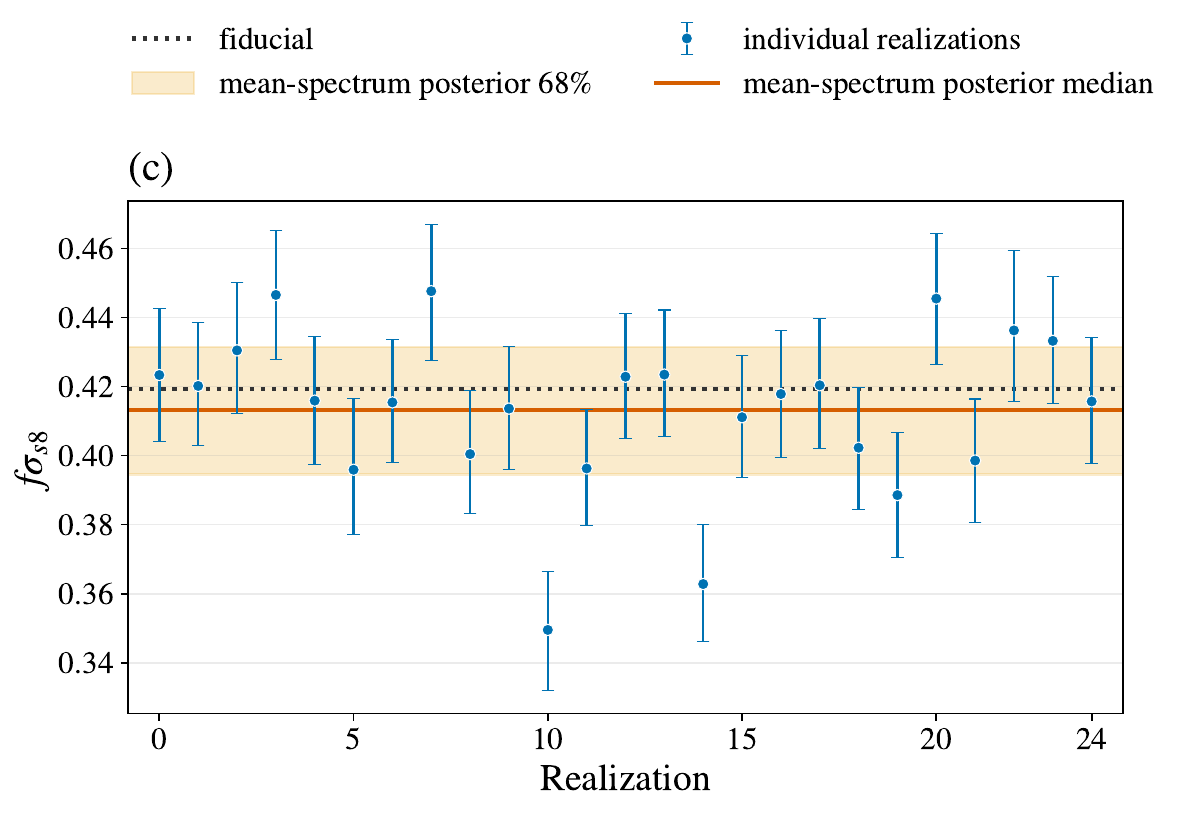}
\includegraphics[width=.47\textwidth]{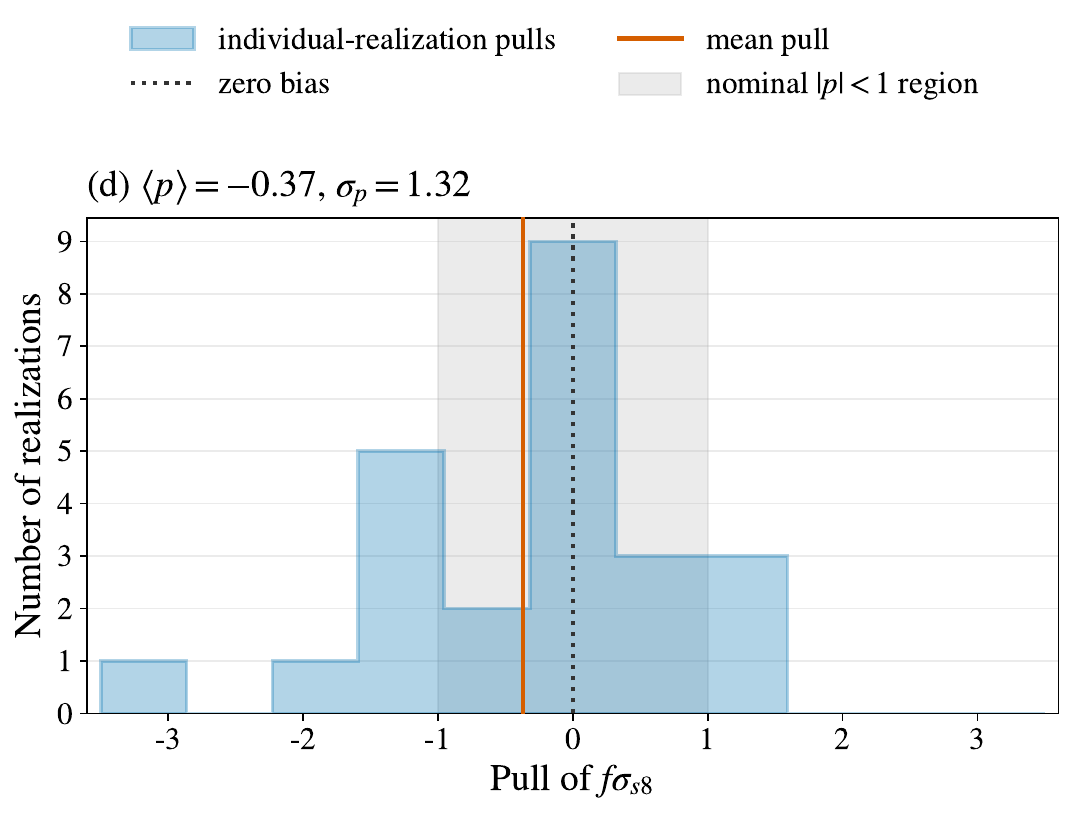}

\caption{
Validation of the fiducial large-scale rescaling $R_\ell^{\mathrm{obs}}$ and the resulting $f\sigma_{s8}$ constraints across the 25 realizations. 
Panels (a) and (b) show the scaling amplitudes $R_0^{\mathrm{obs}}$ and $R_2^{\mathrm{obs}}$ measured separately from each of the 25 realizations. The horizontal lines and shaded regions indicate the realization means and standard deviations, respectively.
Panel (c) shows the marginalized posterior median and 68\% credible interval of $f\sigma_{s8}$ for each realization. The black dotted line indicates the fiducial value, while the orange line and shaded region show the posterior median and 68\% credible interval obtained from the chain fitted to the mean power spectrum. Panel (d) shows the corresponding pull distribution. The black dotted line marks zero bias, the gray shaded region indicates $|p|<1$, and the orange line shows the mean pull.
}
\label{fig:rescaling_realizations}
\end{figure*}

\begin{figure*}
\centering
\includegraphics[width=.49\textwidth]{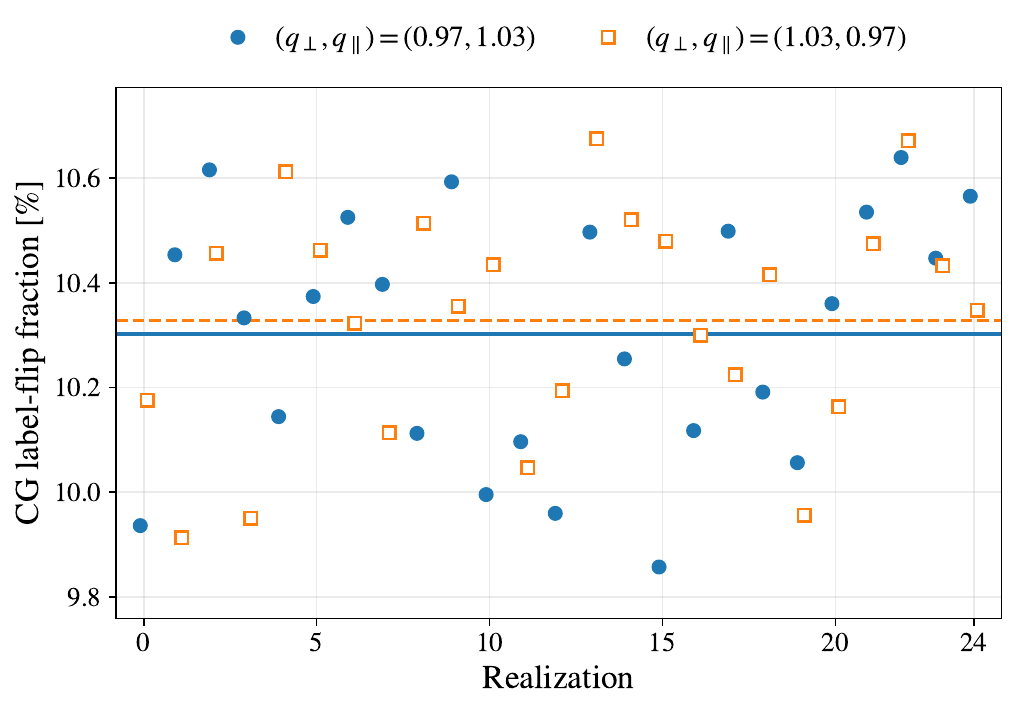}
\hfill
\includegraphics[width=.49\textwidth]{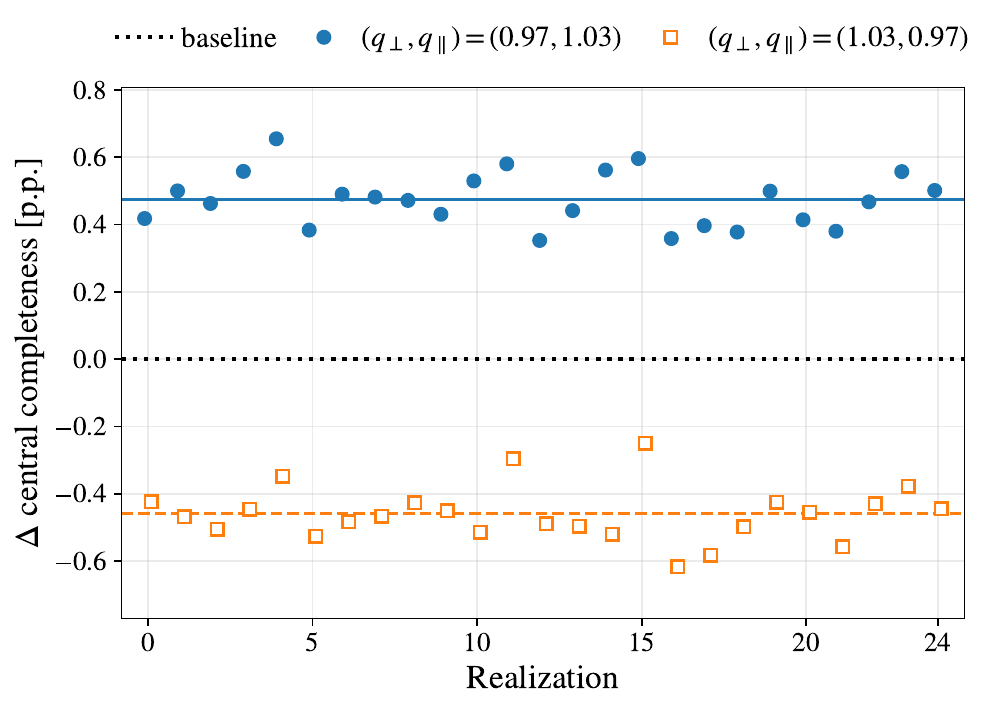}
\caption{
Catalog-level response of the CG reconstruction to imposed AP distortions across the 25 realizations.
The left panel shows the fraction of input galaxies whose CG labels change relative to the baseline reconstruction.
The right panel shows the corresponding change in central completeness.
Filled blue circles and open orange squares represent $(q_\perp,q_\parallel)=(0.97,1.03)$ and $(1.03,0.97)$, respectively.
The points are slightly offset horizontally for clarity, and the colored horizontal lines indicate the means across the realizations.
The black dotted line in the right panel marks the baseline value.
Changes in central completeness are expressed in percentage points.
}
\label{fig:ap_reconstruction_diagnostics}
\end{figure*}

In this appendix we test the dependence of the reconstructed-halo results on the treatment of the large-scale, multipole-dependent rescaling introduced in Sec.~\ref{subsec:incorp_recon_sys}. Our fiducial analysis adopts the directly measurable ratio $R_\ell^{\mathrm{obs}}=[\tilde P_{cc,\ell}(k)/P_{gg,\ell}(k)]|_{k\to 0}$.
Because this ratio is defined relative to the parent galaxy sample, it contains both the response induced by the CG reconstruction and the large-scale difference between galaxy and true-central clustering. In the simulations, the host-halo sample serves as the true-central reference, and, following the notation adopted in Sec.~\ref{subsec:catalog_number}, its power-spectrum multipoles are denoted by $P_{cc,\ell}$. 
We can therefore define the host-halo-calibrated ratio $R_\ell^{\mathrm{halo}}=[\tilde{P}_{cc,\ell}(k)/P_{cc,\ell}(k)]|_{k\to 0}$ and directly test whether the choice of reference sample affects the inferred parameters.

As shown in panels (a) and (b) of Fig.~\ref{fig:rescaling_realizations}, across the 25 realizations we find $R_0^{\mathrm{obs}}=0.76154\pm0.00333$ and $R_2^{\mathrm{obs}}=0.96269\pm0.03882$ for the galaxy-calibrated rescaling. For the host-halo-calibrated rescaling, $R_0^{\mathrm{halo}}=0.85818\pm0.00293$ and $R_2^{\mathrm{halo}}=1.00987\pm0.02810$. Here, the quoted uncertainties denote the standard deviations among the 25 realizations, rather than errors on the mean. To compare directly with our fiducial analysis, we perform the host-halo-calibrated parameter inference for the mean power spectrum using the fiducial scale cut, $k_{\max}=0.2\hmpcinv$.

Figure~\ref{fig:rescaling_posterior} compares the resulting parameter constraints. Using the fiducial galaxy-calibrated rescaling gives $f\sigma_{s8}=0.412\pm0.017$, while the host-halo-calibrated rescaling gives $f\sigma_{s8}=0.417\pm0.018$. The two constraints are mutually consistent. The primary change is instead absorbed by the linear bias: $b_{1\mathrm{E}}\sigma_8=1.218\pm0.021$ for the galaxy-calibrated case and $b_{1\mathrm{E}}\sigma_8=1.138\pm0.022$ for the host-halo-calibrated case.
The latter is close to and statistically consistent with the value inferred directly from the true host-halo sample, $b_{1\mathrm{E}}\sigma_8=1.127 \pm 0.023$. This behavior is expected: the inferred effective bias is closer to the galaxy value when $R_\ell$ is calibrated relative to galaxies (see Sec.~\ref{subsec:degeneracy_params}), and closer to the host-halo value when it is calibrated relative to host halos.
The higher-order bias, counterterm, and stochastic parameters remain nearly unchanged. The largest shift among $q_\parallel$, $q_\perp$, and $dm$ occurs for $dm$ and is approximately $0.1\sigma$, while the AP-parameter central values are unchanged at the quoted precision. These results show that, for the mock sample considered here, the observable $\tilde P_{cc}/P_{gg}$ ratio provides a suitable practical proxy for the response defined relative to host halos in cosmological inference, although the inferred value of $b_{1\mathrm{E}}\sigma_8$ depends on the adopted normalization.

As an additional control, we repeat the fit without applying the multipole-dependent rescaling. Setting $R_\ell=1$ changes the inferred growth constraint to $f\sigma_{s8}=0.453\pm0.020$, producing the visible shift shown in Fig.~\ref{fig:rescaling_posterior}. Thus, although the cosmological constraints are insensitive to whether the rescaling is calibrated relative to galaxies or host halos, the multipole-dependent correction itself cannot simply be omitted.

We further test the effect of statistical fluctuations in the measured rescaling $R_{\ell}^{\mathrm{obs}}$ by analyzing all 25 realizations separately. For realization $i$, we use the values $R_{0,i}^{\mathrm{obs}}$ and $R_{2,i}^{\mathrm{obs}}$ measured from the same realization as the reconstructed-halo power-spectrum data vector. This is therefore a test of the full procedure using each realization separately. It directly captures the joint effect of realization-to-realization fluctuations in the power spectrum and in the measured rescaling.
For each realization $i$, we define the pull as $p_i=[(f\sigma_{s8})_{i,\mathrm{med}}-(f\sigma_{s8})_{\rm fid}]/\sigma_i$, where $\sigma_i$ is half the width of the central $68\%$ credible interval. The empirical coverage is the fraction of realizations whose central $68\%$ interval contains the fiducial value.

As shown in Fig.~\ref{fig:rescaling_realizations}, the ensemble mean of the $f\sigma_{s8}$ pull is $\langle p\rangle=-0.37$, with a standard deviation $\sigma_p=1.32$, while the empirical coverage of the nominal 68\% credible intervals is 60\%.
The mean pull is consistent with no significant ensemble bias, although its dispersion is somewhat broader than unit width, suggesting that the
individual posterior uncertainties may be mildly underestimated. Fluctuations in $R_2$ are correlated with the inferred $f\sigma_{s8}$ residual, with a Pearson coefficient $r=-0.76$, as expected because $R_2$ directly affects the relative amplitude of the monopole and quadrupole. Nevertheless, this correlation does not produce a significant shift in the ensemble mean. The remaining compressed parameters, $q_\parallel$, $q_\perp$, and $dm$, likewise show no significant mean bias. The mean goodness of fit across the individual realizations is $\langle\chi^2_{\min}/\mathrm{dof}\rangle=1.034$.

Together, the calibration comparison and the closure test across the 25 realizations show that the cosmological results are not driven by the use of the galaxy power spectrum as the observable reference for $R_\ell$. They also confirm that accounting for the reconstruction-induced, multipole-dependent
change in the large-scale amplitude is necessary for unbiased inference.

\section{Robustness to Alcock--Paczynski distortions}
\label{app:ap_distortions}

The CG reconstruction is performed in fiducial comoving coordinates, so an AP distortion can alter the galaxies included within the grouping cylinder.
To test this catalog-level dependence, we map the simulation coordinates to distorted fiducial coordinates according to
\begin{equation}
(x,y,z)\mapsto
\left(
\frac{x}{q_\perp},
\frac{y}{q_\perp},
\frac{z}{q_\parallel}
\right).
\end{equation}
The dimensions of the periodic box are transformed by the same factors.
We apply this transformation before computing the environmental ranking and performing the CG reconstruction.
We consider $(q_\perp,q_\parallel)=(0.97,1.03)$ and $(1.03,0.97)$ and repeat the reconstruction for all 25 realizations, keeping the numerical cylinder parameters fixed in the distorted coordinates.
The resulting catalogs are compared object by object with the baseline reconstruction at $(q_\perp,q_\parallel)=(1,1)$.

We define the CG label-flip fraction as the fraction of input galaxies whose status changes between being retained as a reconstructed-halo tracer and being removed by the grouping.
As shown in the left panel of Fig.~\ref{fig:ap_reconstruction_diagnostics}, both imposed distortions produce a mean label-flip fraction of approximately $10.3\%$.
Thus, the detailed group membership is sensitive to modest changes in the assumed coordinate mapping.

Nevertheless, these changes in individual assignments largely cancel in the bulk reconstruction diagnostics.
Relative to the baseline central completeness of approximately $80.16\%$, the mean changes are only about $+0.47$ and $-0.46$ percentage points for $(q_\perp,q_\parallel)=(0.97,1.03)$ and $(1.03,0.97)$, respectively.
We also find similarly small changes in the central purity, residual satellite fraction, and abundance of the reconstructed sample.
The overall performance of the reconstruction is therefore stable under the AP distortions considered here, despite changes in the assignments of individual galaxies.

This test isolates the sensitivity of the CG assignments to the assumed coordinate mapping. A fully self-consistent test away from the fiducial cosmology, including repeated parameter inference, is left for future work.

\begin{figure*}
\centering
\hspace{-1.5em}
\includegraphics[
  height=0.37\textheight
]{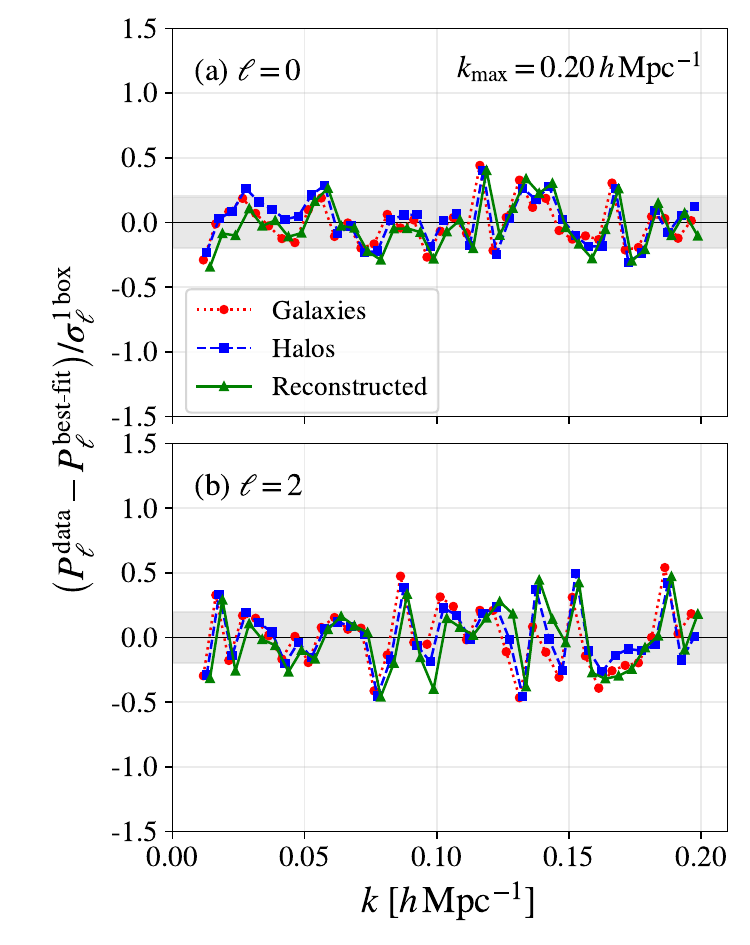}
\hspace{-1.0em}
\includegraphics[
  height=0.37\textheight,
  trim=60 0 0 0,
  clip
]{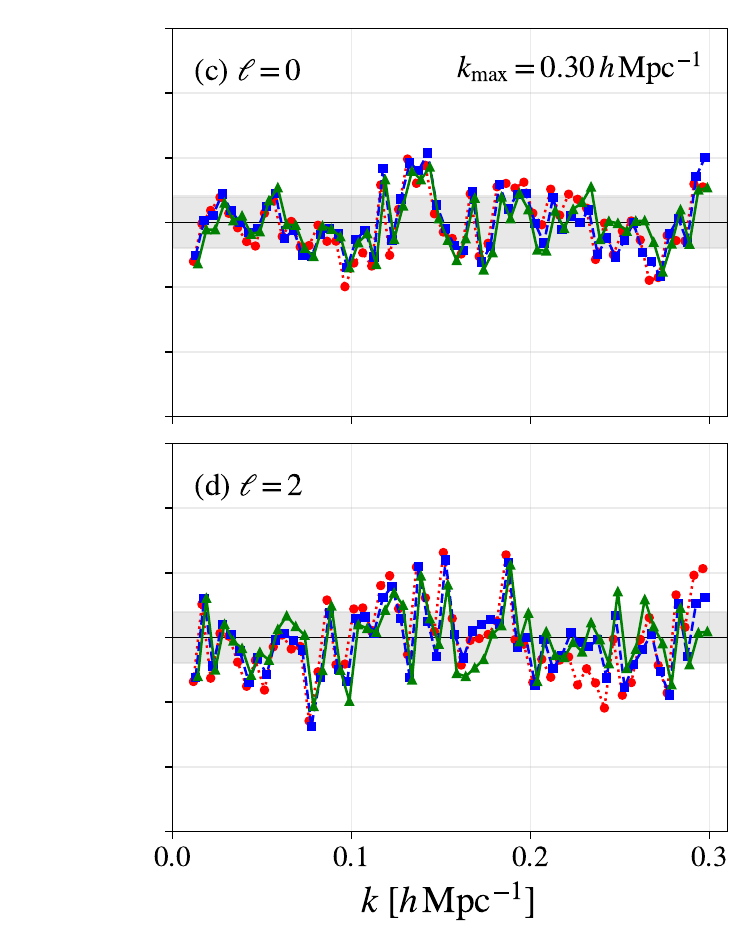}
\hspace{-1.0em}
\includegraphics[
  height=0.37\textheight,
  trim=60 0 0 0,
  clip
]{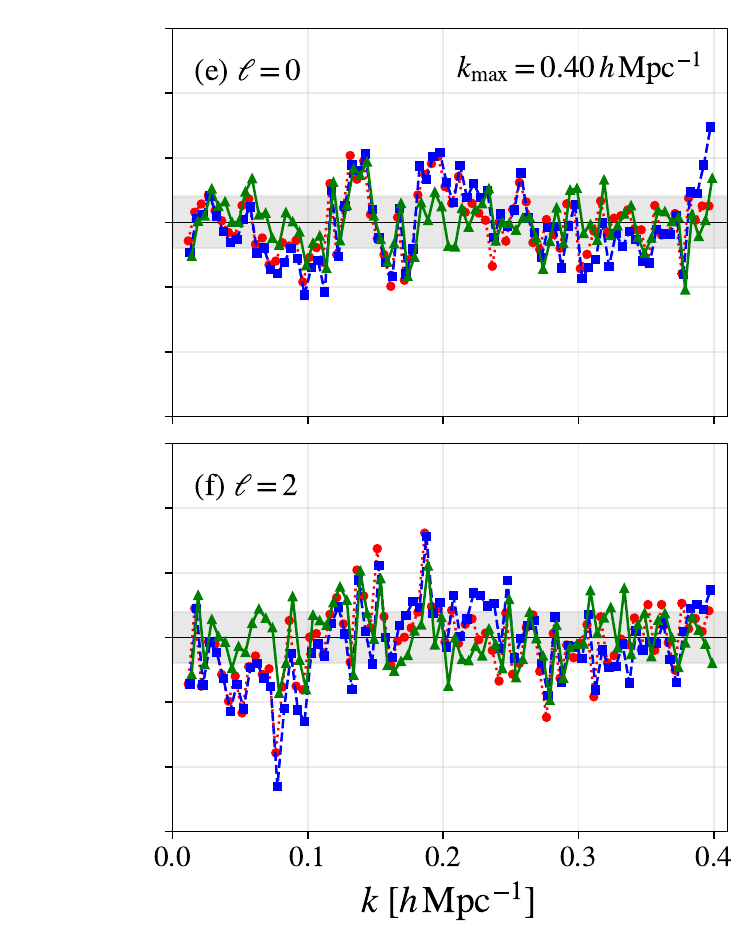}
\caption{
Residuals of the best-fit monopole (upper panels) and quadrupole (lower panels) for galaxies (red), host halos (blue), and reconstructed halos (green). From left to right, the fits use $k_{\max}=0.2$, $0.3$, and $0.4\hmpcinv$.
The residuals are normalized by the standard deviation of one simulation box, $\sigma_{\ell}^{\rm 1\,box}$. The gray bands show $\pm 1/\sqrt{N_{\rm sim}}$, corresponding to the nominal $1\sigma$ scatter of the mean over the $N_{\rm sim}=25$ realizations.
}
\label{fig:highk_residuals}
\end{figure*}

\begin{figure*}
\centering
\includegraphics[width=0.65\textwidth]
{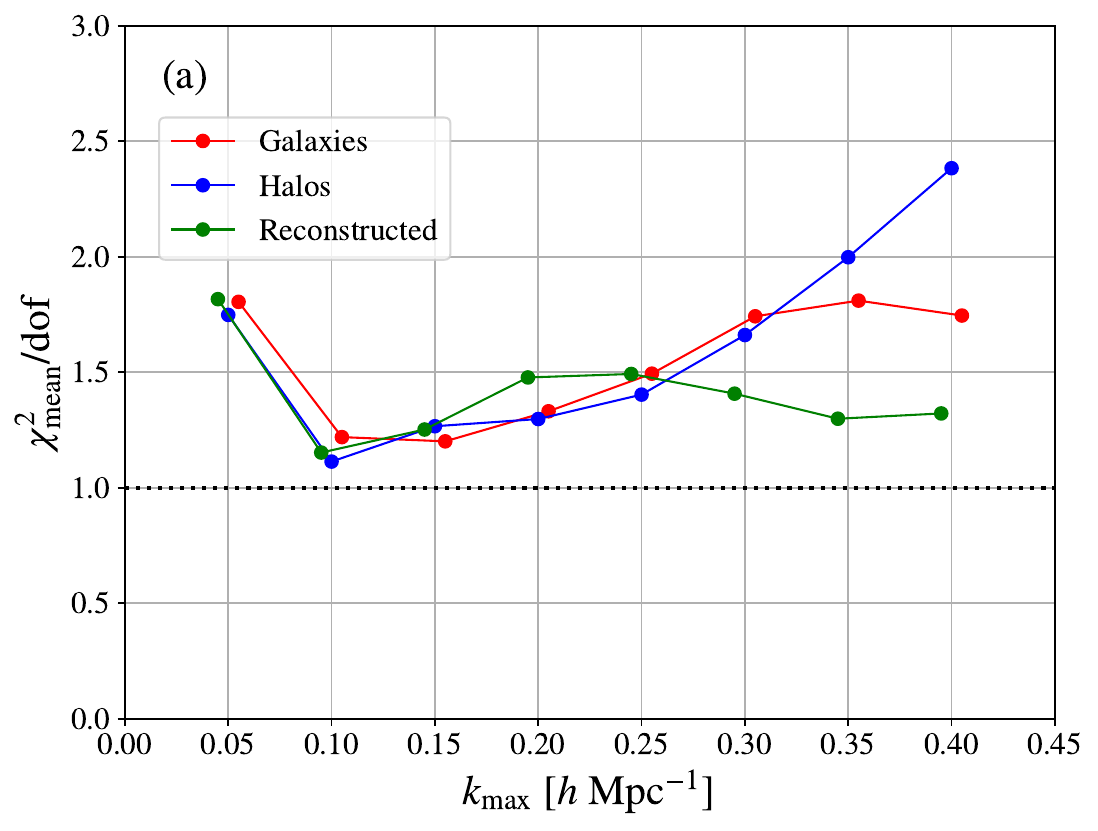}
\hfill

\vspace{2ex}

\includegraphics[width=0.49\textwidth]
{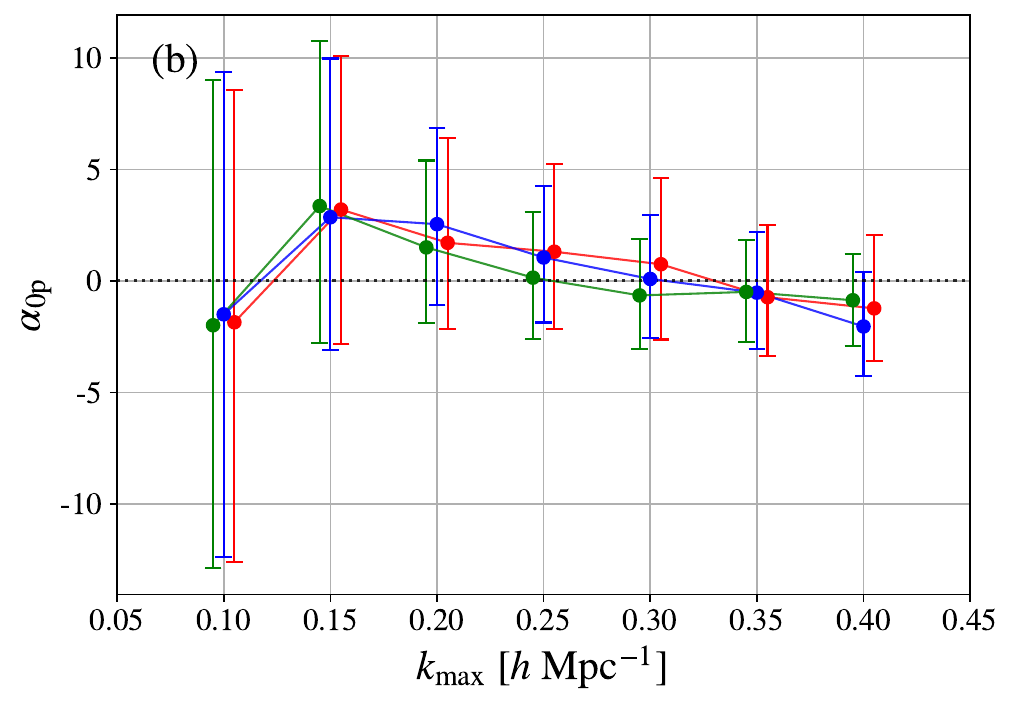}
\includegraphics[width=0.49\textwidth]
{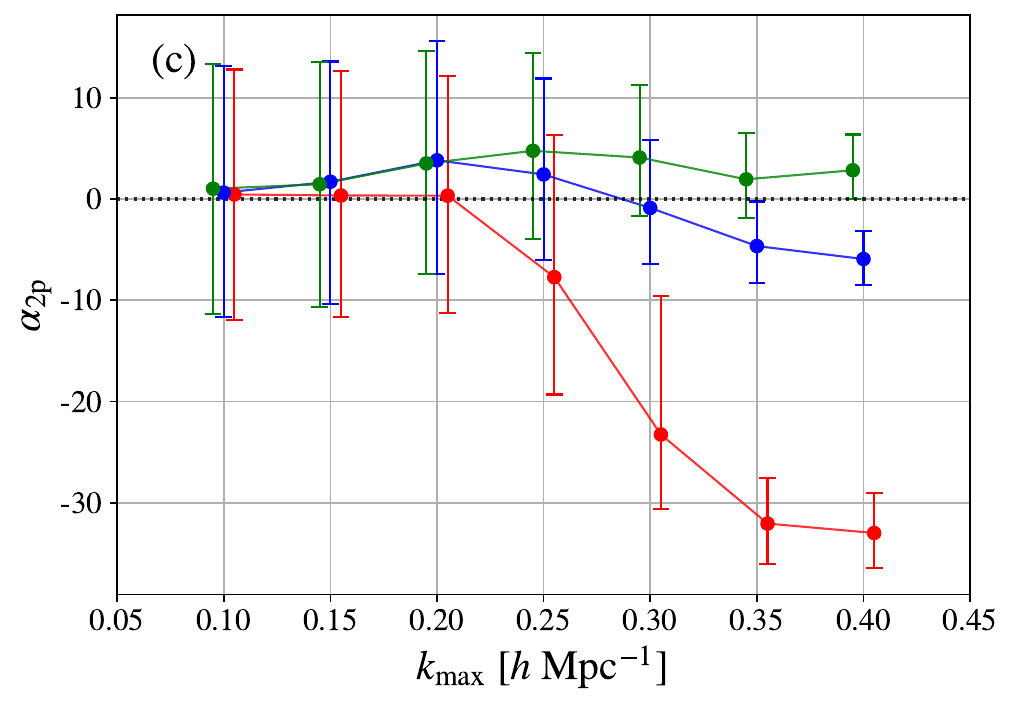}

\vspace{1ex}

\hspace{-0.5em}
\includegraphics[width=0.50\textwidth]
{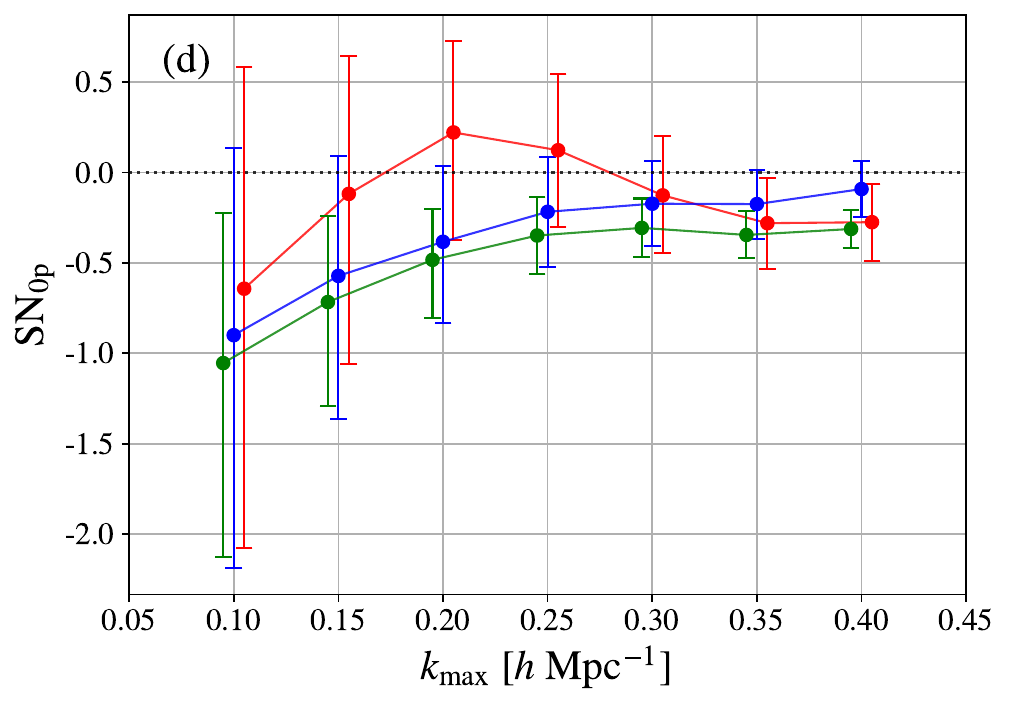}
\hspace{-0.3em}
\includegraphics[width=0.49\textwidth]
{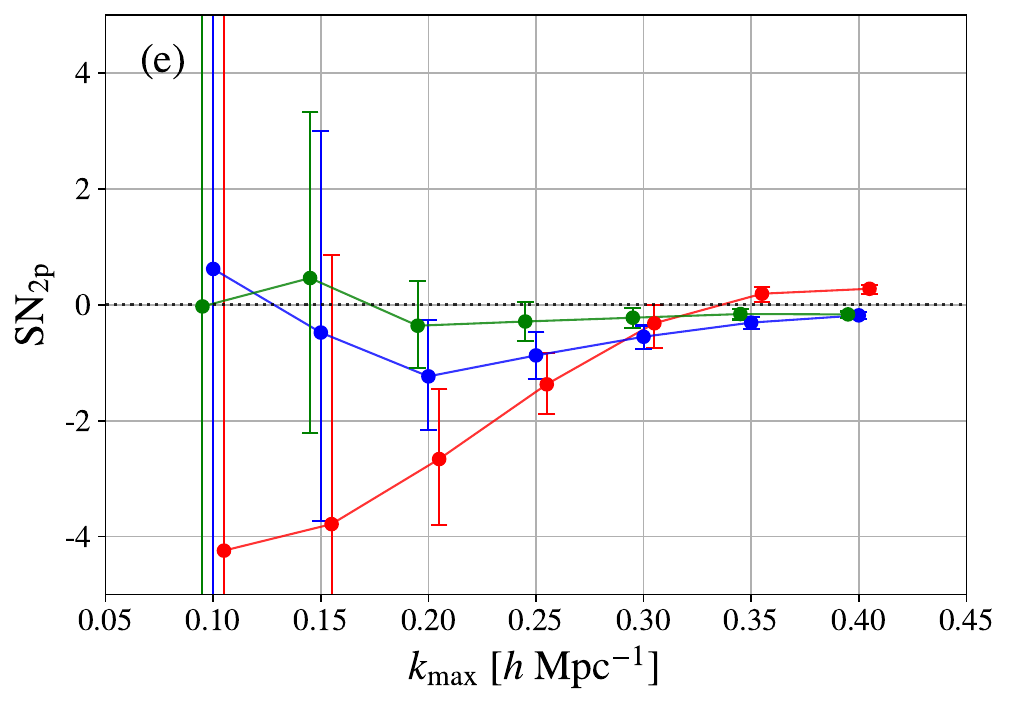}
\caption{
Goodness of fit and constraints on the nuisance parameters as a function of $k_{\max}$ for galaxies (red), host halos (blue), and reconstructed halos (green).
Panel (a) shows $\chi^2_{\rm mean}/\mathrm{dof}$, evaluated using the covariance matrix of the mean, $\boldsymbol{C}_{\rm mean}=\boldsymbol{C}_{\rm 1\,box}/N_{\rm sim}$; the horizontal dotted line marks the expectation value of unity.
Panels (b)--(e) show the marginalized constraints on $\alpha_{0\mathrm{p}}$, $\alpha_{2\mathrm{p}}$, ${\rm SN}_{0\mathrm{p}}$, and ${\rm SN}_{2\mathrm{p}}$, respectively.
Points denote posterior medians and error bars show the central $68\%$ credible intervals.
In panel (e), some of the credible intervals at low $k_{\max}$ extend beyond the displayed vertical range.
The dotted lines in panels (b)--(e) indicate zero reference values.
The tracer points are slightly displaced horizontally for visual clarity.
}
\label{fig:highk_diagnostics}

\end{figure*}

\section{High-\texorpdfstring{$k_{\max}$}{kmax} model diagnostics}
\label{app:highk}

In this appendix, we examine the goodness of fit and the behavior of the EFT nuisance parameters as the maximum wavenumber is extended beyond the fiducial choice of $k_{\max}=0.2\hmpcinv$.
The purpose of these tests is to identify the origin of the apparent stability of the cosmological parameters at high $k_{\max}$, rather than to establish an extended range of validity for the perturbative model.

Figure~\ref{fig:highk_residuals} shows the residuals of the monopole and quadrupole for fits with $k_{\max}=0.2$, $0.3$, and $0.4\hmpcinv$.
For each tracer and each value of $k_{\max}$, the model is evaluated at its corresponding best-fitting parameter point.
The residuals are normalized by the standard deviation measured from a single simulation box, $\sigma_{\ell}^{\rm 1\,box}(k)$,
\begin{equation}
\Delta_{\ell}(k)
\equiv
\frac{
P_{\ell}^{\rm data}(k)-P_{\ell}^{\rm best\mbox{-}fit}(k)
}{
\sigma_{\ell}^{\rm 1\,box}(k)
}.
\end{equation}
Because the data vector is the mean of $N_{\rm sim}=25$ realizations, its expected statistical uncertainty corresponds to $\sigma_{\ell}^{\rm 1\,box}/\sqrt{N_{\rm sim}}$.
The gray bands in Fig.~\ref{fig:highk_residuals} therefore indicate $\lvert\Delta_{\ell}\rvert < 1/\sqrt{N_{\rm sim}}$, i.e., the nominal $1\sigma$ range for the mean spectrum.

At the fiducial limit, $k_{\max}=0.2\hmpcinv$, the residuals are generally small and show no pronounced scale-dependent trend for any of the three tracers.
As the fitting range is extended, coherent residual structures become more visible, although no single Fourier bin dominates the discrepancy.
This behavior indicates that the departures at high $k$ are primarily broadband and can consequently be partially absorbed by the counterterm and stochastic sectors of the model.

We quantify the same behavior in Fig.~\ref{fig:highk_diagnostics}.
Panel (a) shows the reduced chi square,
\begin{equation}
\frac{\chi_{\rm mean}^{2}}{\mathrm{dof}},
\end{equation}
where $\mathrm{dof}=N_{\rm data}-N_{\rm par}$. Here $N_{\rm data}$ is the total number of monopole and quadrupole data points included in the fit, and $N_{\rm par}$ is the number of free model parameters, including the analytically marginalized nuisance parameters.
The quantity $\chi_{\rm mean}^{2}$ is evaluated for the mean spectrum using the covariance of the mean, $\boldsymbol{C}_{\rm mean}=\boldsymbol{C}_{\rm 1\,box}/N_{\rm sim}$. Equivalently, it is obtained using the one-box covariance with the resulting chi square multiplied by $N_{\rm sim}$; this factor is therefore already included in the definition of $\chi_{\rm mean}^{2}$. We use the minimum chi square recovered from the chain, excluding prior contributions, with the analytically marginalized nuisance parameters fixed to their conditional maximum-likelihood values. For a statistically consistent model, the expectation value of $\chi_{\rm mean}^{2}/\mathrm{dof}$ is approximately unity, as indicated by the horizontal dotted line.

Around the fiducial fitting limit, the reduced chi square remains of order unity for all three tracers.
Toward higher $k_{\max}$, it increases for the galaxy and especially the host-halo samples, whereas the reconstructed-halo result remains comparatively stable.
This comparison suggests that extending the fits progressively exposes broadband contributions that are not fully described by the perturbative template.

Panels (b)--(e) show the corresponding marginalized constraints on the counterterm parameters $\alpha_{0\mathrm{p}}$ and $\alpha_{2\mathrm{p}}$, and the stochastic parameters ${\rm SN}_{0\mathrm{p}}$ and ${\rm SN}_{2\mathrm{p}}$.
Since these parameters are analytically marginalized in the likelihood, their marginalized distributions are reconstructed by sampling their joint conditional distribution for every MCMC sample.
The points and error bars denote the posterior medians and central $68\%$ credible intervals, respectively.

The isotropic counterterm $\alpha_{0\mathrm{p}}$ displays only a mild dependence on $k_{\max}$.
In contrast, the anisotropic counterterm $\alpha_{2\mathrm{p}}$ develops a substantial negative shift for the galaxy sample once scales beyond
$k_{\max}\simeq0.2\hmpcinv$ are included.
A weaker high-$k_{\max}$ shift is also visible for the host halos, while the reconstructed-halo result remains substantially more stable.
The stochastic monopole parameter ${\rm SN}_{0\mathrm{p}}$ shows comparatively smooth evolution, whereas ${\rm SN}_{2\mathrm{p}}$ exhibits a stronger dependence on the fitting range for galaxies.
The reconstructed sample again shows the weakest variation.

These trends demonstrate that the apparent high-$k_{\max}$ stability of the cosmological parameters does not by itself imply that the perturbative model remains equally accurate on those scales.
For the galaxy sample in particular, part of the additional small-scale contribution is absorbed through shifts in the anisotropic counterterm and stochastic parameters.
Halo reconstruction reduces these shifts, consistent with its suppression of satellite-induced FoG contributions, but does not eliminate the need for a conservative modeling limit.
We therefore retain $k_{\max}=0.2\hmpcinv$ as the fiducial choice for the main cosmological results and regard the higher-$k_{\max}$ fits as diagnostic tests of the relative behavior of the three tracers.


\bibliography{apssamp}

@ARTICLE{PFS:2014,
       author = {{Takada}, Masahiro and {Ellis}, Richard S. and {Chiba}, Masashi and {Greene}, Jenny E. and {Aihara}, Hiroaki and {Arimoto}, Nobuo and {Bundy}, Kevin and {Cohen}, Judith and {Dor{\'e}}, Olivier and {Graves}, Genevieve and {Gunn}, James E. and {Heckman}, Timothy and {Hirata}, Christopher M. and {Ho}, Paul and {Kneib}, Jean-Paul and {Le F{\`e}vre}, Olivier and {Lin}, Lihwai and {More}, Surhud and {Murayama}, Hitoshi and {Nagao}, Tohru and {Ouchi}, Masami and {Seiffert}, Michael and {Silverman}, John D. and {Sodr{\'e}}, Laerte and {Spergel}, David N. and {Strauss}, Michael A. and {Sugai}, Hajime and {Suto}, Yasushi and {Takami}, Hideki and {Wyse}, Rosemary},
        title = "{Extragalactic science, cosmology, and Galactic archaeology with the Subaru Prime Focus Spectrograph}",
      journal = {\pasj},
         year = 2014,
        month = feb,
       volume = {66},
       number = {1},
          eid = {R1},
        pages = {R1},
          doi = {10.1093/pasj/pst019},
archivePrefix = {arXiv},
       eprint = {1206.0737},
 primaryClass = {astro-ph.CO},
       adsurl = {https://ui.adsabs.harvard.edu/abs/2014PASJ...66R...1T}
}

@ARTICLE{DESI:2016,
       author = {{DESI Collaboration} and {Aghamousa}, Amir and {Aguilar}, Jessica and {Ahlen}, Steve and {Alam}, Shadab and {Allen}, Lori E. and {Allende Prieto}, Carlos and {Annis}, James and {Bailey}, Stephen and {Balland}, Christophe and {Ballester}, Otger and {Baltay}, Charles and {Beaufore}, Lucas and {Bebek}, Chris and {Beers}, Timothy C. and {Bell}, Eric F. and {Bernal}, Jos{\'e} Luis and {Besuner}, Robert and {Beutler}, Florian and {Blake}, Chris and {Bleuler}, Hannes and {Blomqvist}, Michael and {Blum}, Robert and {Bolton}, Adam S. and {Briceno}, Cesar and {Brooks}, David and {Brownstein}, Joel R. and {Buckley-Geer}, Elizabeth and {Burden}, Angela and {Burtin}, Etienne and {Busca}, Nicolas G. and {Cahn}, Robert N. and {Cai}, Yan-Chuan and {Cardiel-Sas}, Laia and {Carlberg}, Raymond G. and {Carton}, Pierre-Henri and {Casas}, Ricard and {Castander}, Francisco J. and {Cervantes-Cota}, Jorge L. and {Claybaugh}, Todd M. and {Close}, Madeline and {Coker}, Carl T. and {Cole}, Shaun and {Comparat}, Johan and {Cooper}, Andrew P. and {Cousinou}, M.-C. and {Crocce}, Martin and {Cuby}, Jean-Gabriel and {Cunningham}, Daniel P. and {Davis}, Tamara M. and {Dawson}, Kyle S. and {de la Macorra}, Axel and {De Vicente}, Juan and {Delubac}, Timoth{\'e}e and {Derwent}, Mark and {Dey}, Arjun and {Dhungana}, Govinda and {Ding}, Zhejie and {Doel}, Peter and {Duan}, Yutong T. and {Ealet}, Anne and {Edelstein}, Jerry and {Eftekharzadeh}, Sarah and {Eisenstein}, Daniel J. and {Elliott}, Ann and {Escoffier}, St{\'e}phanie and {Evatt}, Matthew and {Fagrelius}, Parker and {Fan}, Xiaohui and {Fanning}, Kevin and {Farahi}, Arya and {Farihi}, Jay and {Favole}, Ginevra and {Feng}, Yu and {Fernandez}, Enrique and {Findlay}, Joseph R. and {Finkbeiner}, Douglas P. and {Fitzpatrick}, Michael J. and {Flaugher}, Brenna and {Flender}, Samuel and {Font-Ribera}, Andreu and {Forero-Romero}, Jaime E. and {Fosalba}, Pablo and {Frenk}, Carlos S. and {Fumagalli}, Michele and {Gaensicke}, Boris T. and {Gallo}, Giuseppe and {Garcia-Bellido}, Juan and {Gaztanaga}, Enrique and {Pietro Gentile Fusillo}, Nicola and {Gerard}, Terry and {Gershkovich}, Irena and {Giannantonio}, Tommaso and {Gillet}, Denis and {Gonzalez-de-Rivera}, Guillermo and {Gonzalez-Perez}, Violeta and {Gott}, Shelby and {Graur}, Or and {Gutierrez}, Gaston and {Guy}, Julien and {Habib}, Salman and {Heetderks}, Henry and {Heetderks}, Ian and {Heitmann}, Katrin and {Hellwing}, Wojciech A. and {Herrera}, David A. and {Ho}, Shirley and {Holland}, Stephen and {Honscheid}, Klaus and {Huff}, Eric and {Hutchinson}, Timothy A. and {Huterer}, Dragan and {Hwang}, Ho Seong and {Illa Laguna}, Joseph Maria and {Ishikawa}, Yuzo and {Jacobs}, Dianna and {Jeffrey}, Niall and {Jelinsky}, Patrick and {Jennings}, Elise and {Jiang}, Linhua and {Jimenez}, Jorge and {Johnson}, Jennifer and {Joyce}, Richard and {Jullo}, Eric and {Juneau}, St{\'e}phanie and {Kama}, Sami and {Karcher}, Armin and {Karkar}, Sonia and {Kehoe}, Robert and {Kennamer}, Noble and {Kent}, Stephen and {Kilbinger}, Martin and {Kim}, Alex G. and {Kirkby}, David and {Kisner}, Theodore and {Kitanidis}, Ellie and {Kneib}, Jean-Paul and {Koposov}, Sergey and {Kovacs}, Eve and {Koyama}, Kazuya and {Kremin}, Anthony and {Kron}, Richard and {Kronig}, Luzius and {Kueter-Young}, Andrea and {Lacey}, Cedric G. and {Lafever}, Robin and {Lahav}, Ofer and {Lambert}, Andrew and {Lampton}, Michael and {Landriau}, Martin and {Lang}, Dustin and {Lauer}, Tod R. and {Le Goff}, Jean-Marc and {Le Guillou}, Laurent and {Le Van Suu}, Auguste and {Lee}, Jae Hyeon and {Lee}, Su-Jeong and {Leitner}, Daniela and {Lesser}, Michael and {Levi}, Michael E. and {L'Huillier}, Benjamin and {Li}, Baojiu and {Liang}, Ming and {Lin}, Huan and {Linder}, Eric and {Loebman}, Sarah R. and {Luki{\'c}}, Zarija and {Ma}, Jun and {MacCrann}, Niall and {Magneville}, Christophe and {Makarem}, Laleh and {Manera}, Marc and {Manser}, Christopher J. and {Marshall}, Robert and {Martini}, Paul and {Massey}, Richard and {Matheson}, Thomas and {McCauley}, Jeremy and {McDonald}, Patrick and {McGreer}, Ian D. and {Meisner}, Aaron and {Metcalfe}, Nigel and {Miller}, Timothy N. and {Miquel}, Ramon and {Moustakas}, John and {Myers}, Adam and {Naik}, Milind and {Newman}, Jeffrey A. and {Nichol}, Robert C. and {Nicola}, Andrina and {Nicolati da Costa}, Luiz and {Nie}, Jundan and {Niz}, Gustavo and {Norberg}, Peder and {Nord}, Brian and {Norman}, Dara and {Nugent}, Peter and {O'Brien}, Thomas and {Oh}, Minji and {Olsen}, Knut A.~G.},
        title = "{The DESI Experiment Part I: Science,Targeting, and Survey Design}",
      journal = {arXiv e-prints},
         year = 2016,
        month = oct,
          eid = {arXiv:1611.00036},
        pages = {arXiv:1611.00036},
          doi = {10.48550/arXiv.1611.00036},
archivePrefix = {arXiv},
       eprint = {1611.00036},
 primaryClass = {astro-ph.IM},
       adsurl = {https://ui.adsabs.harvard.edu/abs/2016arXiv161100036D}
}

@ARTICLE{Kobayashi:2022,
       author = {{Kobayashi}, Yosuke and {Nishimichi}, Takahiro and {Takada}, Masahiro and {Miyatake}, Hironao},
        title = "{Full-shape cosmology analysis of the SDSS-III BOSS galaxy power spectrum using an emulator-based halo model: A 5\% determination of {\ensuremath{\sigma}}$_{8}$}",
      journal = {\prd},
         year = 2022,
        month = apr,
       volume = {105},
       number = {8},
          eid = {083517},
        pages = {083517},
          doi = {10.1103/PhysRevD.105.083517},
archivePrefix = {arXiv},
       eprint = {2110.06969},
 primaryClass = {astro-ph.CO},
       adsurl = {https://ui.adsabs.harvard.edu/abs/2022PhRvD.105h3517K}
}

@ARTICLE{Philcox:2020,
       author = {{Philcox}, Oliver H.~E. and {Ivanov}, Mikhail M. and {Simonovi{\'c}}, Marko and {Zaldarriaga}, Matias},
        title = "{Combining full-shape and BAO analyses of galaxy power spectra: a 1.6\% CMB-independent constraint on H$_{0}$}",
      journal = {\jcap},
         year = 2020,
        month = may,
       volume = {05},
       number = {5},
          eid = {032},
        pages = {032},
          doi = {10.1088/1475-7516/2020/05/032},
archivePrefix = {arXiv},
       eprint = {2002.04035},
 primaryClass = {astro-ph.CO},
       adsurl = {https://ui.adsabs.harvard.edu/abs/2020JCAP...05..032P}
}

@ARTICLE{dAmico:2020,
       author = {{d'Amico}, Guido and {Gleyzes}, J{\'e}r{\^o}me and {Kokron}, Nickolas and {Markovic}, Katarina and {Senatore}, Leonardo and {Zhang}, Pierre and {Beutler}, Florian and {Gil-Mar{\'\i}n}, H{\'e}ctor},
        title = "{The cosmological analysis of the SDSS/BOSS data from the Effective Field Theory of Large-Scale Structure}",
      journal = {\jcap},
         year = 2020,
        month = may,
       volume = {05},
       number = {5},
          eid = {005},
        pages = {005},
          doi = {10.1088/1475-7516/2020/05/005},
archivePrefix = {arXiv},
       eprint = {1909.05271},
 primaryClass = {astro-ph.CO},
       adsurl = {https://ui.adsabs.harvard.edu/abs/2020JCAP...05..005D}
}

@ARTICLE{Ivanov:2020,
       author = {{Ivanov}, Mikhail M. and {Simonovi{\'c}}, Marko and {Zaldarriaga}, Matias},
        title = "{Cosmological parameters from the BOSS galaxy power spectrum}",
      journal = {\jcap},
         year = 2020,
        month = may,
       volume = {05},
       number = {5},
          eid = {042},
        pages = {042},
          doi = {10.1088/1475-7516/2020/05/042},
archivePrefix = {arXiv},
       eprint = {1909.05277},
 primaryClass = {astro-ph.CO},
       adsurl = {https://ui.adsabs.harvard.edu/abs/2020JCAP...05..042I}
}

@article{Okumura:2008,
	adsurl = {http://adsabs.harvard.edu/abs/2008ApJ...676..889O},
	archiveprefix = {arXiv},
	author = {{Okumura}, T. and {Matsubara}, T. and {Eisenstein}, D.~J. and {Kayo}, I. and {Hikage}, C. and {Szalay}, A.~S. and {Schneider}, D.~P.},
	doi = {10.1086/528951},
	eid = {889-898},
	eprint = {0711.3640},
	journal = {\apj},
	month = apr,
	pages = {889-898},
	title = {{Large-Scale Anisotropic Correlation Function of SDSS Luminous Red Galaxies}},
	volume = 676,
	year = 2008}

@article{Alam:2017,
	adsurl = {https://ui.adsabs.harvard.edu/abs/2017MNRAS.470.2617A},
	archiveprefix = {arXiv},
	author = {{Alam}, Shadab and {Ata}, Metin and {Bailey}, Stephen and {Beutler}, Florian and {Bizyaev}, Dmitry and {Blazek}, Jonathan A. and {Bolton}, Adam S. and {Brownstein}, Joel R. and {Burden}, Angela and {Chuang}, Chia-Hsun and {Comparat}, Johan and {Cuesta}, Antonio J. and {Dawson}, Kyle S. and {Eisenstein}, Daniel J. and {Escoffier}, Stephanie and {Gil-Mar{\'\i}n}, H{\'e}ctor and {Grieb}, Jan Niklas and {Hand}, Nick and {Ho}, Shirley and {Kinemuchi}, Karen and {Kirkby}, David and {Kitaura}, Francisco and {Malanushenko}, Elena and {Malanushenko}, Viktor and {Maraston}, Claudia and {McBride}, Cameron K. and {Nichol}, Robert C. and {Olmstead}, Matthew D. and {Oravetz}, Daniel and {Padmanabhan}, Nikhil and {Palanque-Delabrouille}, Nathalie and {Pan}, Kaike and {Pellejero-Ibanez}, Marcos and {Percival}, Will J. and {Petitjean}, Patrick and {Prada}, Francisco and {Price-Whelan}, Adrian M. and {Reid}, Beth A. and {Rodr{\'\i}guez-Torres}, Sergio A. and {Roe}, Natalie A. and {Ross}, Ashley J. and {Ross}, Nicholas P. and {Rossi}, Graziano and {Rubi{\~n}o-Mart{\'\i}n}, Jose Alberto and {Saito}, Shun and {Salazar-Albornoz}, Salvador and {Samushia}, Lado and {S{\'a}nchez}, Ariel G. and {Satpathy}, Siddharth and {Schlegel}, David J. and {Schneider}, Donald P. and {Sc{\'o}ccola}, Claudia G. and {Seo}, Hee-Jong and {Sheldon}, Erin S. and {Simmons}, Audrey and {Slosar}, An{\v{z}}e and {Strauss}, Michael A. and {Swanson}, Molly E.~C. and {Thomas}, Daniel and {Tinker}, Jeremy L. and {Tojeiro}, Rita and {Maga{\~n}a}, Mariana Vargas and {Vazquez}, Jose Alberto and {Verde}, Licia and {Wake}, David A. and {Wang}, Yuting and {Weinberg}, David H. and {White}, Martin and {Wood-Vasey}, W. Michael and {Y{\`e}che}, Christophe and {Zehavi}, Idit and {Zhai}, Zhongxu and {Zhao}, Gong-Bo},
	doi = {10.1093/mnras/stx721},
	eprint = {1607.03155},
	journal = {\mnras},
	month = sep,
	number = {3},
	pages = {2617-2652},
	primaryclass = {astro-ph.CO},
	title = {{The clustering of galaxies in the completed SDSS-III Baryon Oscillation Spectroscopic Survey: cosmological analysis of the DR12 galaxy sample}},
	volume = {470},
	year = 2017}

@article{Okumura:2016,
	adsurl = {https://ui.adsabs.harvard.edu/abs/2016PASJ...68...38O},
	archiveprefix = {arXiv},
	author = {{Okumura}, Teppei and {Hikage}, Chiaki and {Totani}, Tomonori and {Tonegawa}, Motonari and {Okada}, Hiroyuki and {Glazebrook}, Karl and {Blake}, Chris and {Ferreira}, Pedro G. and {More}, Surhud and {Taruya}, Atsushi and {Tsujikawa}, Shinji and {Akiyama}, Masayuki and {Dalton}, Gavin and {Goto}, Tomotsugu and {Ishikawa}, Takashi and {Iwamuro}, Fumihide and {Matsubara}, Takahiko and {Nishimichi}, Takahiro and {Ohta}, Kouji and {Shimizu}, Ikkoh and {Takahashi}, Ryuichi and {Takato}, Naruhisa and {Tamura}, Naoyuki and {Yabe}, Kiyoto and {Yoshida}, Naoki},
	doi = {10.1093/pasj/psw029},
	eid = {38},
	eprint = {1511.08083},
	journal = {\pasj},
	month = jun,
	number = {3},
	pages = {38},
	primaryclass = {astro-ph.CO},
	title = {{The Subaru FMOS galaxy redshift survey (FastSound). IV. New constraint on gravity theory from redshift space distortions at z {\ensuremath{\sim}} 1.4}},
	volume = {68},
	year = 2016}

@inbook{Hamilton:1998,
	adsurl = {https://ui.adsabs.harvard.edu/abs/1998ASSL..231..185H},
	author = {{Hamilton}, A.~J.~S.},
	booktitle = {The Evolving Universe},
	doi = {10.1007/978-94-011-4960-0_17},
	editor = {{Hamilton}, Donald},
	pages = {185},
	series = {Astrophysics and Space Science Library},
	title = {{Linear Redshift Distortions: a Review}},
	volume = {231},
	year = {1998}}

@article{Ballinger:1996,
	adsurl = {http://adsabs.harvard.edu/abs/1996MNRAS.282..877B},
	author = {{Ballinger}, W.~E. and {Peacock}, J.~A. and {Heavens}, A.~F.},
	eprint = {arXiv:astro-ph/9605017},
	journal = {\mnras},
	month = oct,
	pages = {877-+},
	title = {{Measuring the cosmological constant with redshift surveys}},
	volume = 282,
	year = 1996}

@article{Alcock:1979,
	adsurl = {https://ui.adsabs.harvard.edu/abs/1979Natur.281..358A},
	author = {{Alcock}, C. and {Paczynski}, B.},
	doi = {10.1038/281358a0},
	journal = {\nat},
	month = oct,
	pages = {358},
	title = {{An evolution free test for non-zero cosmological constant}},
	volume = {281},
	year = 1979}

@article{Okumura:2015,
	adsurl = {http://adsabs.harvard.edu/abs/2015PhRvD..92j3516O},
	archiveprefix = {arXiv},
	author = {{Okumura}, T. and {Hand}, N. and {Seljak}, U. and {Vlah}, Z. and {Desjacques}, V.},
	doi = {10.1103/PhysRevD.92.103516},
	eid = {103516},
	eprint = {1506.05814},
	journal = {\prd},
	month = nov,
	number = 10,
	pages = {103516},
	title = {{Galaxy power spectrum in redshift space: Combining perturbation theory with the halo model}},
	volume = 92,
	year = 2015}

@ARTICLE{Jackson:1972,
       author = {{Jackson}, J.~C.},
        title = "{A critique of Rees's theory of primordial gravitational radiation}",
      journal = {\mnras},
         year = 1972,
        month = jan,
       volume = {156},
        pages = {1P},
          doi = {10.1093/mnras/156.1.1P},
archivePrefix = {arXiv},
       eprint = {0810.3908},
 primaryClass = {astro-ph},
       adsurl = {https://ui.adsabs.harvard.edu/abs/1972MNRAS.156P...1J}
}

@article{Reid:2009,
	adsurl = {http://adsabs.harvard.edu/abs/2009ApJ...698..143R},
	archiveprefix = {arXiv},
	author = {{Reid}, B.~A. and {Spergel}, D.~N.},
	doi = {10.1088/0004-637X/698/1/143},
	eprint = {0809.4505},
	journal = {\apj},
	month = jun,
	pages = {143-154},
	title = {{Constraining the Luminous Red Galaxy Halo Occupation Distribution Using Counts-In-Cylinders}},
	volume = 698,
	year = 2009}

@ARTICLE{Okumura:2017,
       author = {{Okumura}, Teppei and {Takada}, Masahiro and {More}, Surhud and {Masaki}, Shogo},
        title = "{Reconstruction of halo power spectrum from redshift-space galaxy distribution: cylinder-grouping method and halo exclusion effect}",
      journal = {\mnras},
         year = 2017,
        month = jul,
       volume = {469},
       number = {1},
        pages = {459-475},
          doi = {10.1093/mnras/stx850},
archivePrefix = {arXiv},
       eprint = {1611.04165},
 primaryClass = {astro-ph.CO},
       adsurl = {https://ui.adsabs.harvard.edu/abs/2017MNRAS.469..459O}
}

@ARTICLE{Maus:2025,
       author = {{Maus}, M. and {Chen}, S. and {White}, M. and {Aguilar}, J. and {Ahlen}, S. and {Aviles}, A. and {Brieden}, S. and {Brooks}, D. and {Claybaugh}, T. and {Cole}, S. and {de la Macorra}, A. and {Dey}, Arjun and {Doel}, P. and {Ferraro}, S. and {Findlay}, N. and {Forero-Romero}, J.~E. and {Gazta{\~n}aga}, E. and {Gil-Mar{\'\i}n}, H. and {Gontcho}, S. Gontcho A. and {Hahn}, C. and {Honscheid}, K. and {Howlett}, C. and {Ishak}, M. and {Juneau}, S. and {Kremin}, A. and {Lai}, Y. and {Landriau}, M. and {Levi}, M.~E. and {Manera}, M. and {Miquel}, R. and {Mueller}, E. and {Myers}, A.~D. and {Nadathur}, S. and {Nie}, J. and {Noriega}, H.~E. and {Palanque-Delabrouille}, N. and {Percival}, W.~J. and {Poppett}, C. and {Ramirez-Solano}, S. and {Rezaie}, M. and {Rocher}, A. and {Rossi}, G. and {Sanchez}, E. and {Schlegel}, D. and {Schubnell}, M. and {Seo}, H. and {Sprayberry}, D. and {Tarl{\'e}}, G. and {Vargas-Maga{\~n}a}, M. and {Weaver}, B.~A. and {Yuan}, S. and {Zarrouk}, P. and {Zhang}, H. and {Zhou}, R. and {Zou}, H.},
        title = "{An analysis of parameter compression and Full-Modeling techniques with Velocileptors for DESI 2024 and beyond}",
      journal = {\jcap},
         year = 2025,
        month = jan,
       volume = {01},
       number = {1},
          eid = {138},
        pages = {138},
          doi = {10.1088/1475-7516/2025/01/138},
archivePrefix = {arXiv},
       eprint = {2404.07312},
 primaryClass = {astro-ph.CO},
       adsurl = {https://ui.adsabs.harvard.edu/abs/2025JCAP...01..138M}
}

@ARTICLE{Chen:2020,
       author = {{Chen}, Shi-Fan and {Vlah}, Zvonimir and {White}, Martin},
        title = "{Consistent modeling of velocity statistics and redshift-space distortions in one-loop perturbation theory}",
      journal = {\jcap},
         year = 2020,
        month = jul,
       volume = {07},
       number = {7},
          eid = {062},
        pages = {062},
          doi = {10.1088/1475-7516/2020/07/062},
archivePrefix = {arXiv},
       eprint = {2005.00523},
 primaryClass = {astro-ph.CO},
       adsurl = {https://ui.adsabs.harvard.edu/abs/2020JCAP...07..062C}
}

@ARTICLE{Chen:2021,
       author = {{Chen}, Shi-Fan and {Vlah}, Zvonimir and {Castorina}, Emanuele and {White}, Martin},
        title = "{Redshift-space distortions in Lagrangian perturbation theory}",
      journal = {\jcap},
         year = 2021,
        month = mar,
       volume = {03},
       number = {3},
          eid = {100},
        pages = {100},
          doi = {10.1088/1475-7516/2021/03/100},
archivePrefix = {arXiv},
       eprint = {2012.04636},
 primaryClass = {astro-ph.CO},
       adsurl = {https://ui.adsabs.harvard.edu/abs/2021JCAP...03..100C}
}

@ARTICLE{Carrasco:2012,
       author = {{Carrasco}, John Joseph M. and {Hertzberg}, Mark P. and {Senatore}, Leonardo},
        title = "{The effective field theory of cosmological large scale structures}",
      journal = {Journal of High Energy Physics},
         year = 2012,
        month = sep,
       volume = {2012},
          eid = {82},
        pages = {82},
          doi = {10.1007/JHEP09(2012)082},
archivePrefix = {arXiv},
       eprint = {1206.2926},
 primaryClass = {astro-ph.CO},
       adsurl = {https://ui.adsabs.harvard.edu/abs/2012JHEP...09..082C}
}

@ARTICLE{Porto:2014,
       author = {{Porto}, Rafael A. and {Senatore}, Leonardo and {Zaldarriaga}, Matias},
        title = "{The Lagrangian-space Effective Field Theory of large scale structures}",
      journal = {\jcap},
         year = 2014,
        month = may,
       volume = {05},
       number = {5},
          eid = {022},
        pages = {022},
          doi = {10.1088/1475-7516/2014/05/022},
archivePrefix = {arXiv},
       eprint = {1311.2168},
 primaryClass = {astro-ph.CO},
       adsurl = {https://ui.adsabs.harvard.edu/abs/2014JCAP...05..022P}
}

@ARTICLE{Vlah:2015,
       author = {{Vlah}, Zvonimir and {White}, Martin and {Aviles}, Alejandro},
        title = "{A Lagrangian effective field theory}",
      journal = {\jcap},
         year = 2015,
        month = sep,
       volume = {09},
       number = {9},
        pages = {014-014},
          doi = {10.1088/1475-7516/2015/09/014},
archivePrefix = {arXiv},
       eprint = {1506.05264},
 primaryClass = {astro-ph.CO},
       adsurl = {https://ui.adsabs.harvard.edu/abs/2015JCAP...09..014V}
}

@ARTICLE{Maksimova:2021,
       author = {{Maksimova}, Nina A. and {Garrison}, Lehman H. and {Eisenstein}, Daniel J. and {Hadzhiyska}, Boryana and {Bose}, Sownak and {Satterthwaite}, Thomas P.},
        title = "{ABACUSSUMMIT: a massive set of high-accuracy, high-resolution N-body simulations}",
      journal = {\mnras},
         year = 2021,
        month = dec,
       volume = {508},
       number = {3},
        pages = {4017-4037},
          doi = {10.1093/mnras/stab2484},
archivePrefix = {arXiv},
       eprint = {2110.11398},
 primaryClass = {astro-ph.CO},
       adsurl = {https://ui.adsabs.harvard.edu/abs/2021MNRAS.508.4017M}
}

@ARTICLE{Brieden:2021,
       author = {{Brieden}, Samuel and {Gil-Mar{\'\i}n}, H{\'e}ctor and {Verde}, Licia},
        title = "{ShapeFit: extracting the power spectrum shape information in galaxy surveys beyond BAO and RSD}",
      journal = {\jcap},
         year = 2021,
        month = dec,
       volume = {12},
       number = {12},
          eid = {054},
        pages = {054},
          doi = {10.1088/1475-7516/2021/12/054},
archivePrefix = {arXiv},
       eprint = {2106.07641},
 primaryClass = {astro-ph.CO},
       adsurl = {https://ui.adsabs.harvard.edu/abs/2021JCAP...12..054B}
}

@ARTICLE{Hamilton:1992,
       author = {{Hamilton}, A.~J.~S.},
        title = "{Measuring Omega and the Real Correlation Function from the Redshift Correlation Function}",
      journal = {\apjl},
         year = 1992,
        month = jan,
       volume = {385},
        pages = {L5},
          doi = {10.1086/186264},
       adsurl = {https://ui.adsabs.harvard.edu/abs/1992ApJ...385L...5H}
}

@ARTICLE{Cole:1994,
       author = {{Cole}, S. and {Fisher}, K.~B. and {Weinberg}, D.~H.},
        title = "{Fourier Analysis of Redshift Space Distortions and the Determination of Omega}",
      journal = {\mnras},
         year = 1994,
        month = apr,
       volume = {267},
        pages = {785},
          doi = {10.1093/mnras/267.3.785},
archivePrefix = {arXiv},
       eprint = {astro-ph/9308003},
 primaryClass = {astro-ph},
       adsurl = {https://ui.adsabs.harvard.edu/abs/1994MNRAS.267..785C}
}

@ARTICLE{Baldauf:2013,
       author = {{Baldauf}, Tobias and {Seljak}, Uro{\v{s}} and {Smith}, Robert E. and {Hamaus}, Nico and {Desjacques}, Vincent},
        title = "{Halo stochasticity from exclusion and nonlinear clustering}",
      journal = {\prd},
         year = 2013,
        month = oct,
       volume = {88},
       number = {8},
          eid = {083507},
        pages = {083507},
          doi = {10.1103/PhysRevD.88.083507},
archivePrefix = {arXiv},
       eprint = {1305.2917},
 primaryClass = {astro-ph.CO},
       adsurl = {https://ui.adsabs.harvard.edu/abs/2013PhRvD..88h3507B}
}

@ARTICLE{vdBosch:2013,
       author = {{van den Bosch}, Frank C. and {More}, Surhud and {Cacciato}, Marcello and {Mo}, Houjun and {Yang}, Xiaohu},
        title = "{Cosmological constraints from a combination of galaxy clustering and lensing - I. Theoretical framework}",
      journal = {\mnras},
         year = 2013,
        month = apr,
       volume = {430},
       number = {2},
        pages = {725-746},
          doi = {10.1093/mnras/sts006},
archivePrefix = {arXiv},
       eprint = {1206.6890},
 primaryClass = {astro-ph.CO},
       adsurl = {https://ui.adsabs.harvard.edu/abs/2013MNRAS.430..725V}
}

@ARTICLE{Yuan:2022,
       author = {{Yuan}, Sihan and {Hadzhiyska}, Boryana and {Bose}, Sownak and {Eisenstein}, Daniel J.},
        title = "{Illustrating galaxy-halo connection in the DESI era with ILLUSTRISTNG}",
      journal = {\mnras},
         year = 2022,
        month = jun,
       volume = {512},
       number = {4},
        pages = {5793-5811},
          doi = {10.1093/mnras/stac830},
archivePrefix = {arXiv},
       eprint = {2202.12911},
 primaryClass = {astro-ph.CO},
       adsurl = {https://ui.adsabs.harvard.edu/abs/2022MNRAS.512.5793Y}
}

@ARTICLE{Zheng:2007,
       author = {{Zheng}, Zheng and {Coil}, Alison L. and {Zehavi}, Idit},
        title = "{Galaxy Evolution from Halo Occupation Distribution Modeling of DEEP2 and SDSS Galaxy Clustering}",
      journal = {\apj},
         year = 2007,
        month = oct,
       volume = {667},
       number = {2},
        pages = {760-779},
          doi = {10.1086/521074},
archivePrefix = {arXiv},
       eprint = {astro-ph/0703457},
 primaryClass = {astro-ph},
       adsurl = {https://ui.adsabs.harvard.edu/abs/2007ApJ...667..760Z}
}

@ARTICLE{Simpson:2013,
       author = {{Simpson}, Fergus and {Heavens}, Alan F. and {Heymans}, Catherine},
        title = "{Clipping the cosmos. II. Cosmological information from nonlinear scales}",
      journal = {\prd},
         year = 2013,
        month = oct,
       volume = {88},
       number = {8},
          eid = {083510},
        pages = {083510},
          doi = {10.1103/PhysRevD.88.083510},
archivePrefix = {arXiv},
       eprint = {1306.6349},
 primaryClass = {astro-ph.CO},
       adsurl = {https://ui.adsabs.harvard.edu/abs/2013PhRvD..88h3510S}
}

@ARTICLE{DESI:2024VII,
       author = {{Adame}, A.~G. and {Aguilar}, J. and {Ahlen}, S. and {Alam}, S. and {Alexander}, D.~M. and {Allende Prieto}, C. and {Alvarez}, M. and {Alves}, O. and {Anand}, A. and {Andrade}, U. and {Armengaud}, E. and {Avila}, S. and {Aviles}, A. and {Awan}, H. and {Bahr-Kalus}, B. and {Bailey}, S. and {Baltay}, C. and {Bault}, A. and {Behera}, J. and {BenZvi}, S. and {Beutler}, F. and {Bianchi}, D. and {Blake}, C. and {Blum}, R. and {Bonici}, M. and {Brieden}, S. and {Brodzeller}, A. and {Brooks}, D. and {Buckley-Geer}, E. and {Burtin}, E. and {Calderon}, R. and {Canning}, R. and {Carnero Rosell}, A. and {Cereskaite}, R. and {Cervantes-Cota}, J.~L. and {Chabanier}, S. and {Chaussidon}, E. and {Chaves-Montero}, J. and {Chebat}, D. and {Chen}, S. and {Chen}, X. and {Claybaugh}, T. and {Cole}, S. and {Cuceu}, A. and {Davis}, T.~M. and {Dawson}, K. and {de la Macorra}, A. and {de Mattia}, A. and {Deiosso}, N. and {Dey}, A. and {Dey}, B. and {Ding}, Z. and {Doel}, P. and {Edelstein}, J. and {Eftekharzadeh}, S. and {Eisenstein}, D.~J. and {Elbers}, W. and {Elliott}, A. and {Fagrelius}, P. and {Fanning}, K. and {Ferraro}, S. and {Ereza}, J. and {Findlay}, N. and {Flaugher}, B. and {Font-Ribera}, A. and {Forero-S{\'a}nchez}, D. and {Forero-Romero}, J.~E. and {Frenk}, C.~S. and {Garcia-Quintero}, C. and {Garrison}, L.~H. and {Gazta{\~n}aga}, E. and {Gil-Mar{\'\i}n}, H. and {Gontcho}, S. Gontcho A. and {Gonzalez-Morales}, A.~X. and {Gonzalez-Perez}, V. and {Gordon}, C. and {Green}, D. and {Gruen}, D. and {Gsponer}, R. and {Gutierrez}, G. and {Guy}, J. and {Hadzhiyska}, B. and {Hahn}, C. and {Hanif}, M.~M.~S. and {Herrera-Alcantar}, H.~K. and {Honscheid}, K. and {Howlett}, C. and {Huterer}, D. and {Ir{\v{s}}i{\v{c}}}, V. and {Ishak}, M. and {Joyce}, R. and {Juneau}, S. and {Kara{\c{c}}ayl{\i}}, N.~G. and {Kehoe}, R. and {Kent}, S. and {Kirkby}, D. and {Kong}, H. and {Koposov}, S.~E. and {Kremin}, A. and {Krolewski}, A. and {Lahav}, O. and {Lai}, Y. and {Lan}, T.-W. and {Landriau}, M. and {Lang}, D. and {Lasker}, J. and {Le Goff}, J.~M. and {Le Guillou}, L. and {Leauthaud}, A. and {Levi}, M.~E. and {Li}, T.~S. and {Lodha}, K. and {Magneville}, C. and {Manera}, M. and {Margala}, D. and {Martini}, P. and {Matthewson}, W. and {Maus}, M. and {McDonald}, P. and {Medina-Varela}, L. and {Meisner}, A. and {Mena-Fern{\'a}ndez}, J. and {Miquel}, R. and {Moon}, J. and {Moore}, S. and {Moustakas}, J. and {Mudur}, N. and {Mueller}, E. and {Mu{\~n}oz-Guti{\'e}rrez}, A. and {Myers}, A.~D. and {Nadathur}, S. and {Napolitano}, L. and {Neveux}, R. and {Newman}, J.~A. and {Nguyen}, N.~M. and {Nie}, J. and {Niz}, G. and {Noriega}, H.~E. and {Padmanabhan}, N. and {Paillas}, E. and {Palanque-Delabrouille}, N. and {Pan}, J. and {Penmetsa}, S. and {Percival}, W.~J. and {Pieri}, M.~M. and {Pinon}, M. and {Poppett}, C. and {Porredon}, A. and {Prada}, F. and {P{\'e}rez-Fern{\'a}ndez}, A. and {P{\'e}rez-R{\`a}fols}, I. and {Rabinowitz}, D. and {Raichoor}, A. and {Ram{\'\i}rez-P{\'e}rez}, C. and {Ramirez-Solano}, S. and {Rashkovetskyi}, M. and {Ravoux}, C. and {Rezaie}, M. and {Rich}, J. and {Rocher}, A. and {Rockosi}, C. and {Roe}, N.~A. and {Rosado-Marin}, A. and {Ross}, A.~J. and {Rossi}, G. and {Ruggeri}, R. and {Ruhlmann-Kleider}, V. and {Samushia}, L. and {Sanchez}, E. and {Saulder}, C. and {Schlafly}, E.~F. and {Schlegel}, D. and {Schubnell}, M. and {Seo}, H. and {Shafieloo}, A. and {Sharples}, R. and {Silber}, J. and {Slosar}, A. and {Smith}, A. and {Sprayberry}, D. and {Tan}, T. and {Tarl{\'e}}, G. and {Taylor}, P. and {Trusov}, S. and {Vaisakh}, R. and {Valcin}, D. and {Valdes}, F. and {Valogiannis}, G. and {Vargas-Maga{\~n}a}, M. and {Verde}, L. and {Walther}, M. and {Wang}, B. and {Wang}, M.~S. and {Weaver}, B.~A. and {Weaverdyck}, N. and {Wechsler}, R.~H. and {Weinberg}, D.~H. and {White}, M. and {Wilson}, M.~J. and {Yi}, L.},
        title = "{DESI 2024 VII: cosmological constraints from the full-shape modeling of clustering measurements}",
      journal = {\jcap},
         year = 2025,
        month = jul,
       volume = {07},
       number = {7},
          eid = {028},
        pages = {028},
          doi = {10.1088/1475-7516/2025/07/028},
archivePrefix = {arXiv},
       eprint = {2411.12022},
 primaryClass = {astro-ph.CO},
       adsurl = {https://ui.adsabs.harvard.edu/abs/2025JCAP...07..028A}
}

@ARTICLE{Kaiser:1987,
       author = {{Kaiser}, Nick},
        title = "{Clustering in real space and in redshift space}",
      journal = {\mnras},
         year = 1987,
        month = jul,
       volume = {227},
        pages = {1-21},
          doi = {10.1093/mnras/227.1.1},
       adsurl = {https://ui.adsabs.harvard.edu/abs/1987MNRAS.227....1K}
}

@ARTICLE{Guzzo:2008,
       author = {{Guzzo}, L. and {Pierleoni}, M. and {Meneux}, B. and {Branchini}, E. and {Le F{\`e}vre}, O. and {Marinoni}, C. and {Garilli}, B. and {Blaizot}, J. and {De Lucia}, G. and {Pollo}, A. and {McCracken}, H.~J. and {Bottini}, D. and {Le Brun}, V. and {Maccagni}, D. and {Picat}, J.~P. and {Scaramella}, R. and {Scodeggio}, M. and {Tresse}, L. and {Vettolani}, G. and {Zanichelli}, A. and {Adami}, C. and {Arnouts}, S. and {Bardelli}, S. and {Bolzonella}, M. and {Bongiorno}, A. and {Cappi}, A. and {Charlot}, S. and {Ciliegi}, P. and {Contini}, T. and {Cucciati}, O. and {de la Torre}, S. and {Dolag}, K. and {Foucaud}, S. and {Franzetti}, P. and {Gavignaud}, I. and {Ilbert}, O. and {Iovino}, A. and {Lamareille}, F. and {Marano}, B. and {Mazure}, A. and {Memeo}, P. and {Merighi}, R. and {Moscardini}, L. and {Paltani}, S. and {Pell{\`o}}, R. and {Perez-Montero}, E. and {Pozzetti}, L. and {Radovich}, M. and {Vergani}, D. and {Zamorani}, G. and {Zucca}, E.},
        title = "{A test of the nature of cosmic acceleration using galaxy redshift distortions}",
      journal = {\nat},
         year = 2008,
        month = jan,
       volume = {451},
       number = {7178},
        pages = {541-544},
          doi = {10.1038/nature06555},
archivePrefix = {arXiv},
       eprint = {0802.1944},
 primaryClass = {astro-ph},
       adsurl = {https://ui.adsabs.harvard.edu/abs/2008Natur.451..541G}
}

@ARTICLE{Handley:2019,
       author = {{Handley}, Will and {Lemos}, Pablo},
        title = "{Quantifying tensions in cosmological parameters: Interpreting the DES evidence ratio}",
      journal = {\prd},
         year = 2019,
        month = aug,
       volume = {100},
       number = {4},
          eid = {043504},
        pages = {043504},
          doi = {10.1103/PhysRevD.100.043504},
archivePrefix = {arXiv},
       eprint = {1902.04029},
 primaryClass = {astro-ph.CO},
       adsurl = {https://ui.adsabs.harvard.edu/abs/2019PhRvD.100d3504H}
}

@ARTICLE{Lemos:2021,
       author = {{Lemos}, P. and {Raveri}, M. and {Campos}, A. and {Park}, Y. and {Chang}, C. and {Weaverdyck}, N. and {Huterer}, D. and {Liddle}, A.~R. and {Blazek}, J. and {Cawthon}, R. and {Choi}, A. and {DeRose}, J. and {Dodelson}, S. and {Doux}, C. and {Gatti}, M. and {Gruen}, D. and {Harrison}, I. and {Krause}, E. and {Lahav}, O. and {MacCrann}, N. and {Muir}, J. and {Prat}, J. and {Rau}, M.~M. and {Rollins}, R.~P. and {Samuroff}, S. and {Zuntz}, J. and {Aguena}, M. and {Allam}, S. and {Annis}, J. and {Avila}, S. and {Bacon}, D. and {Bernstein}, G.~M. and {Bertin}, E. and {Brooks}, D. and {Burke}, D.~L. and {Carnero Rosell}, A. and {Carrasco Kind}, M. and {Carretero}, J. and {Castander}, F.~J. and {Conselice}, C. and {Costanzi}, M. and {Crocce}, M. and {Pereira}, M.~E.~S. and {Davis}, T.~M. and {De Vicente}, J. and {Desai}, S. and {Diehl}, H.~T. and {Doel}, P. and {Eckert}, K. and {Eifler}, T.~F. and {Elvin-Poole}, J. and {Everett}, S. and {Evrard}, A.~E. and {Ferrero}, I. and {Fert{\'e}}, A. and {Flaugher}, B. and {Fosalba}, P. and {Frieman}, J. and {Garc{\'\i}a-Bellido}, J. and {Gaztanaga}, E. and {Gerdes}, D.~W. and {Giannantonio}, T. and {Gruendl}, R.~A. and {Gschwend}, J. and {Gutierrez}, G. and {Hartley}, W.~G. and {Hinton}, S.~R. and {Hollowood}, D.~L. and {Honscheid}, K. and {Hoyle}, B. and {Huff}, E.~M. and {James}, D.~J. and {Jarvis}, M. and {Lima}, M. and {Maia}, M.~A.~G. and {March}, M. and {Marshall}, J.~L. and {Martini}, P. and {Melchior}, P. and {Menanteau}, F. and {Miquel}, R. and {Mohr}, J.~J. and {Morgan}, R. and {Myles}, J. and {Ogando}, R.~L.~C. and {Palmese}, A. and {Pandey}, S. and {Paz-Chinch{\'o}n}, F. and {Plazas Malag{\'o}n}, A.~A. and {Rodriguez-Monroy}, M. and {Roodman}, A. and {Sanchez}, E. and {Scarpine}, V. and {Schubnell}, M. and {Secco}, L.~F. and {Serrano}, S. and {Sevilla-Noarbe}, I. and {Smith}, M. and {Soares-Santos}, M. and {Suchyta}, E. and {Swanson}, M.~E.~C. and {Tarle}, G. and {Thomas}, D. and {To}, C. and {Troxel}, M.~A. and {Varga}, T.~N. and {Weller}, J. and {Wester}, W. and {DES Collaboration}},
        title = "{Assessing tension metrics with dark energy survey and Planck data}",
      journal = {\mnras},
         year = 2021,
        month = aug,
       volume = {505},
       number = {4},
        pages = {6179-6194},
          doi = {10.1093/mnras/stab1670},
archivePrefix = {arXiv},
       eprint = {2012.09554},
 primaryClass = {astro-ph.CO},
       adsurl = {https://ui.adsabs.harvard.edu/abs/2021MNRAS.505.6179L}
}

@ARTICLE{Gomez:2022,
       author = {{G{\'o}mez-Valent}, Adri{\`a}},
        title = "{Fast test to assess the impact of marginalization in Monte Carlo analyses and its application to cosmology}",
      journal = {\prd},
         year = 2022,
        month = sep,
       volume = {106},
       number = {6},
          eid = {063506},
        pages = {063506},
          doi = {10.1103/PhysRevD.106.063506},
archivePrefix = {arXiv},
       eprint = {2203.16285},
 primaryClass = {astro-ph.CO},
       adsurl = {https://ui.adsabs.harvard.edu/abs/2022PhRvD.106f3506G}
}

@ARTICLE{Seljak:2012,
       author = {{Seljak}, Uro{\v{s}}},
        title = "{Bias, redshift space distortions and primordial nongaussianity of nonlinear transformations: application to Ly-{\ensuremath{\alpha}} forest}",
      journal = {\jcap},
         year = 2012,
        month = mar,
       volume = {03},
       number = {3},
          eid = {004},
        pages = {004},
          doi = {10.1088/1475-7516/2012/03/004},
archivePrefix = {arXiv},
       eprint = {1201.0594},
 primaryClass = {astro-ph.CO},
       adsurl = {https://ui.adsabs.harvard.edu/abs/2012JCAP...03..004S}
}

@ARTICLE{Brieden:2023,
       author = {{Brieden}, Samuel and {Gil-Mar{\'\i}n}, H{\'e}ctor and {Verde}, Licia},
        title = "{A tale of two (or more) h's}",
      journal = {\jcap},
         year = 2023,
        month = apr,
       volume = {2023},
       number = {4},
          eid = {023},
        pages = {023},
          doi = {10.1088/1475-7516/2023/04/023},
archivePrefix = {arXiv},
       eprint = {2212.04522},
 primaryClass = {astro-ph.CO},
       adsurl = {https://ui.adsabs.harvard.edu/abs/2023JCAP...04..023B}
}

\end{document}